\documentclass[nofootinbib,aps,11pt]{revtex4-1}
\usepackage{graphicx}
\usepackage{subfigure} 
\usepackage{hyperref}
\usepackage{cancel}
\usepackage{amssymb}
\usepackage{textcomp}
\usepackage{amsmath}
\usepackage{bm}
\usepackage{times}
\usepackage{epsfig}
\usepackage{color}
\usepackage{mathrsfs}
\newcommand{\hc}{{\rm h.c.}}
\newcommand{\eV}{{\rm eV}}

\newcommand{\GeV}{{\rm GeV}}
\newcommand{\TeV}{{\rm TeV}}

\newcommand{\BR}{{\rm BR}}

\newcommand{\SM}{{\rm SM}}

\newcommand{\eq}{{\rm eq}}

\begin{document}
\title{\Large  Coscattering Dark Matter in the Inverse Scotogenic Model}
\bigskip
\author{Ang Liu$^{1,4}$}
\email{AL@jnxy.edu.cn}
\author{Zhi-Long Han$^2$}
\email{sps\_hanzl@ujn.edu.cn}
\author{Fei Huang$^{2,3}$}
\email{sps\_huangf@ujn.edu.cn}
\author{Feng-Lan Shao$^4$}
\author{Wei Wang$^{3,5}$}
\affiliation{$^1$School of Physical Science and Electronic Engineering, Jining University, Shandong 273155, China}
\affiliation{$^2$School of Physics and Technology, University of Jinan, Jinan, Shandong 250022, China}
\affiliation{$^3$State Key Laboratory of Dark Matter Physics, School of Physics and Astronomy, Shanghai Jiao Tong University, Shanghai 200240, China}	
\affiliation{$^4$School of Physics and Physical Engineering, Qufu Normal University, Qufu, Shandong 273165, China}  
\affiliation{$^5$Southern Center for Nuclear-Science Theory (SCNT), Institute of Modern Physics, Chinese Academy of Sciences, Huizhou, Guangdong 516000, China}  
\date{\today}
\begin{abstract} 
The Scotogenic mechanism is an appealing pathway to naturally explain the common origin of dark matter and tiny neutrino mass. However, the conventional scotogenic dark matter usually suffers stringent constraints from the non-observation of lepton flavor violation and direct detection. To generate the non-zero neutrino masses, at least two generations of dark particles are required. For example, two real scalar singlets $\phi_1$ and $\phi_2$ are involved in the inverse scotogenic  model, which are odd under the $Z_2$ symmetry. In this paper, we consider the masses of dark scalars are nearly degenerate $m_{\phi_1}\lesssim m_{\phi_2}$, which opens new viable pathway for the generation of dark matter $\phi_1$, such as the coscattering process $\phi_1\text{SM}\to \phi_2 \text{SM}$ and coannihilation processes $\phi_1 \phi_2 \to \text{SM~SM}$ via the Higgs portal or Yukawa portal interactions. We explore the parameter space to produce the correct relic density through coscattering, as well as the contrastive coannihilation channel. We then comprehensively study the constraints of dark matter from Higgs decay, direct detection, and indirect detection. For the heavier dark scalar, the three-body decay $\phi_2\to\phi_1 f\bar{f}$ not only alerts the predictions of big bang nucleosynthesis and cosmic microwave background, but also leads to the observable displaced vertex signature at colliders.
\end{abstract}
\maketitle

\section{Introduction}	

Even the nearly perfect theory of the standard model (SM) has its limitations in addressing certain issues related to neutrino mass and dark matter (DM).   Observations of neutrino oscillations~\cite{Super-Kamiokande:1998kpq, SNO:2002tuh} indicate that neutrinos have tiny masses under the constraint from cosmology $\sum m_\nu<0.12$ eV ~\cite{Planck:2018vyg}. Meanwhile, various astrophysical and cosmological observations support the existence of particle dark matter \cite{Bertone:2004pz,Cirelli:2024ssz}. To obtain an unified theory of new physics beyond SM, the common origin of tiny neutrino mass and dark matter is extensively studied \cite{Dodelson:1993je,Krauss:2002px,Asaka:2005an,Ma:2007gq,Aoki:2008av,Gustafsson:2012vj,Restrepo:2013aga,AristizabalSierra:2014wal,Ma:2015xla,Escudero:2016tzx,Escudero:2016ksa,Cai:2017jrq,Yao:2017vtm,Becker:2018rve,CentellesChulia:2019xky,DeGouvea:2019wpf,Kelly:2020aks,Liu:2022rst,Liu:2022cct,Liu:2023kil,Liu:2023zah,Yang:2025ouc,Avila:2025qsc,Guo:2025xmz,Roy:2025moo}.

Scotogenic mechanism \cite{Tao:1996vb,Ma:2006km} is a fascinating scenario, where tiny neutrino masses are mediated by the dark matter at the loop level. Typically, when the dark matter pair annihilates via the Yukawa portal interactions, it suffers stringent constraints from lepton flavor violation \cite{Kubo:2006yx,Toma:2013zsa}. To satisfy the observed relic density, a hierarchy structure of the dark Yukawa couplings $|y_{1e}|\ll |y_{1\mu}|\lesssim|y_{1\tau}|\sim\mathcal{O}(1)$ is required \cite{Vicente:2014wga,Guo:2020qin}, which can be fully tested at the future muon collider \cite{Liu:2022byu}. It should be noted that such a hierarchy structure heavily depends on the current relatively loose constraints from lepton flavor violation  $\tau$ decays, and also needs fine-tuning of certain parameters to reproduce the neutrino oscillation data \cite{Esteban:2024eli}. 

On the other hand, the scalar dark matter candidates could also annihilate via the additional Higgs or gauge boson portal interactions \cite{GAMBIT:2018eea,Arcadi:2021mag,Avila:2021mwg,Abouabid:2023cdz}. With correct relic density, the scalar dark matter usually induces a relatively large dark matter-nucleon scattering cross section, thus most parameter space is already excluded by current direct detection experiments \cite{XENON:2023cxc,PandaX:2024qfu,LZ:2024zvo}. To avoid the tight constraints of thermal dark matter from lepton flavor violation and direct detection, non-thermal dark matter produced through the freeze-in mechanism is also considered \cite{Molinaro:2014lfa,Borah:2017dfn,Baumholzer:2018sfb,Baumholzer:2019twf}. 

Recently, the coscattering \cite{DAgnolo:2017dbv} or conversion \cite{Garny:2017rxs} mechanism was proposed, where the dark matter is generated through the inelastic scattering of the dark particles \cite{Garny:2018icg,DAgnolo:2018wcn}. With relatively small couplings of the coscattering dark matter to SM particles for relic density, the various constraints can be naturally satisfied, hence this scenario receives increasing interest \cite{Cheng:2018vaj,Junius:2019dci,DAgnolo:2019zkf,Brummer:2019inq,Garny:2021qsr,Filimonova:2022pkj,Acaroglu:2023phy,Heisig:2024xbh,DiazSaez:2024dzx,Zhang:2024sox,Paul:2024prs,Belanger:2025wjh}.  Motivated by the study of coscattering fermion dark matter in Scotogenic model \cite{Heeck:2022rep,Heisig:2024mwr,Sahoo:2026xlh}, we consider the scalar option. Provided scalar dark matter $\phi_1$ coscattering with the fermion doublet partner $\Psi$, the required tiny Yukawa coupling would be similar with the fermion case \cite{Heeck:2022rep}. Therefore, we investigate a new kind of coscattering process, i.e., the coscattering of scalar dark matter $\phi_1$ with the scalar partner $\phi_2$.  Such coscattering scalar case has both Yukawa and Higgs portal interactions, which is different from the fermion case with only Yukawa interactions.

Scalar singlet dark matter exists in various Scotogenic models \cite{Fraser:2014yha,Fraser:2015mhb,Mandal:2019oth}. Although  the explicit phenomenology is model dependent, the results of dark matter are quite similar for these models. The benchmark model selected for this paper is the inverse Scotogenic  model \cite{Fraser:2014yha}. It is well known that to generate the observed two squared mass gaps of light neutrinos, at least two generations of new particles are required \cite{Ma:1998dn}. Therefore, the inverse  Scotogenic  model introduces three scalar singlets $\phi_i(i=1,2,3)$, doublet fermion $\Psi=(\psi^0,\psi^-)^T$, and Majorana fermion $\chi$ into the dark sector \cite{Fraser:2014yha}.
When consider the  nearly degenerate dark scalars $m_{\phi_1}\lesssim m_{\phi_2}$,  the relic density of dark matter $\phi_1$ is determined by the coscattering of dark partner $\phi_2$. Additionally, the coscattering regime of scalar dark matter in the Scotogenic model is distinguishable from the pure Higgs portal models \cite{Ghorbani:2014gka,DiazSaez:2024nrq,Hooper:2025fda,Guo:2025qes} due to more complex interactions.

The structure of this paper is organized as follows. In Section \ref{SEC:TF}, we briefly  explain the theoretical framework. The calculation of  relic density as well as the associated constraints in the Higgs portal scenario are discussed in Section \ref{SEC:HPS}. Next, we investigate the Yukawa portal scenario in Section \ref{SEC:YS}. Finally, we summarize the results in Section \ref{SEC:CL}.

\section{The framework}\label{SEC:TF}
\begin{table}
	\begin{center}\large
		\begin{tabular}{|c|c c c | c c |} 
			\hline
			\hline
			&$\ell_\alpha$ &~$\Psi$~& $\chi$ &~$\phi_i$~ & $H$ \\ \hline
			$SU(2)_L$ &2 &2 & 1&1 &2  \\ \hline
			$U(1)_Y$ &$-\frac{1}{2}$ &$-\frac{1}{2}$ & 0&0 &$\frac{1}{2}$  \\ \hline
			$U(1)_\ell$ &1 &1 & 1&0 &0  \\ \hline
			$Z_2$ &$+$ &$-$ & $-$&$-$ &$+$  \\ \hline
			\hline
		\end{tabular}
	\end{center}
	\caption{Relevant particle contents and the corresponding charge assignments, where $U(1)_\ell$ denotes the lepton number.
		\label{Tab:Particle}}
\end{table}

In this paper, we take the inverse Scotogenic  model \cite{Fraser:2014yha} as the benchmark model to illustrate the coscattering scalar dark matter. The framework contains new particles as: doublet fermion $\Psi\equiv(\psi^0,\psi^-)^T$ with  hypercharge $Y=-1/2$, singlet fermion $\chi$ and three real singlet scalar $\phi_{i}(i=1,2,3)$ with zero vacuum expectation value. These new particles are all charged under the $Z_2$ symmetry, while the SM particles transform trivially.  The dark fermions $\Psi$ and $\chi$ also have lepton number $+1$. The particle contents and the corresponding charge assignments are listed in Table \ref{Tab:Particle}. We assume that $\phi_1$ is the dark matter candidate. To realize the coscattering regime, we further consider the scalar mass spectrum $m_{\phi_1}\lesssim m_{\phi_2} \ll m_{\phi_3}$ for simplicity.  In this way, the dark scalar $\phi_3$ has a negligible impact on the dark matter phenomenon, as it is dynamically irrelevant during freeze-out.

Under the $Z_2$ symmetry, the most general scalar potential of the two nearly degenerate dark scalars $\phi_{1,2}$ can be written as \cite{Casas:2017jjg,Bhattacharya:2017fid}
\begin{eqnarray}
	-\mathcal{L}_{V}&=&
	\left(\frac{\lambda_{1}}{2}\phi_1^2+ \frac{\lambda_{2}}{2}\phi_2^2+\lambda_{12}\phi_1\phi_2\right)H^\dagger H +\frac{\lambda_{22}}{4} \phi_1^2\phi_2^2 \\ \nonumber
	&&+\frac{\lambda_{13}}{6}\phi_1 \phi_2^3 +\frac{\lambda_{31}}{6}\phi_1^3 \phi_2 + \frac{\lambda_{14}}{12}\phi_1^4 +\frac{\lambda_{24}}{12}\phi_2^4,
\end{eqnarray}
where $H$ is the SM Higgs doublet. In following studies, we focus on the SM Higgs portal interactions, and assume vanishing self-interactions of dark scalar for simplicity, i.e., $\lambda_{22}=\lambda_{13}=\lambda_{31}=0$. Including these self-interactions would contribute to the conversion processes \cite{Maity:2019hre}, thus weakening the effect of the Higgs portal interaction.

The Yukawa interaction of the dark scalars is
\begin{eqnarray}\label{Eqn:Yuk}
	\mathcal{L}_{Y}=y_{i\alpha} \phi_i \bar{\Psi} \ell_{\alpha} + \hc,
\end{eqnarray}
where $\ell_\alpha=(\nu_\alpha,\ell^-_\alpha)^T$ is the SM lepton doublets. Supposing small doublet-singlet fermion mixing term $y_\chi \bar{\Psi} \tilde{H} \chi$ \cite{Konar:2020wvl},  we denote $m_F=m_{\psi^\pm}\simeq m_{\psi^0}$ in the following discussion.

The loop induced light neutrino mass is calculated as \cite{Esch:2016jyx}
\begin{equation}\label{Eqn:Nm}
	m^{\nu}_{\alpha \beta} = \sum_{i,k} \frac{y_{i \alpha} y_{i \beta}}{16\pi^2} m_{\chi_k} (\xi_{2k})^2 \frac{m_{\chi_k}^2}{m_{\chi_k}^2-m_{\phi_i}^2}\log\left(\frac{m_{\chi_k}^2}{m_{\phi_i}^2}\right),
\end{equation}
where $\chi_k$ is the mass eigenstate of dark fermions,  and $\xi$ is the corresponding mixing matrix which accurate results are performed by numerical computations. Approximately, the light neutrino mass could be evaluated as 
\begin{equation}\label{Eqn:mves}
	m^{\nu}\sim 0.1~\eV \times \left( \frac{m_F}{1~\text{TeV}} \right) \left(\frac{y_{i\alpha} y_\chi}{10^{-5}}\right)^2.
\end{equation}
For instance, the Yukawa coupling $y_{i\alpha} \sim \mathcal{O}(10^{-5}), y_\chi\sim \mathcal{O}(1)$ and $m_F\sim \mathcal{O}(1)$ TeV can naturally accommodate the tiny neutrino masses,  which corresponds to the typical benchmark of the Higgs portal scenario. 

Then the neutrino mass matrix could be diagonalized  by an unitary matrix $U$  like $\hat{m}_\nu=U^T m^\nu U$, where  $U$ could be  identified as the standard PMNS neutrino mixing matrix, and $\hat{m}_\nu=\text{diag}(m_{\nu1},m_{\nu2},m_{\nu3})$ is the diagonalized neutrino mass matrix. Additionally, by utilizing the Casas-Ibarra parametrization\cite{Casas:2001sr}, the Yukawa coupling $y_{i\alpha}$ can be expressed as 
	\begin{equation}\label{Eq:CIy}
		y_{i\alpha}^T = i U^*\sqrt{\hat{m}_\nu} R \sqrt{F_\phi^{-1}},
	\end{equation}
where $F_\phi=\text{diag}(F_{\phi_1},F_{\phi_2},F_{\phi_3})$ with 
	\begin{equation}
			F_{\phi i}=\sum_{k}\frac{m_{\chi_k}}{16\pi^2}(\xi_{2k})^2\left(\frac{m_{\chi_k}^2}{m_{\chi_k}^2-m_{\phi_{i}^2}}\log\left(\frac{m_{\chi_k}^2}{m_{\phi_{i}}^2}\right)\right),
	\end{equation}
and $v_h=246$ GeV. In this paper, we fix the neutrino oscillation parameters to the best fit values in Ref. \cite{Esteban:2024eli} with vanishing  Majorana phases.  Provided normal mass hierarchy, the neutrino masses are fixed as
	\begin{equation}\label{Eqn:Bfnm}
	m_{\nu1}=10^{-4}~\eV,m_{\nu2}=8.6\times10^{-3}~\eV,m_{\nu3}=5\times10^{-2}~\eV.
	\end{equation} 
$R$ is a complex orthogonal rotation matrix, which can be parametrized through three arbitrary mixing angles $(\omega_{12},\omega_{13},\omega_{23})$. Such arbitrariness allows us to achieve a hierarchical Yukawa coupling $y_{i\alpha}$  by fine-tuning these angles, which is particularly important for implementing the coscattering  mechanism in the Yukawa portal scenario to satisfy the constraints from lepton flavor violation.

\section{Higgs portal scenario}\label{SEC:HPS}

Generally speaking, the scalar dark matter could annihilate via both the Higgs portal and Yukawa portal interactions at the same time. To seek the distinct features of these portals, we discuss them separately. Firstly, we focus on the Higgs portal interactions, which favor the  relation of couplings as $\lambda_{ij}\gg y_{i\alpha}$. The free parameters involved in this situation are
\begin{eqnarray}
	\{m_{\phi_1},\Delta{m_\phi}\equiv m_{\phi_2}-m_{\phi_1},\lambda_1,\lambda_2,\lambda_{12}\}.
\end{eqnarray}

\allowdisplaybreaks

\subsection{Relic Density}\label{SEC:HRD}

In this scenario, the SM Higgs $h$ mediates all annihilation and transformation processes related to $\phi_1$ and $\phi_2$,  the relevant Boltzmann equations can be expressed as
\begin{eqnarray}\label{Eqn:HBE1}
	\frac{dY_{\phi_1}}{dx} &= & -\frac{s}{\mathcal{H}x}\bigg[ \left<\sigma v\right>_{\phi_1\phi_1\to\SM\SM}\left(Y_{\phi_1}^2 -(Y_{\phi_1}^{\eq})^2\right)+\left<\sigma v\right>_{\phi_1\phi_2\to\SM\SM}\left(Y_{\phi_1}Y_{\phi_2}-Y_{\phi_1}^{\eq}Y_{\phi_2}^{\eq}\right)\nonumber \\
	&-& \left<\sigma v\right>_{\phi_2\SM\to\phi_1\SM}\left(Y_{\phi_2}Y_{\SM}^{\eq}-\frac{Y_{\phi_2}^{\eq}}{Y_{\phi_1}^{\eq}}Y_{\phi_1}Y_{\SM}^{\eq}\right)-\left<\sigma v\right>_{\phi_2\phi_2\to\phi_1\phi_1}\left(Y_{\phi_2}^2-\frac{(Y_{\phi_2}^{\eq})^2}{(Y_{\phi_1}^{\eq})^2}Y_{\phi_1}^2
	\right)\nonumber \\
	&-&\left<\sigma v\right>_{\phi_1\phi_2\to\phi_1\phi_1}\left(Y_{\phi_1} Y_{\phi_2}-\frac{Y_{\phi_2}^{\eq}}{Y_{\phi_1}^{\eq}}Y_{\phi_1}^2\right)-\left<\sigma v\right>_{\phi_2\phi_2\to\phi_1\phi_2}\left(Y_{\phi_2}^2-\frac{Y_{\phi_2}^{\eq}}{Y_{\phi_1}^{\eq}}Y_{\phi_1}Y_{\phi_2}\right)
	\nonumber \\
	&-&\frac{\Gamma_{\phi_2\to\phi_1\bar{f}f}}{s}\left(Y_{\phi_2}-\frac{Y_{\phi_2}^{\eq}}{Y_{\phi_1}^{\eq}}Y_{\phi_1}\right)\bigg],
\end{eqnarray}
\begin{eqnarray}\label{Eqn:HBE2}
	\frac{dY_{\phi_2}}{dx} &= & -\frac{s}{\mathcal{H}x}\bigg[ \left<\sigma v\right>_{\phi_2\phi_2\to\SM\SM}\left(Y_{\phi_2}^2 -(Y_{\phi_2}^{\eq})^2\right)+\left<\sigma v\right>_{\phi_1\phi_2\to\SM\SM}\left(Y_{\phi_1}Y_{\phi_2}-Y_{\phi_1}^{\eq}Y_{\phi_2}^{\eq}\right)\nonumber \\
	&+& \left<\sigma v\right>_{\phi_2\SM\to\phi_1\SM}\left(Y_{\phi_2}Y_{\SM}^{\eq}-\frac{Y_{\phi_2}^{\eq}}{Y_{\phi_1}^{\eq}}Y_{\phi_1}Y_{\SM}^{\eq}\right)+\left<\sigma v\right>_{\phi_2\phi_2\to\phi_1\phi_1}\left(Y_{\phi_2}^2-\frac{(Y_{\phi_2}^{\eq})^2}{(Y_{\phi_1}^{\eq})^2}Y_{\phi_1}^2
	\right)\nonumber \\
	&+&\left<\sigma v\right>_{\phi_1\phi_2\to\phi_1\phi_1}\left(Y_{\phi_1} Y_{\phi_2}-\frac{Y_{\phi_2}^{\eq}}{Y_{\phi_1}^{\eq}}Y_{\phi_1}^2\right)+\left<\sigma v\right>_{\phi_2\phi_2\to\phi_1\phi_2}\left(Y_{\phi_2}^2-\frac{Y_{\phi_2}^{\eq}}{Y_{\phi_1}^{\eq}}Y_{\phi_1}Y_{\phi_2}\right)
	\nonumber \\
	&+&\frac{\Gamma_{\phi_2\to\phi_1\bar{f}f}}{s}\left(Y_{\phi_2}-\frac{Y_{\phi_2}^{\eq}}{Y_{\phi_1}^{\eq}}Y_{\phi_1}\right)\bigg],
\end{eqnarray}
where $x=m_{\phi_1}/T$, the entropy density $s=2\pi^2 g_s T^3/45$. The
Hubble expansion rate is defined as $\mathcal{H}=\sqrt{4\pi^3g_*/45} T^2/m_{pl}$ with the Planck mass $m_{pl}=1.22\times10^{19}~\GeV$. Here, $g_s$  and $g_\star$ are the number of relativistic degrees of freedom for the entropy density and energy density, respectively. The thermal average cross sections $\left<\sigma v\right>$ of various channels are calculated numerically by micrOMEGAs~\cite{Belanger:2013oya,Alguero:2022inz}. The thermal decay width is denoted as:
\begin{eqnarray}
	\Gamma_{\phi_2\to\phi_1\bar{f}f}&=&\frac{\mathcal{K}_1\left(\frac{m_{\phi_2}}{m_{\phi_1}}x\right)}{\mathcal{K}_2\left(\frac{m_{\phi_2}}{m_{\phi_1}}x\right)}\tilde{\Gamma}_{\phi_2\to\phi_1\bar{f}f}.
\end{eqnarray}
where the decay width mediated by the SM Higgs is ~\cite{Guo:2021vpb}
\begin{eqnarray}\label{Eqn:HG2}
\tilde{\Gamma}_{\phi_2\to\phi_1\bar{f}f}\simeq\frac{\lambda_{12}^2 m_f^2 (m_{\phi_1}+\Delta{m_\phi})^3 (\Delta{m_\phi})^5}{240\pi^3m_h^4m_{\phi_1}^5}\times\theta^\prime(\Delta{m_\phi}-2m_f),
\end{eqnarray}
and $\mathcal{K}_{1,2}$ are modified Bessel functions of the second kind. Here, $f$ and $h$ stand for the SM fermion and Higgs. $\theta^\prime$ is the Heaviside theta function. The abundance of $\phi_1,\phi_2$ at thermally equilibrium  can be expressed as~\cite{Alguero:2022inz}
\begin{eqnarray}
 Y_{\phi_1}^{\eq}=\frac{45x^2}{4\pi^4g_s}\mathcal{K}_2(x),~Y_{\phi_2}^{\eq}=\frac{45x^2m_{\phi_2}^2}{4\pi^4g_sm_{\phi_1}^2}\mathcal{K}_2\left(\frac{m_{\phi_2}}{m_{\phi_1}}x\right).
\end{eqnarray}
$Y_{\SM}^{\eq}$ take the value of 0.238.  In the Higgs portal scenario, the Boltzmann Equations \eqref{Eqn:HBE1} and \eqref{Eqn:HBE2} are evolved numerically up to $x=10^5$ to make sure the abundance of dark matter $Y_{\phi_1}$ fully converged.

\begin{figure} 
	\begin{center}
		\includegraphics[width=0.45\linewidth]{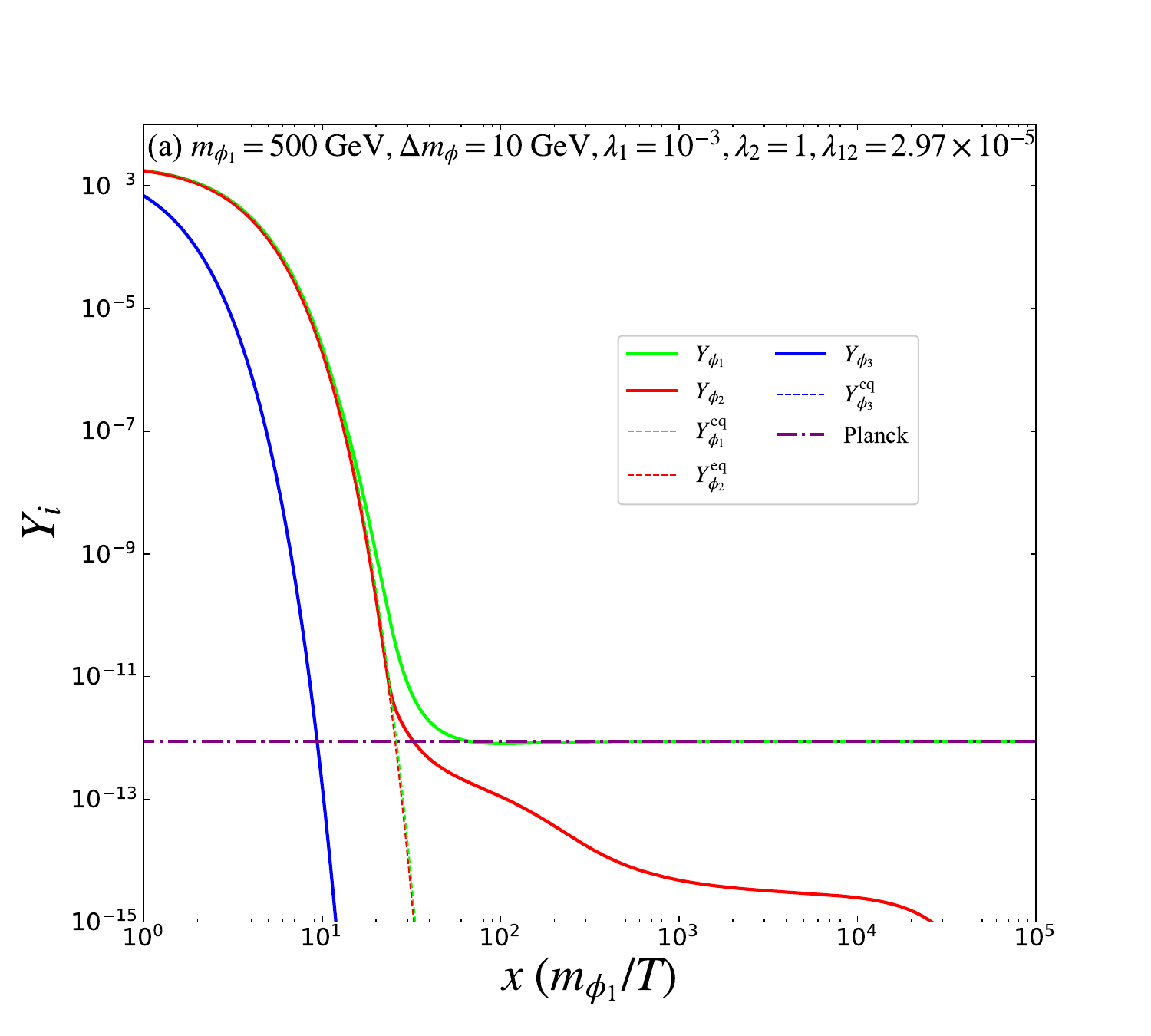}
		\includegraphics[width=0.45\linewidth]{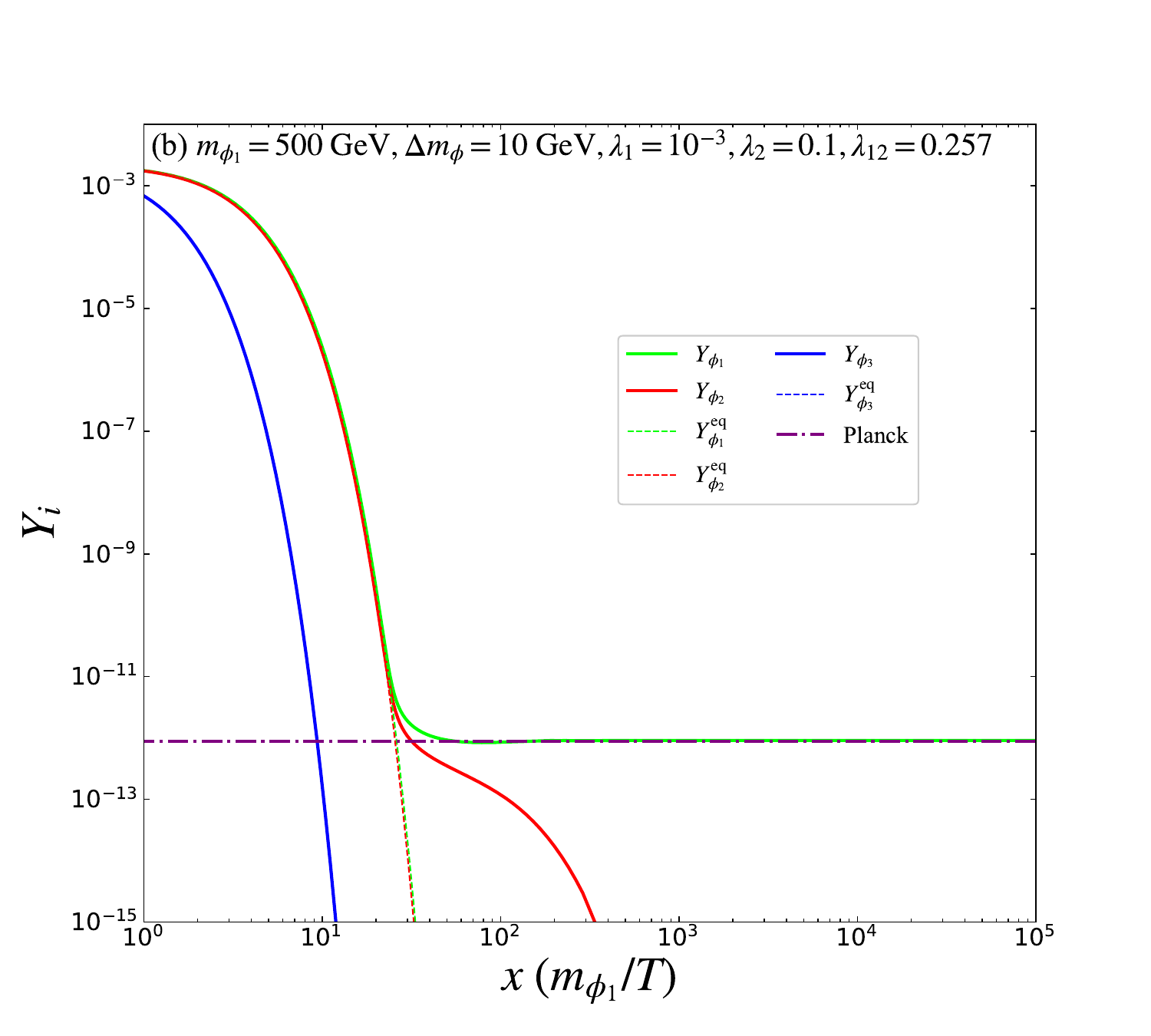}
		\includegraphics[width=0.45\linewidth]{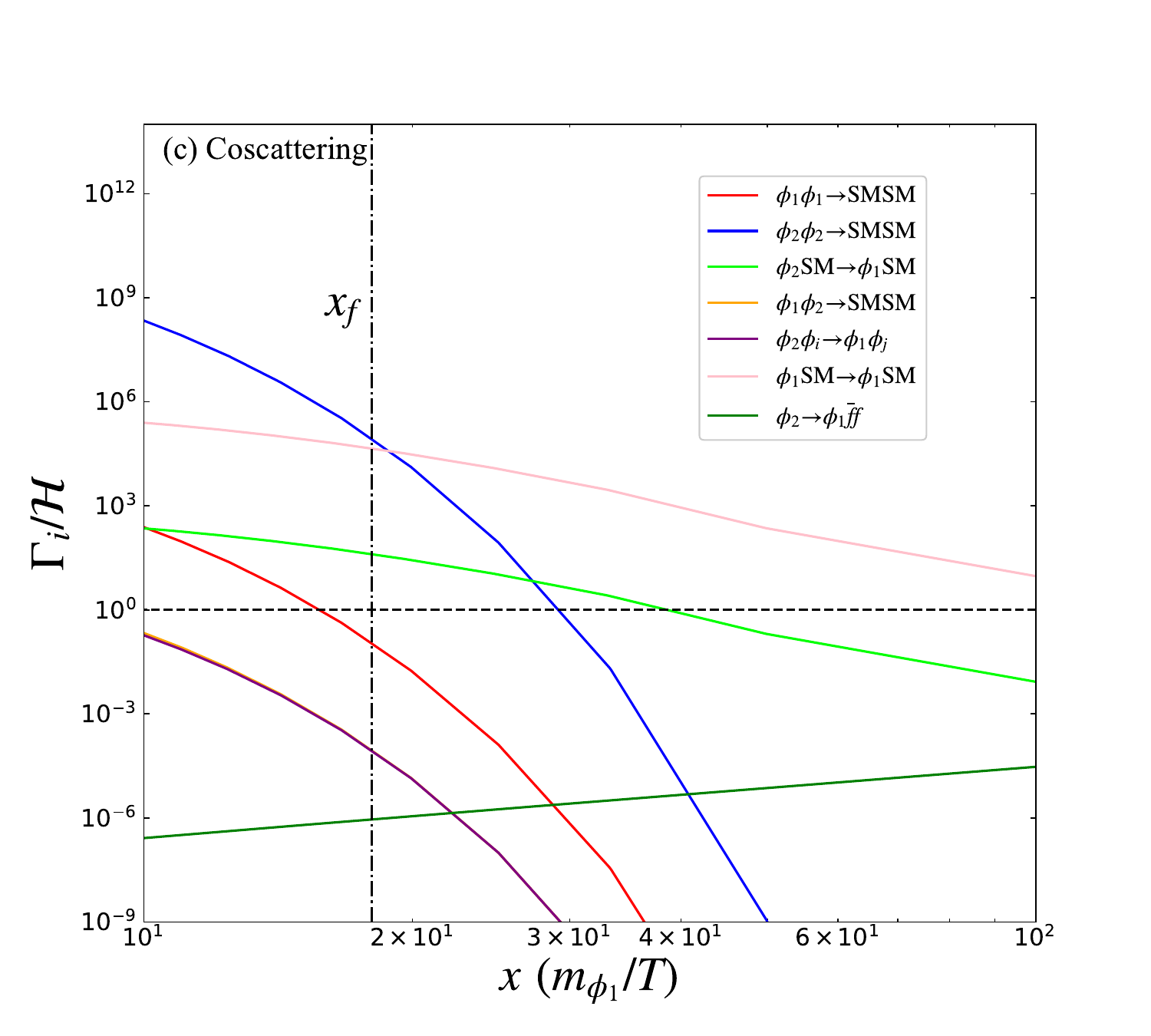}
		\includegraphics[width=0.45\linewidth]{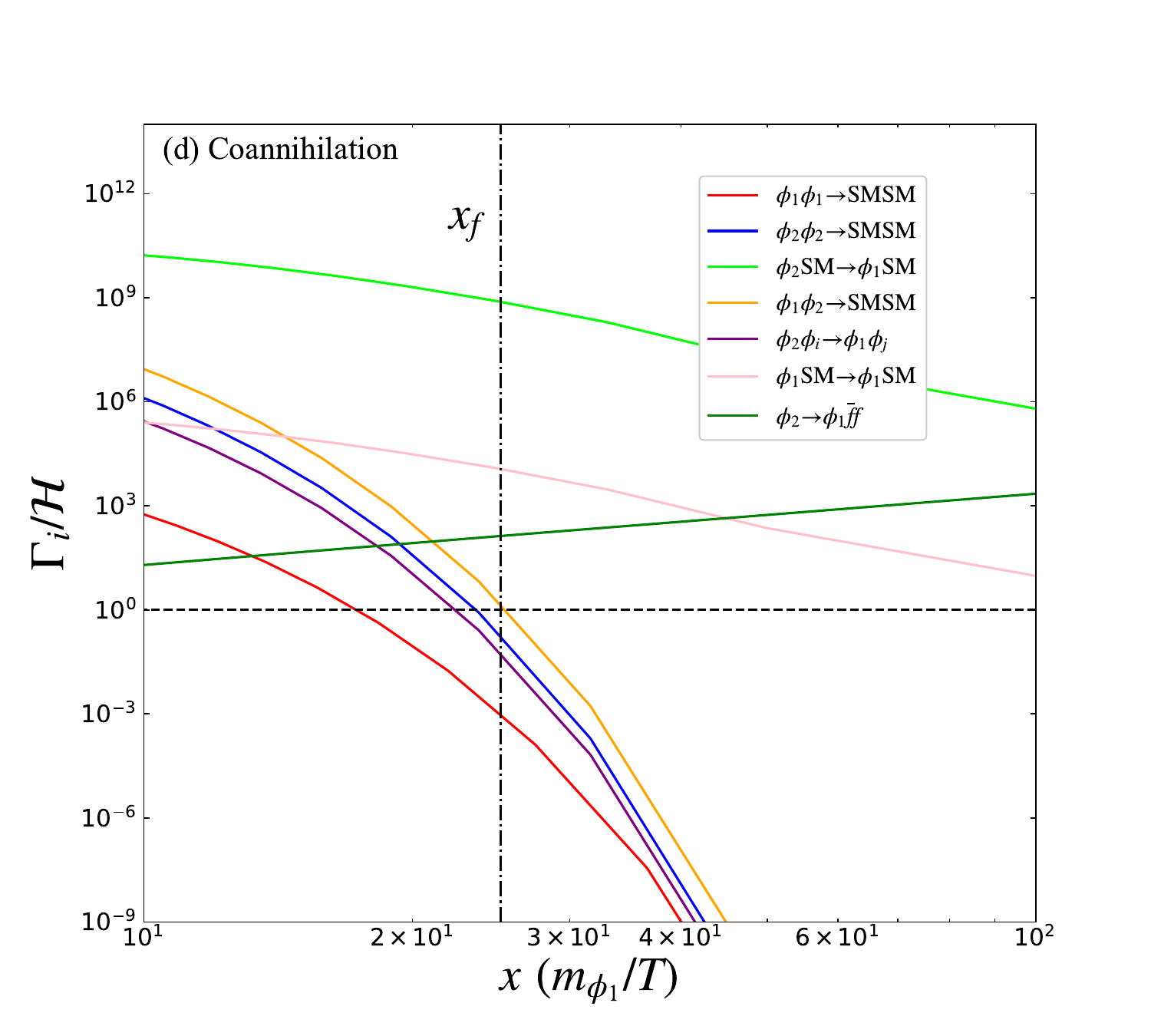}
	\end{center}
	\caption{The evolutions of various abundances $Y_i$ of coscattering (a) and coannihilation (b) benchmarks in the Higgs portal scenario. In panels (a) and (b),  the solid red, green  and blue lines represent the abundance evolution of $\phi_1$, $\phi_2$ and $\phi_3$, while the corresponding dashed lines indicate their respective thermal equilibrium. Purple dot-dashed line stands for the observation of DM~\cite{Planck:2018vyg}. Panels (c) and (d) describe the thermal rates of relevant interactions in panels (a) and  (b), respectively. Additionally, the conversion channels $\phi_2\phi_2\to\phi_1\phi_1$, $\phi_1\phi_2\to\phi_1\phi_1$ and $\phi_2\phi_2\to\phi_1\phi_2$ are added together as $\phi_2\phi_i\to\phi_1\phi_j$.  The black vertical dashed line corresponds to the thermal decoupling temperature when $Y_{\phi_1}/Y_{\phi_1}^\eq=2.5$. The horizontal black line is $\Gamma_i=\mathcal{H}$.
	}
	\label{FIG:fig1}
\end{figure}

In Figure \ref{FIG:fig1}, we illustrate two benchmark points for the coscattering and coannihilation mechanisms in panels (a) and (b), respectively, which could be assessed through the corresponding thermal rates below in panels (c) and (d). The coscattering will occur when the following conditions are met at the freeze-out temperature \cite{DAgnolo:2019zkf}:
(1) there is no chemical potential for $\phi_1$, (2) the last reactions to decouple which changes the number density of $\phi_1$ are exchange reactions between $\phi_1$ and $\phi_2$.  Furthermore, we assume that $\phi_1$ is in kinetic equilibrium at the freeze-out temperature through rapid  energy exchange with SM, which could be quantitatively  parameterized as $\Gamma_{\phi_{1,2}\SM\to\phi_1\SM}\gg\mathcal{H}$. The precise results require solving the full unintegrated Boltzmann equations, which may introduce an $\mathcal{O}(10\%)$ distinction  compared to $\phi_1$ not being in kinetic equilibrium \cite{Garny:2017rxs}. The coscattering scenario shown in panel (a) and (c) of Figure \ref{FIG:fig1} satisfies all these requirements. Specifically, adequate annihilation of $\phi_2$, namely, $\Gamma_{\phi_2\phi_2\to\SM\SM}>\Gamma_{\phi_2\SM\to\phi_1\SM}$, enables condition (1) to be satisfied. As for condition (2),  it suffices that $\Gamma_{\phi_2\SM\to\phi_1\SM}$ is greater than $\Gamma_{\phi_1\phi_1\to\SM\SM}$. The other exchange reactions, such as $\phi_2\phi_i\to\phi_1\phi_j$ and  $\phi_2\to\phi_1\bar{f}f$, provide tiny contributions. In contrast, the reduction of $\lambda_{2}$ leads to a significant increase of $\lambda_{12}$ in order to meet the observation of dark matter. The abundance of dark matter is determined by the process $\phi_1\phi_2\to \SM \SM$, thus the benchmark point in panels (b) and (d) of Figure \ref{FIG:fig1} belongs to the coannihilation scenario.

From the evolutions of abundances in Figure \ref{FIG:fig1}, a notable distinction is that $\phi_1$ decouples prematurely in the coscattering scenario, which occurs roughly at $\Gamma_{\phi_1\phi_1\to\SM\SM}\sim \mathcal{H}$. However, the abundance $Y_{\phi_1}$ subsequently continues to show a rapid decline until the depletion of the coscattering reaction $\phi_2\SM\to\phi_1\SM$.  Such an evolutionary trend of $Y_{\phi_1}$ in the  coscattering case in our work is consistent with the result in Ref.~\cite{Alguero:2022inz}.  Meanwhile, $Y_{\phi_2}$ continuously decreases due to the decline in conversion rate and the decay begins when $x\sim\mathcal{O}(10^4)$, so the suppressed magnitude  has a negligible impact on dark matter. By comparison, the abundance $Y_{\phi_1}$ quickly approaches a constant value after it deviates from thermal equilibrium when $\Gamma_{\phi_1\phi_2\to\SM\SM}\sim \mathcal{H}$ in the coannihilation case,  and the rapid decay of $\phi_{2}$ has little impact on $Y_{\phi_1}$. For large hierarchical  $\phi_3$ with sizable Higgs portal coupling, such as $m_{\phi_3}=1600~\GeV$ and $\lambda_3=1$, it  would render $\phi_3$ nearly to be a thermal bath particle, which naturally does not affect the conversion system composed of $\phi_1$ and $\phi_2$. 

 Moreover, the benchmark points considered here can simultaneously satisfy the neutrino mass constraints by the parametrization of Yukawa coupling in Equation~\eqref{Eq:CIy}. To make sure that the Yukawa portal has a negligible contribution compared to the Higgs portal scenario, a relatively small Yukawa coupling is required. For instance, the following benchmark Yukawa coupling is obtained 
\begin{align}\label{Eqn:y1}
|y_{i\alpha}| \simeq 10^{-6}\begin{pmatrix}
	8.1\times10^{-2} && 6.9\times10^{-2} && 2\times10^{-1} \\
	1.6 && 3.7 && 4.7 \\
	11 && 13 && 14
\end{pmatrix},
\end{align}
with the additional parameters $m_F=1500~\GeV,~y_\chi=0.87$ and  $\omega_{12}=\omega_{13}=\omega_{23}=0.01+0.01i$. 

In fact, within this Higgs portal scenario, for all benchmarks satisfying both dark matter observation and neutrino mass constraints, the resulting coupling $y_{i\alpha}$ can be significantly smaller than $\lambda_{ij}$ when the three mixing angles $(\omega_{12},\omega_{13},\omega_{23})$  have both real and imaginary parts much smaller than 1. The light neutrino mass is typically suppressed by the smallness of the Yukawa coupling $y_{i \alpha}$,   which can be verified by Equations \eqref{Eqn:mves}. Since the benchmark Yukawa $y_{i\alpha}$ in Equation \eqref{Eqn:y1} exhibits no significant hierarchies, we substitute its values into Equation \eqref{Eqn:mves}, yielding an approximate neutrino mass in the range of $\mathcal{O}(0.01)~\eV-\mathcal{O}(0.1)~\eV$. 

\begin{figure} 
	\begin{center}
		\includegraphics[width=0.45\linewidth]{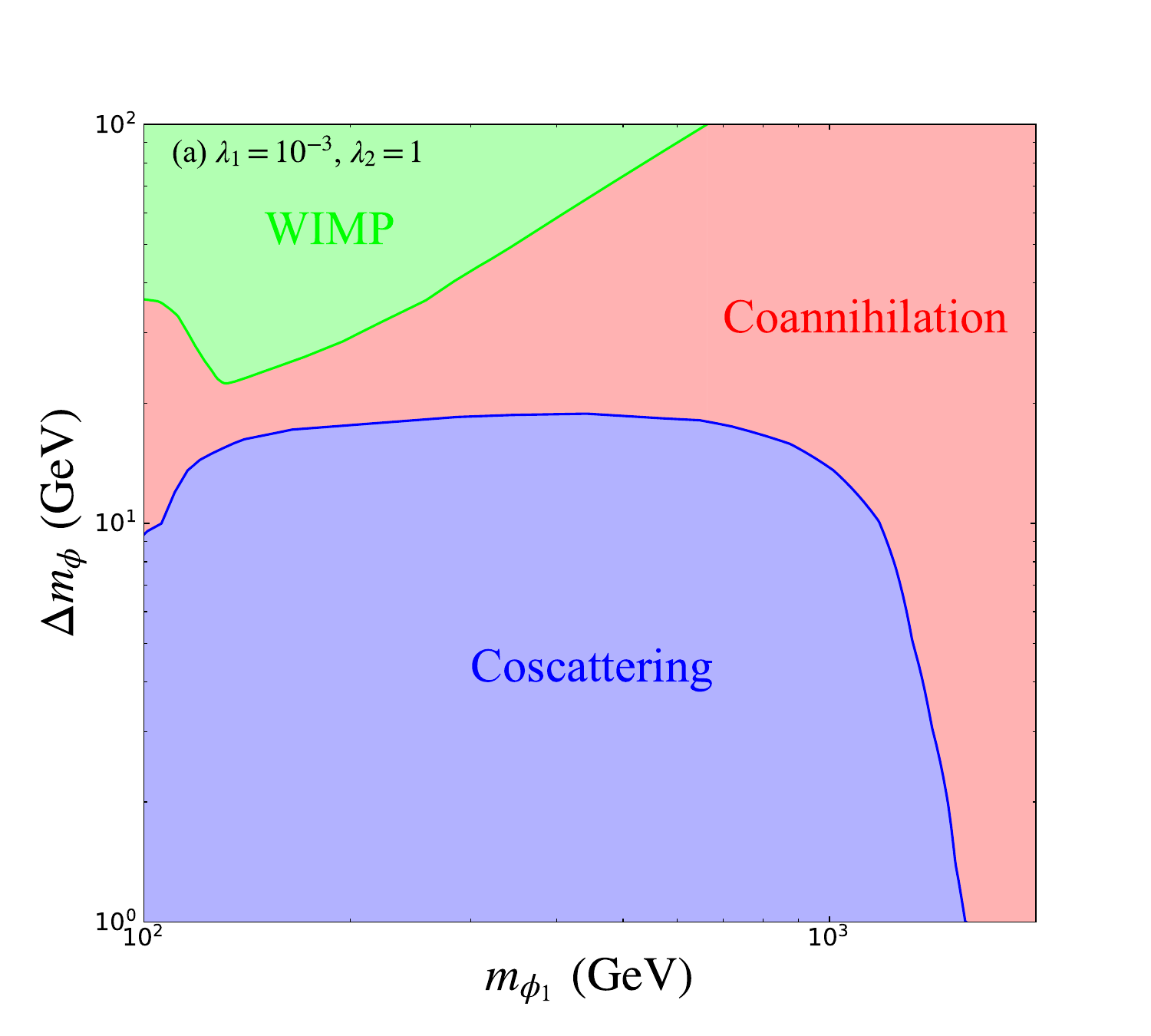}
		\includegraphics[width=0.45\linewidth]{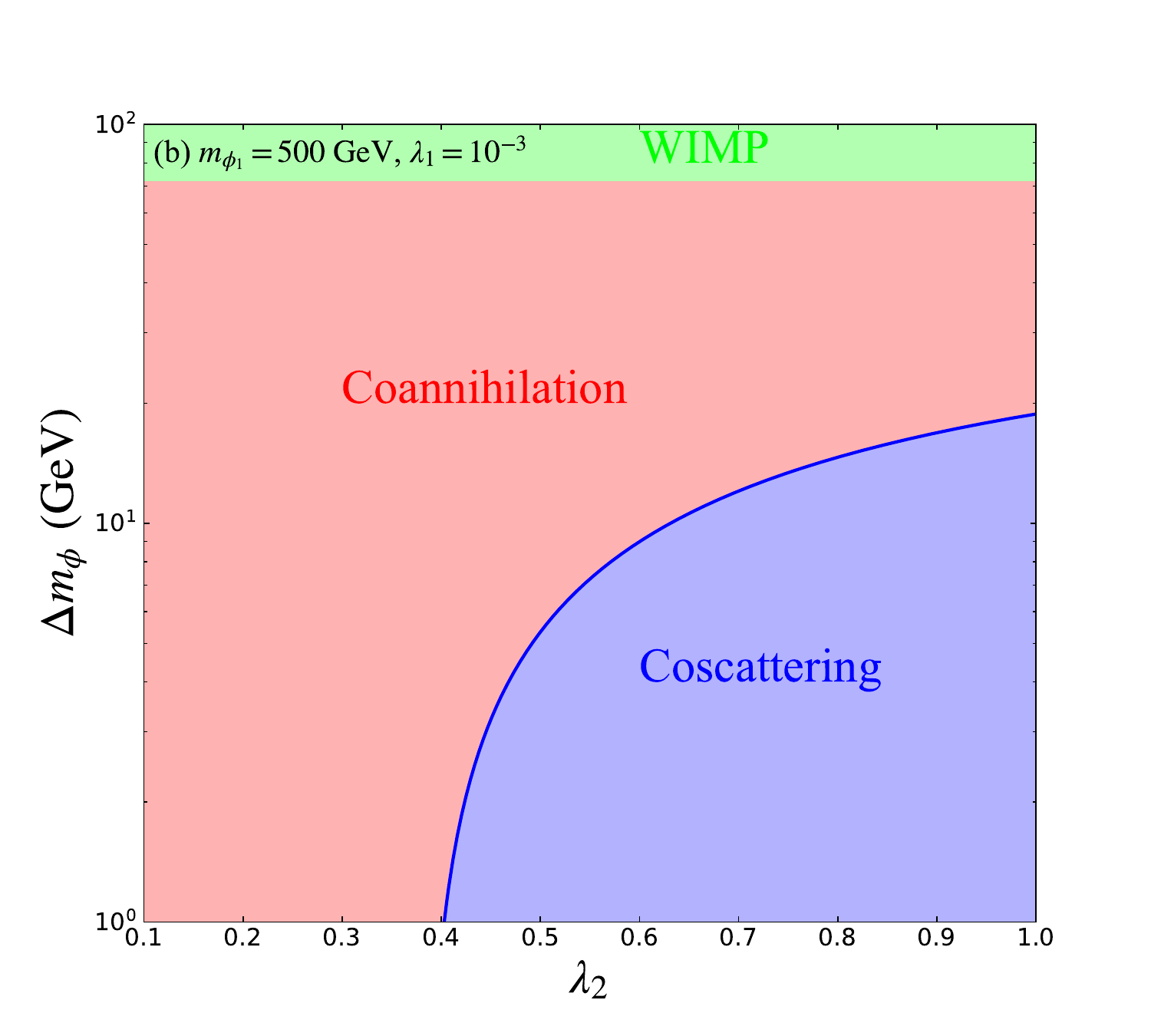}
	\end{center}
	\caption{Freeze-out phase diagrams in the parameter spaces of  $\Delta{m_\phi}-m_{\phi_1}$ in panel (a) and  $\Delta{m_\phi}-\lambda_{2}$ in panel (b). The blue, red, and green regions correspond to the phases of coscattering, coannihilation, and conventional WIMP, respectively.
	}
	\label{FIG:fig2}
\end{figure}

To clarify the parameter space where coscattering takes effect, we present the freeze-out phase diagrams in Figure~\ref{FIG:fig2},  In these two panels, we fix $\lambda_{1}=10^{-3}$ to avoid the stringent direct detection constraints. In panel (a) of Figure~\ref{FIG:fig2},  $\lambda_{2}=1$ is considered to obtain the parameter space of coscattering with $m_{\phi_1}>100$~GeV.  For the parameter spaces with $m_{\phi_1}< 100$ GeV, the presence of SM Higgs resonance phenomena complicates the differentiation among various phases.  We will report some specific results in the following discussion. 

In the coscattering region  dominated by $\phi_2\SM\to\phi_1\SM$, the mass splitting $\Delta m_\phi$ could reach about 19 GeV when $m_{\phi_1}$ equals several hundred GeV.  However, for $m_{\phi_1}$ exceeding 1000 GeV, a significantly increasing of $\lambda_{12}$ is required to prevent an excessive generation of $\phi_1$ from coscattering, then coannihilation dominated by $\phi_1\phi_2\to\SM\SM$ comes into force. Additionally, a sufficiently large mass splitting $\Delta{m_\phi}$, such as greater than 22 GeV,  will inevitably lead to the occurrence of traditional WIMP regime   $\phi_{1}\phi_1\to\SM\SM$. In panel (b) of Figure~\ref{FIG:fig2}, $m_{\phi_1}$ is fixed as 500~GeV. It is clear that coscattering   $\phi_2\SM\to\phi_1\SM$  only occurs when $\lambda_2\gtrsim0.4$. Meanwhile, increasing $\lambda_2$ will lead to a larger mass splitting $\Delta{m_\phi}$ for coscattering. Moreover, $\Delta{m_\phi}=72$ GeV marks the boundary between  coannihilation $\phi_1\phi_2\to\SM\SM$ and WIMP annihilation $\phi_{1}\phi_1\to\SM\SM$.

In summary, the coscattering regime $\phi_2\SM\to\phi_1\SM$ favors the parameter space with small mass splitting $\Delta m_\phi\lesssim \mathcal{O}(10)$ GeV and large coupling $\lambda_2\sim \mathcal{O}(1)$ when $m_{\phi_1}\lesssim 1$ TeV. A moderate mass splitting with proper $\lambda_2$ leads to the coannihilation region $\phi_1\phi_2\to\SM\SM$. While a relatively large mass splitting is required by the WIMP annihilation $\phi_{1}\phi_1\to\SM\SM$.

The typical WIMP phase $\phi_{1}\phi_1\to\SM\SM$ requires a conventional annihilation cross section $\left<\sigma v\right>\sim\mathcal{O}(10^{-26})~\text{cm}^3\text{s}^{-1}$, which determines the coupling $\lambda_{1}\sim[\mathcal{O}(10^{-1}),\mathcal{O}(1)]$ outside the resonance region.  Coannihilation $\phi_{1,2}\phi_2\to\SM\SM$ becomes the dominate channel when $\lambda_{12}\sim[\mathcal{O}(10^{-3}),\mathcal{O}(1)]$ with $\lambda_{2}\sim\mathcal{O}(0.1)$. However, smaller $\lambda_{12}\sim[\mathcal{O}(10^{-5}),\mathcal{O}(10^{-3})]$  will lead to  the coscattering phase $\phi_2\SM\to\phi_1\SM$.  Such a tiny $\lambda_{12}$ in the coscattering regime arises because the number density $n_{\SM}$ is much larger than $n_{\phi_{1,2}}$, as the number densities $n_{\phi_{1,2}}$  are  exponentially suppressed during the freezing-out period. So a comparable  reaction rate of $\phi_2\SM\to\phi_1\SM$  necessitates a significantly smaller $\left<\sigma v\right>_{\phi_2\SM\to\phi_1\SM}$.	While the contributions from other transformation processes are negligible due to small reaction rates.

Based on the results in Figure \ref{FIG:fig2}, we choose four specific scenarios: $\Delta{m_\phi}=1$ GeV or 10 GeV combined with $\lambda_{2}=0.1$ or 1, to illustrate the constraints of different phenomena. We then perform a scan in the following parameter ranges:
\begin{eqnarray}
	\begin{aligned}
		m_{\phi_1}\in[10,2000]~\GeV, \lambda_{1}\in[10^{-4},1], \lambda_{12}\in[10^{-5},1].
	\end{aligned}
\end{eqnarray}
The DM relic density is required within the $3\sigma$ range of the Planck result \cite{Planck:2018vyg}, i.e., $\Omega_{\phi_1}h^2\in[0.117,0.123]$.

\subsection{Phenomenology of Dark Matter $\phi_1$}\label{SEC:HP1}

The elastic scattering of dark matter $\phi_1$ on the nuclei is tightly constrained by the direct detection experiments. The spin-independent scattering cross section is calculated as
\begin{equation}\label{Eq:Hdd}
	\sigma_{\rm SI}=\frac{\lambda_{1}^2C^2 m_n^4}{4\pi (m_{\phi_1}+m_n)^2m_h^4},
\end{equation}
where $C\simeq0.3$ is a nucleon matrix element dependent constant \cite{Belanger:2013oya}, and $m_n$ is the nucleon mass. 

\begin{figure}
	\begin{center}
		\includegraphics[width=0.45\linewidth]{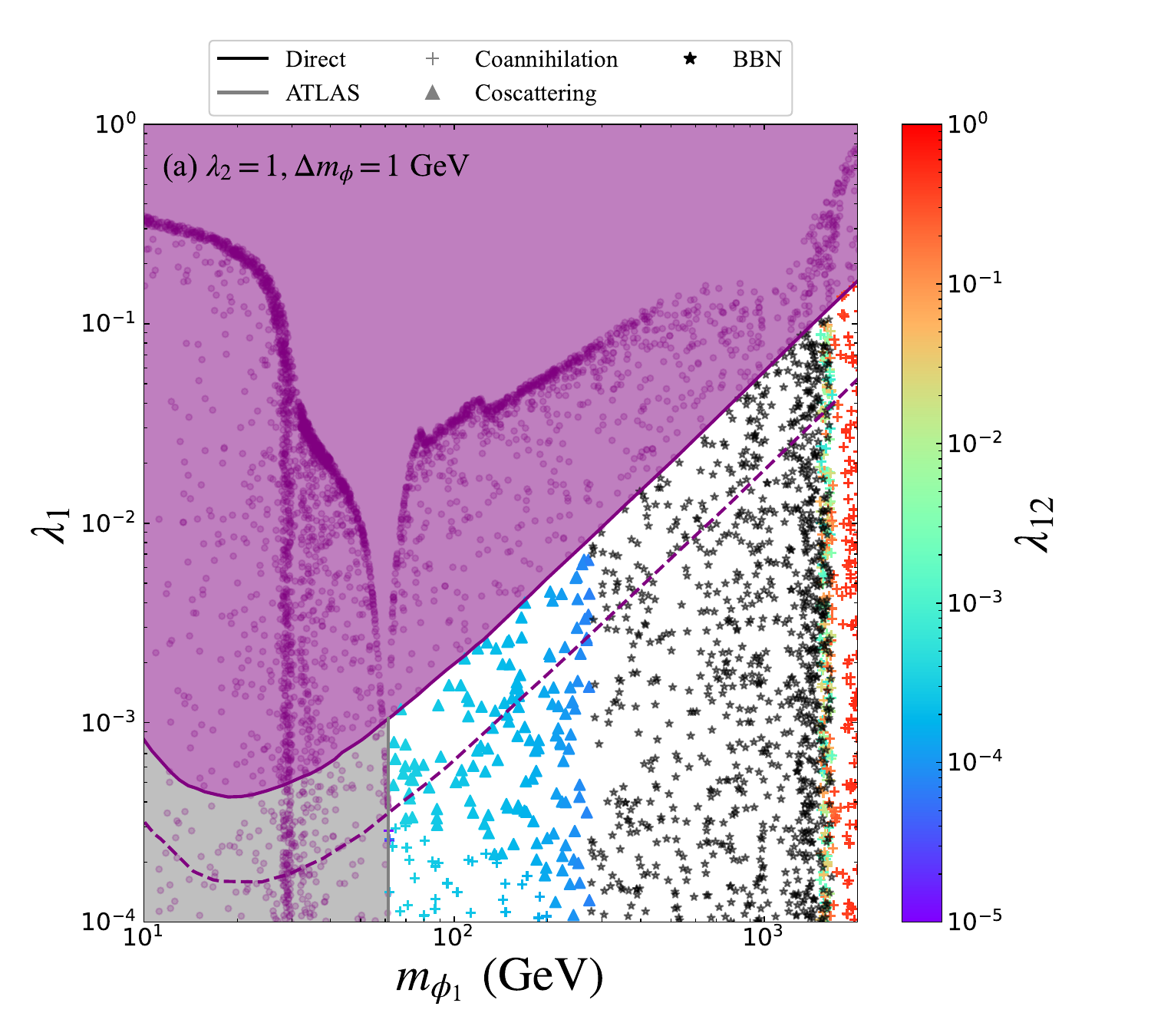}
		\includegraphics[width=0.45\linewidth]{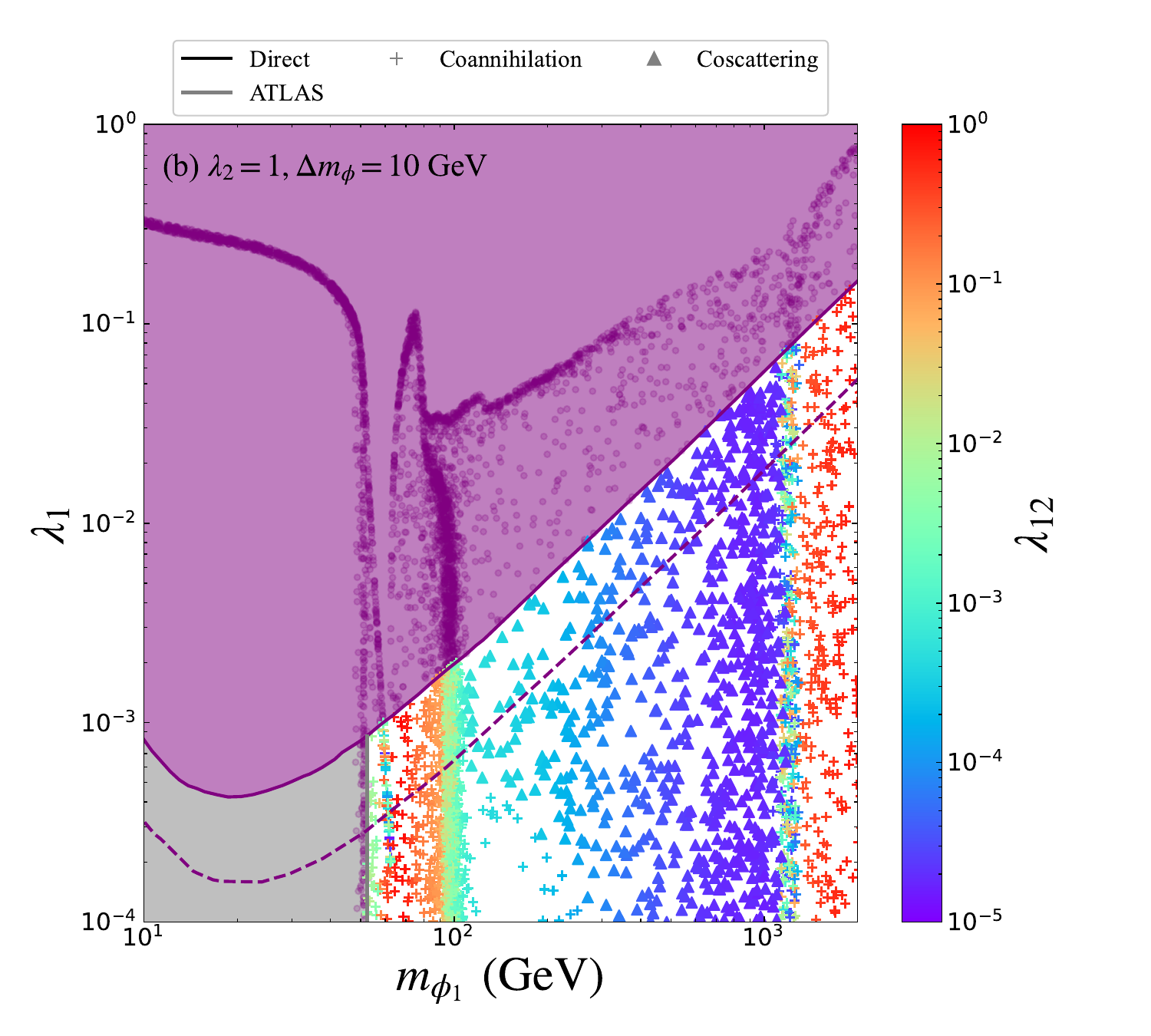}
		\includegraphics[width=0.45\linewidth]{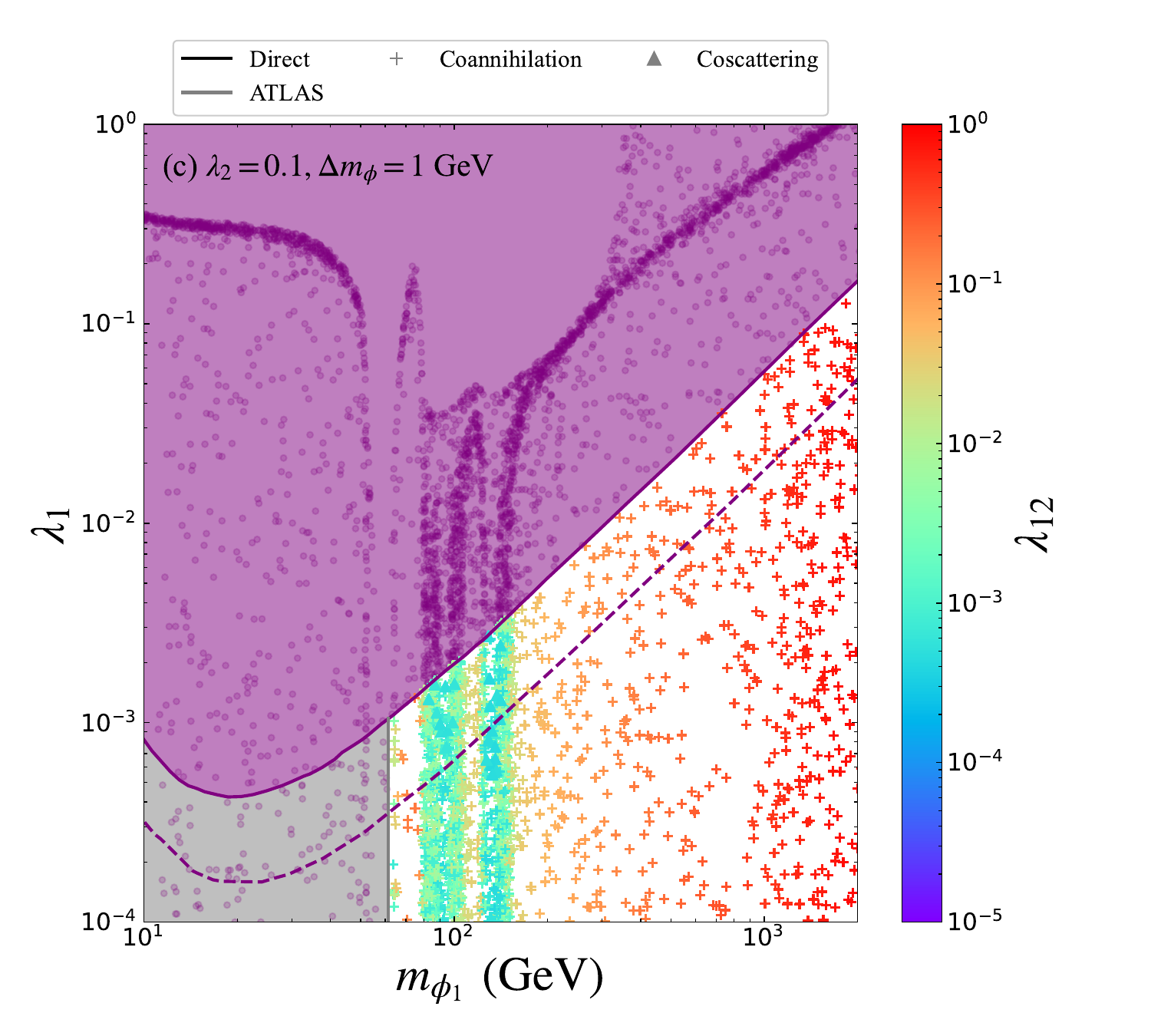}
		\includegraphics[width=0.45\linewidth]{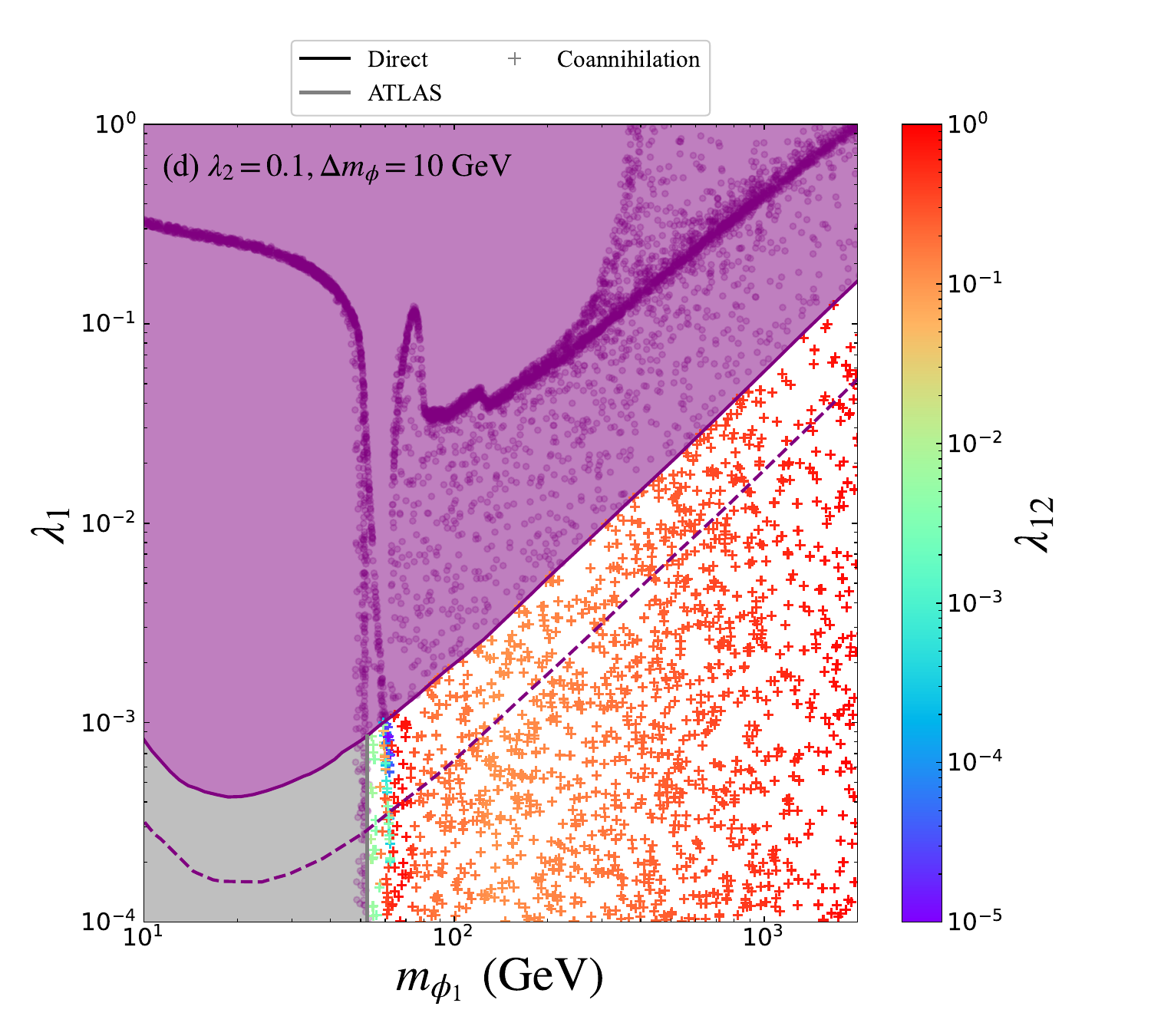}
	\end{center}
	\caption{Direct detection and ATLAS constraints on the $\lambda_{1}-m_{\phi_1}$ parameter space. Panels (a), (b), (c), and (d) correspond to distinct combinations of $\lambda_{2}$ and $\Delta{m_\phi}$, respectively. The present LZ~\cite{LZ:2024zvo} and future DARWIN~\cite{DARWIN:2016hyl} limits are denoted as the purple solid and dashed line. The gray solid curve is the bound of current ATLAS~\cite{ATLAS:2023tkt,ATLAS:2022vkf} searches of Higgs decays. Samples marked as $\bullet$ are already excluded by current experimental searches. For the allowed samples, the coannihilation and coscattering regimes are denoted by the symbols of $+$ and $\blacktriangle$, respectively. The remaining  prominent black star samples $\bigstar$ have been excluded by the BBN constraint~\cite{Kawasaki:2017bqm}, which is elaborated upon in subsequent Section \ref{SEC:HP2}. 
	}
	\label{FIG:fig3}
\end{figure}

In Figure~\ref{FIG:fig3}, we show the current LZ \cite{LZ:2024zvo} and future DARWIN \cite{DARWIN:2016hyl} constraints obtained through Equation~\eqref{Eq:Hdd}. The current LZ bound imposes the strongest constraints when $m_{\phi_1}\sim20$ GeV, where only $\lambda_{1}\lesssim4\times10^{-4}$ can satisfy the limit. The exclusion ability diminishes as  $m_{\phi_1}$ increases, e.g., $\lambda_1$ reaches 0.1 when $m_{\phi_1}\sim \TeV$.  Furthermore, as we consider the mass splitting $\Delta m_\phi\geq1$ GeV, the inelastic scattering process  $\phi_1n\to\phi_2n$ is kinematically forbidden ~\cite{Tucker-Smith:2001myb,Bramante:2016rdh}.

Light scalars inevitably induce new decay modes of the SM Higgs as $h\to \phi_i \phi_j$. The decay widths are calculated as
\begin{eqnarray}
	\Gamma_{h\to\phi_1\phi_1}&=&\frac{\lambda_1^2 v_H^2}{32\pi m_h}\sqrt{1-\frac{4m_{\phi_1}^2}{m_h^2}}, \\
	\Gamma_{h\to\phi_1\phi_2}&=&\frac{\lambda_{12}^2 v_H^2}{16\pi m_h^3}\sqrt{(m_h^2-\Delta{m_\phi}^2)(m_h^2-(\Delta{m_\phi}+2m_{\phi_1})^2)},
	\\
	\Gamma_{h\to\phi_2\phi_2}&=&\frac{\lambda_2^2 v_H^2}{32\pi m_h}\sqrt{1-\frac{4(m_{\phi_1}+\Delta{m_\phi})^2}{m_h^2}},
\end{eqnarray}
where $v_H=246$ GeV is the vacuum expectation value of the SM Higgs doublet. The decay mode $h\to \phi_1\phi_1$ contributes to the invisible decay of the SM Higgs. According to Equation \ref{Eqn:HG2}, the three-body decay width of $\phi_2$ is suppressed by the light fermion mass and small mass splitting, which results in $\phi_2$ being long-lived. As will be shown later in Figure \ref{FIG:fig6}, the decay length of dark partner $c\tau_{\phi_2}$ might be too large for relatively small mass splitting as $\Delta m_\phi=1$ GeV, so $\phi_2$ is also invisible at the collider. On the other hand, when the mass splitting is relatively large as $\Delta m_\phi=10$ GeV, $\phi_2$ could decay inside the detector and lead to the displaced vertex signature. Currently, the ATLAS experiment yields the upper limits of invisible and undetected non-SM Higgs boson decays as BR$_\text{inv.}<0.11$ \cite{ATLAS:2023tkt} and BR$_\text{u.}<0.12$ \cite{ATLAS:2022vkf}. As we focus on the unconventional scenario with relatively large $\lambda_2\geq 0.1$, the region with $m_{\phi_2}<m_h/2$ has a too large new decay width of $h\to \phi_2\phi_2$, therefore, such a region is completely excluded.

It is also essential to analyze the distinct features of the benchmark scenarios. For the scenario with $\lambda_2=1$ and $\Delta m_\phi=1$ GeV in panel (a) of Figure \ref{FIG:fig3}, the allowed samples distributed within $m_{\phi_1}\lesssim270$~GeV and $3\times10^{-4}\lesssim \lambda_1\lesssim7.6\times10^{-3}$ are dominated by coscattering, where the corresponding $\lambda_{12}$ is at the order of $\mathcal{O}(10^{-4})$. However, for an excessively small $\lambda_{1}\lesssim3\times10^{-4}$,  the reaction rate of the elastic scattering $\phi_1\SM\to\phi_1\SM$ is lower than that of the inelastic scattering $\phi_2\SM\to\phi_1\SM$, which contradicts the judgment of the coscattering condition (1). Hence, these samples are classified as coannihilation.  For samples within the range of $m_{\phi_1}\in[270, 1700]$ GeV, BBN brings devastating limitations due to the long-lived decays of dark partner $\phi_2$. As $m_{\phi_1}$ is larger than 1.7 TeV, only $\lambda_{12}$  at the order of $\mathcal{O}(0.1)$ can meet the observation of DM, which signifies a return to coannihilation.  In this scenario, the future DARWIN experiment can examine coscattering samples with $\lambda_{1}\sim\mathcal{O}(10^{-3})$, but demand $\lambda_{1}$ to be close to $5\times10^{-2}$ for TeV scale coannihilation samples.

Just increasing $\Delta{m_\phi}$ to 10 GeV in panel (b) of Figure \ref{FIG:fig3}, the lifetime of dark partner $\tau_{\phi_2}$ becomes small enough to satisfy the BBN constraint, so coscattering samples thrive abundantly within the range of $100~\GeV\lesssim m_{\phi_1}\lesssim 1150~\GeV$.  Then allowed samples at both sides, i.e., 50 GeV $\lesssim m_{\phi_1}\lesssim100$ GeV and $m_{\phi_1}\gtrsim1150$ GeV, are controlled by coannihilation. The current upper limit on $\lambda_{1}$ is determined by the LZ bound. Therefore, the detectable range of the coscattering region increases to $\lambda_{1}\sim\mathcal{O}(10^{-2})$. At the same time, coannihilation samples below 100 GeV with $\lambda_{1}\sim\mathcal{O}(10^{-4})$ are also expected to be discovered by DARWIN.

When the coupling $\lambda_{2}$ drops to 0.1 in panels (c) and (d) of Figure \ref{FIG:fig3}, samples through coscattering have a significant reduction and eventually disappear completely when $\Delta m_\phi=10$ GeV. Coscattering happens sparsely when $m_{\phi_1}$ is near the masses of the SM bosons with $\Delta m_\phi=1$ GeV. It is easy to understand that these boson final states result in the relic density being too small to satisfy the observation, hence the solution involves much smaller $\lambda_{12}$ to reduce the contribution of $\phi_1\phi_2\to\SM\SM$.  With an appropriate value $\lambda_{12}\sim10^{-3}$, the interaction rate $\Gamma_{\phi_2\SM\to\phi_1\SM}$ drops below the threshold $\Gamma_{\phi_2\phi_2\to\SM\SM}$, thereby naturally establishing the determination of coscattering. In panel (d) of Figure \ref{FIG:fig3},  coscattering is no longer present, leaving only coannihilation. Prospective direct detection experiment is only sensitive to $\lambda_{1}\sim\mathcal{O}(10^{-4})$ for coscattering samples in scenario (c). But for coannihilation in these two scenarios, multiple orders of magnitude of $\lambda_{1}$ are promising, which depends on the mass of dark matter $m_{\phi_1}$.

\begin{figure}
	\begin{center}
		\includegraphics[width=0.45\linewidth]{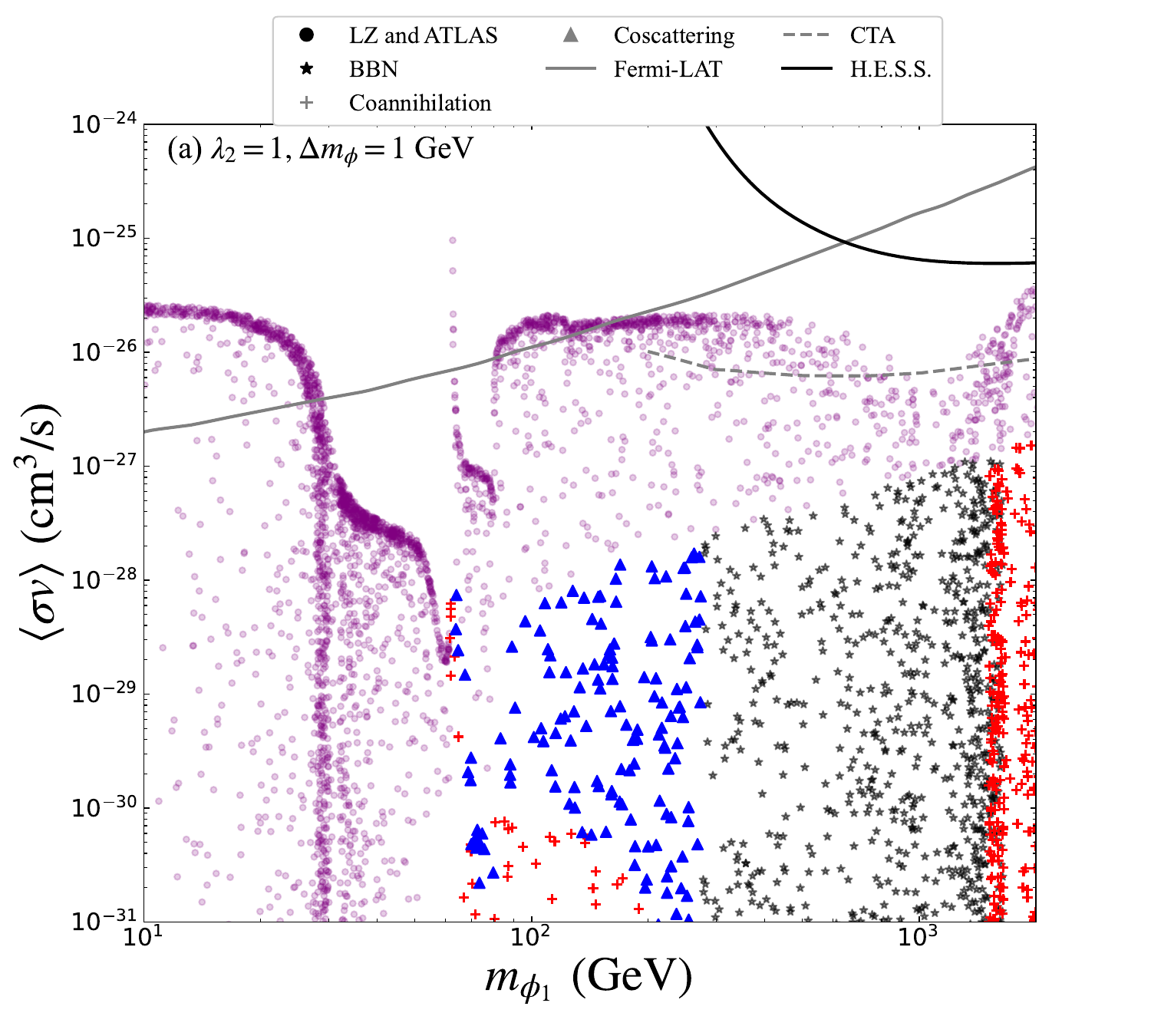}
		\includegraphics[width=0.45\linewidth]{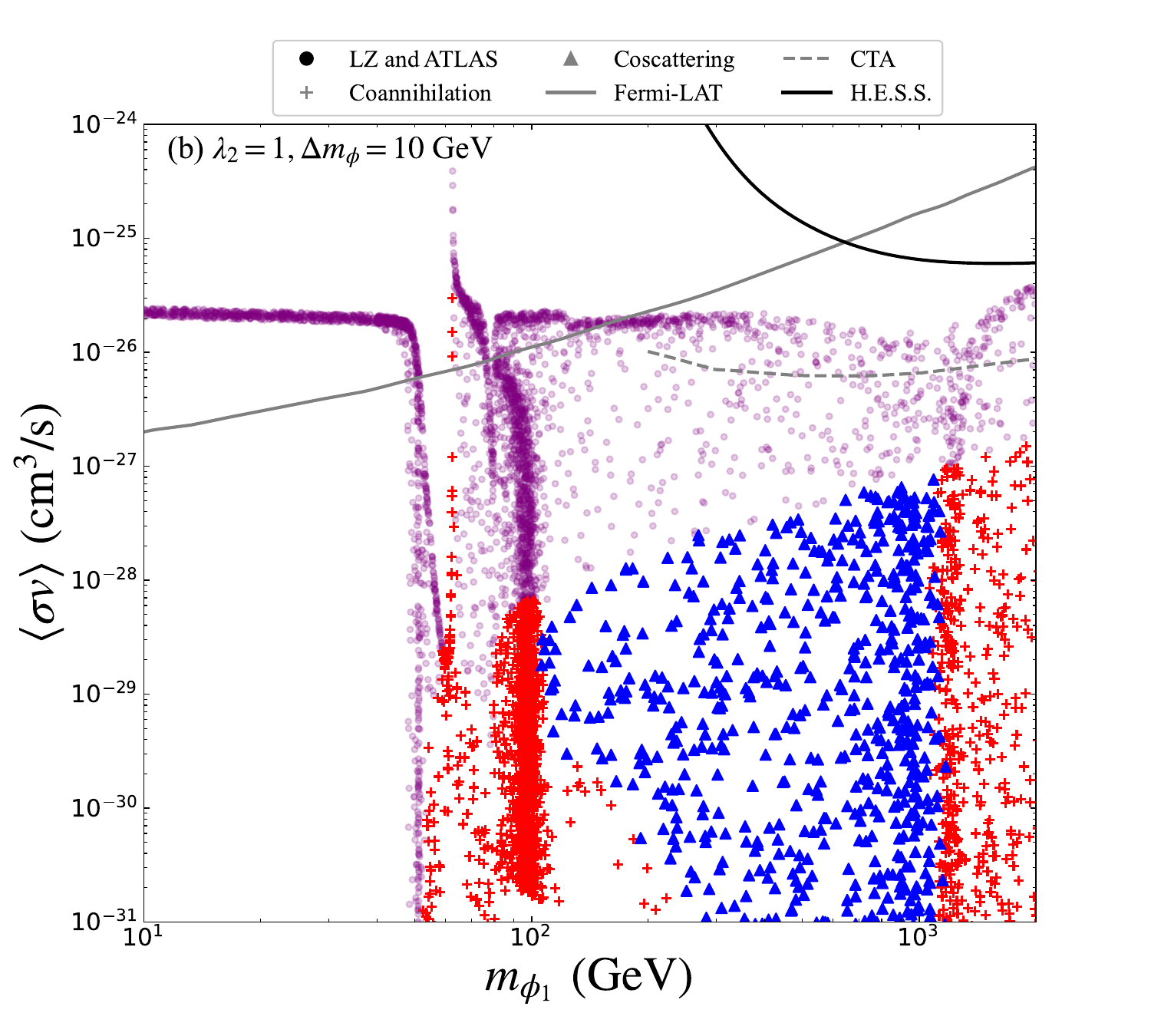}
		\includegraphics[width=0.45\linewidth]{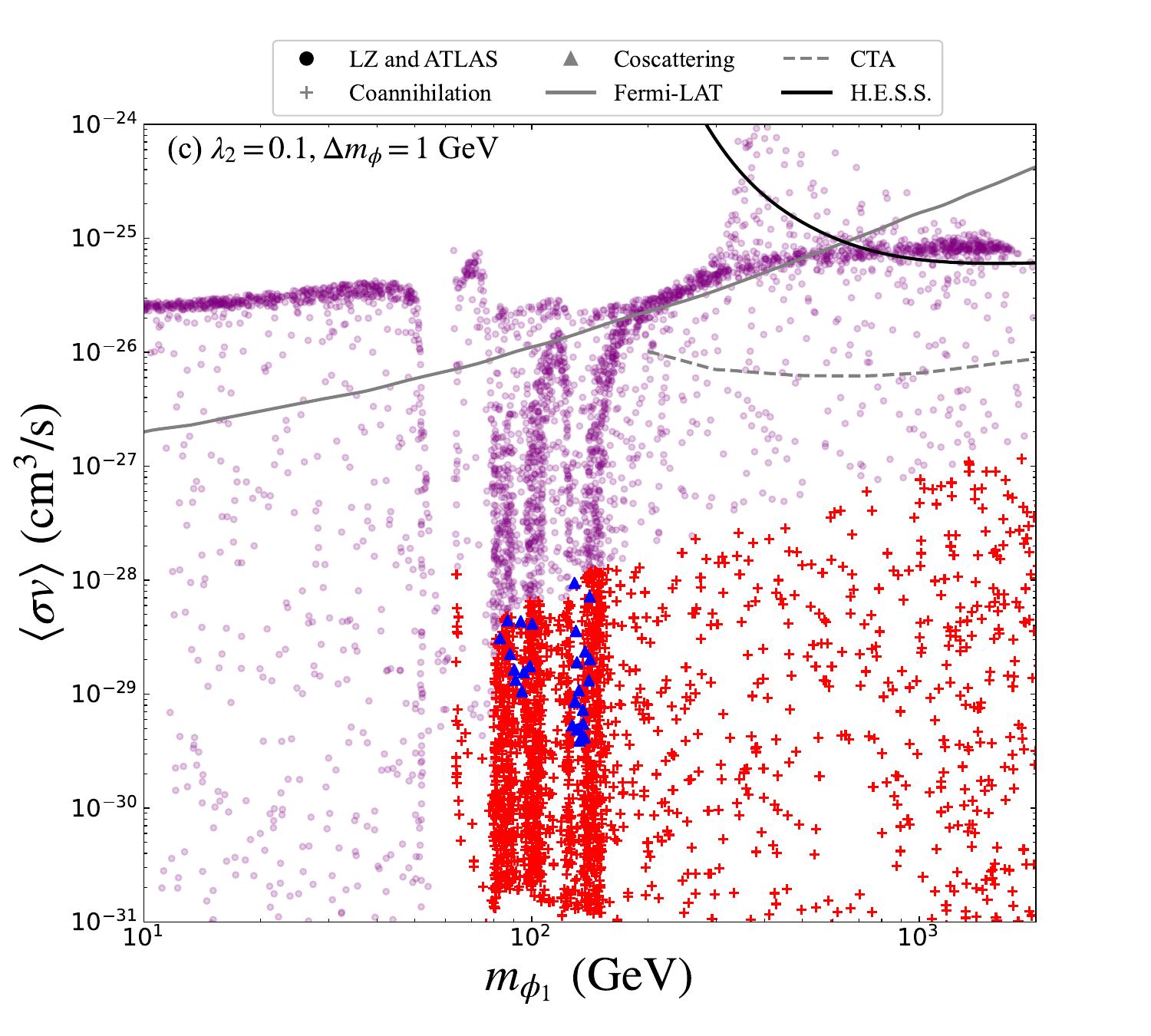}
		\includegraphics[width=0.45\linewidth]{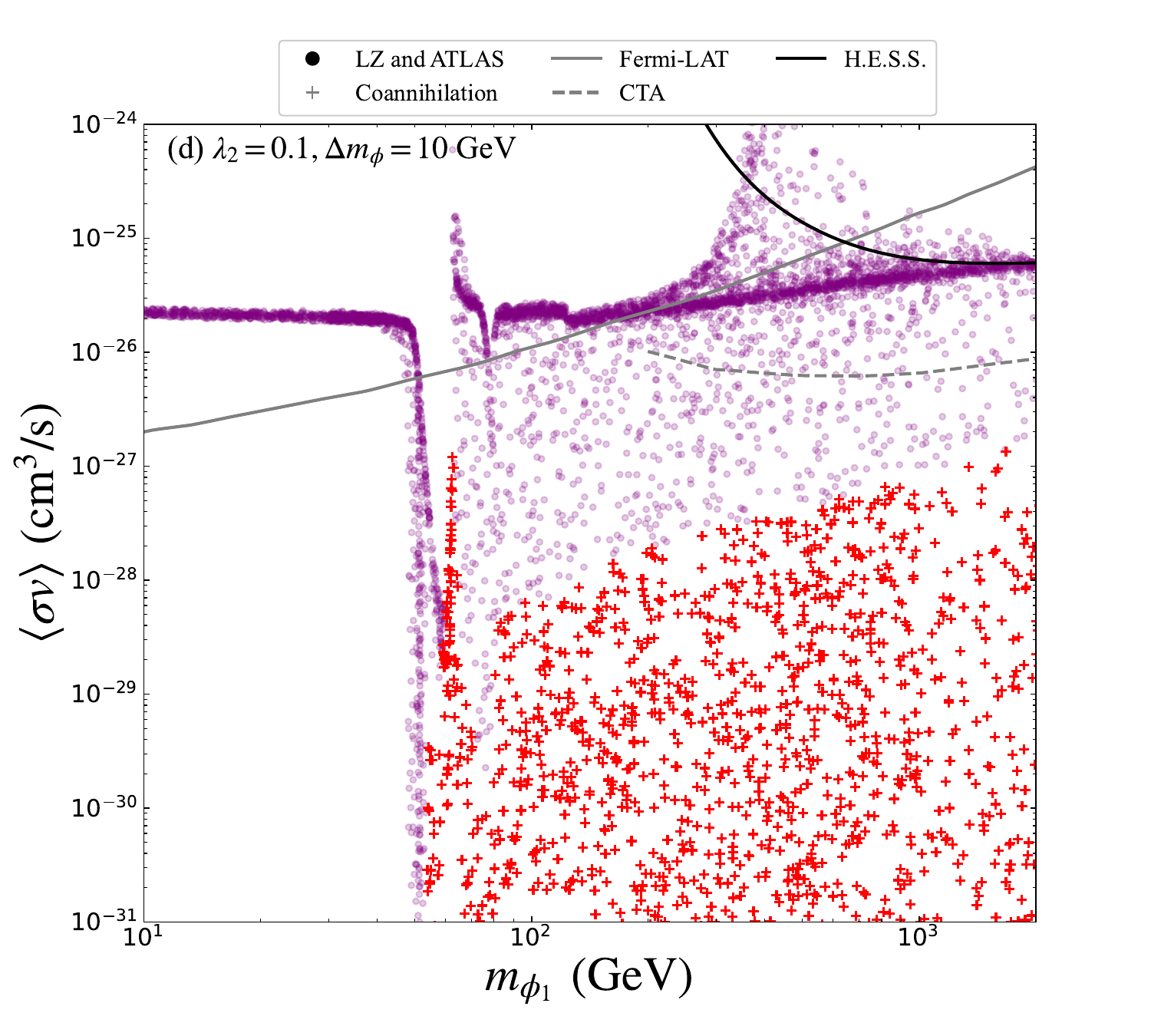}
	\end{center}
	\caption{Constraints from the indirect detection experiments in the Higgs portal scenario. Panels (a)-(d) represent the four scenarios with combinations of $\lambda_{2}$ and $\Delta{m_\phi}$. The current direct detection constraint from LZ and the Higgs decay limit from ATLAS exclude the purple samples. The remaining samples with different shapes share the same representation as those in Figure~\ref{FIG:fig3}.  For improved readability, coannihilation and coscattering are highlighted in red and blue, respectively.  The gray and black solid lines stand for the Fermi-LAT observed limit on the $b\bar{b}$ final state~\cite{Fermi-LAT:2015att} and H.E.S.S. observed limit on the $W^+W^-$ final state \cite{HESS:2016mib}. The gray dashed line represents the predicted results of CTA in the $W^+W^-$ final state \cite{CTA:2020qlo}.}
	\label{FIG:fig4}
\end{figure}

The indirect detection of dark matter aims to search for dark matter via its annihilated SM final states, which sets constraints on the present thermal average cross sections of $\phi_1\phi_1\to\SM\SM$. In the Higgs portal scenario, the dominant channel transitions from $\phi_1\phi_1\to b\bar{b}$ to $W^+W^-$ final state as $m_{\phi_1}$ enlarges. Therefore, the current constraint of Fermi-LAT on the $b\bar{b}$ final state and the projected limit of CTA on the $W^+W^-$ final state are utilized to illustrate in Figure~\ref{FIG:fig4}.

In panel (a) of Figure~\ref{FIG:fig4}, under the strict constraints of LZ, the annihilation cross section $\left<\sigma v\right>$ of the coscattering region is obviously below $2\times10^{-28}~\rm{cm^3/s}$, which is at least  two orders of magnitude smaller than the current indirect detection bounds.  For the TeV scale coannihilation samples, despite $\left<\sigma v\right>$  increases by an order of magnitude, yet it remains challenging for future CTA to capture it.  In the subsequent three panels, whether coscattering or coannihilation, the corresponding very small $\left<\sigma v\right>$is far below the sensitivity of indirect detection experiments. Furthermore, a few allowed points within the Higgs resonance region, namely $m_{\phi_1}$ slightly greater than $m_h/2$,  would be constrained by the Fermi-LAT  $b\bar{b}$ limit as shown in panel (b) of Figure~\ref{FIG:fig4}.

\subsection{Phenomenology of Dark Partner $\phi_2$}\label{SEC:HP2}

The focus of this work is primarily on the scenarios of  $\Delta{m_\phi}=1$ GeV and 10 GeV, which determines that the only decay mode of dark partner $\phi_2$ is $\phi_2\to \phi_1h^\star\to\phi_1\bar{f}f$. This mainly leads to two aspects of phenomenology: (1) the additional energetic injection  will affect the big bang nucleosynthesis (BBN) predictions and the cosmic microwave background (CMB) anisotropy power spectra; (2) the delayed decay products can be reconstructed as a displaced vertex (DV), which can be captured at colliders.

\begin{figure}
	\begin{center}
		\includegraphics[width=0.45\linewidth]{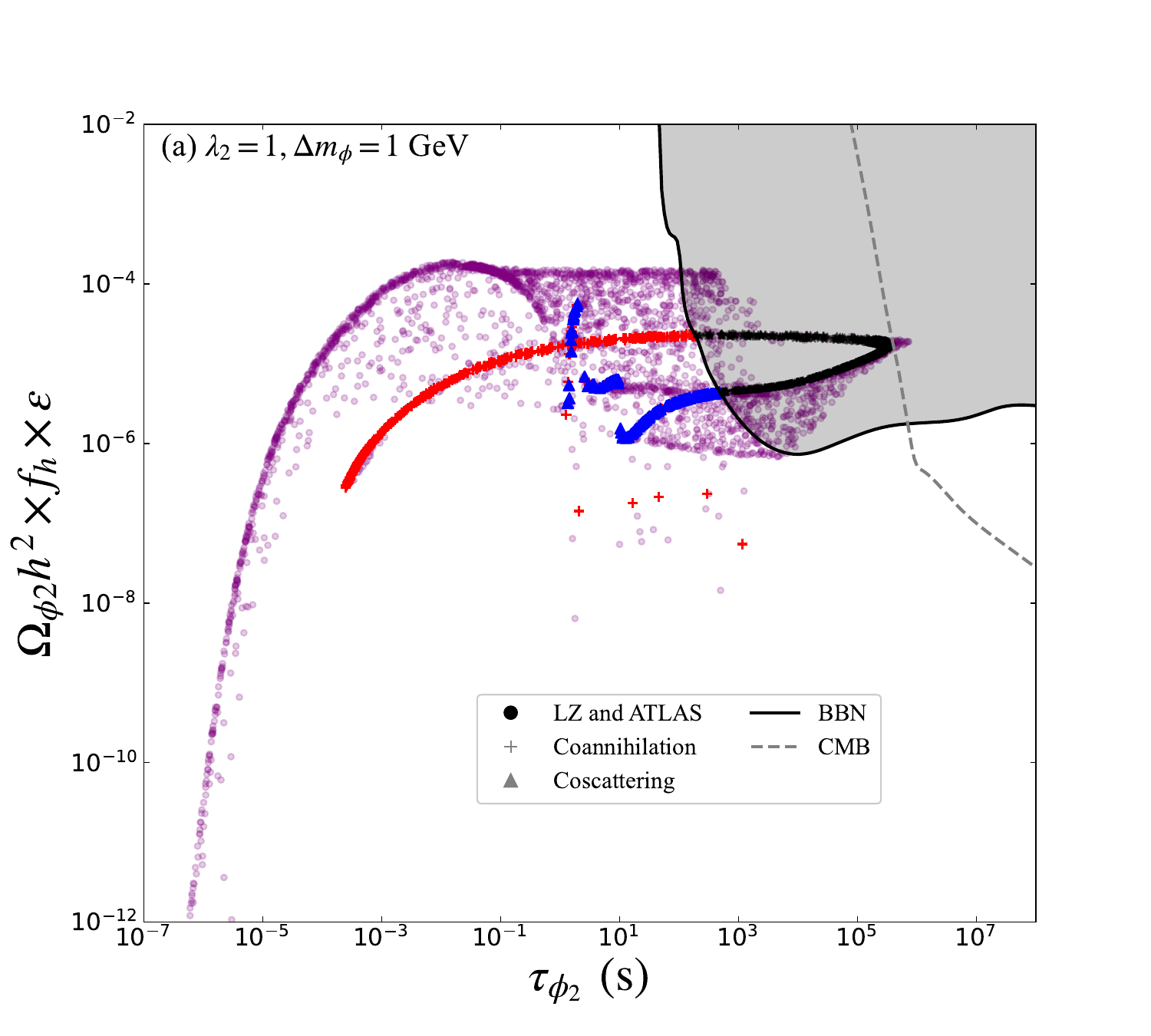}
		\includegraphics[width=0.45\linewidth]{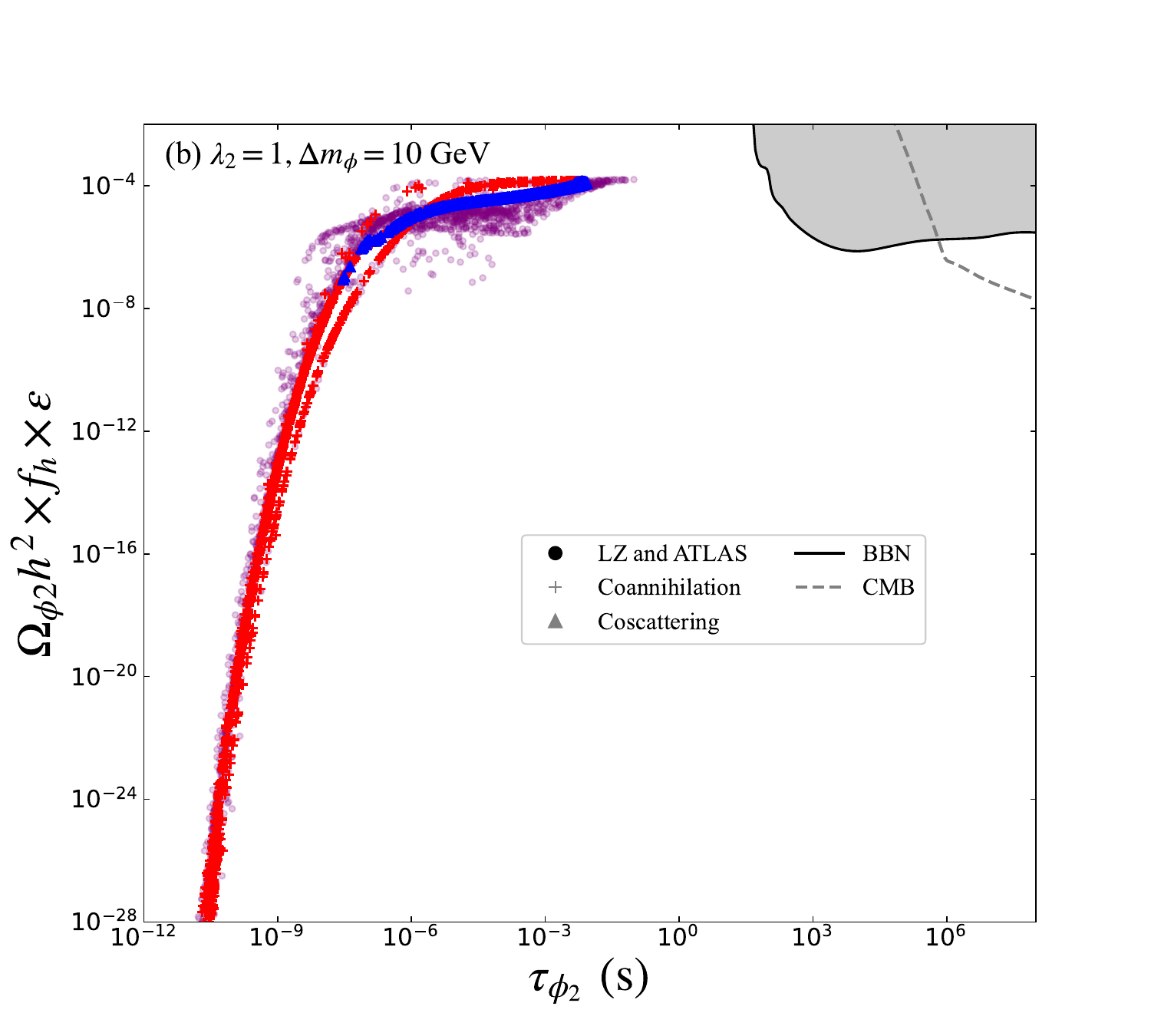}
		\includegraphics[width=0.45\linewidth]{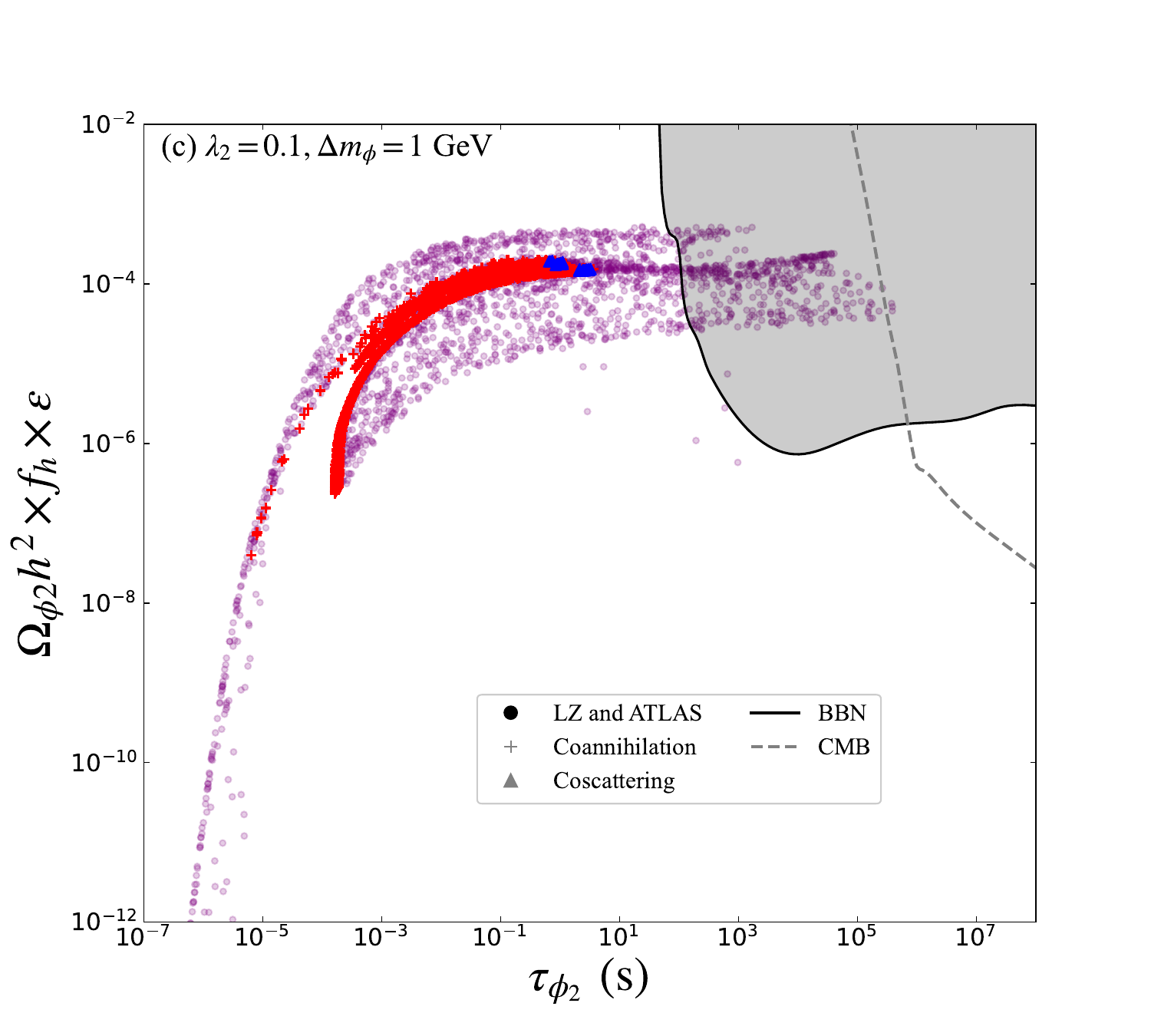}
		\includegraphics[width=0.45\linewidth]{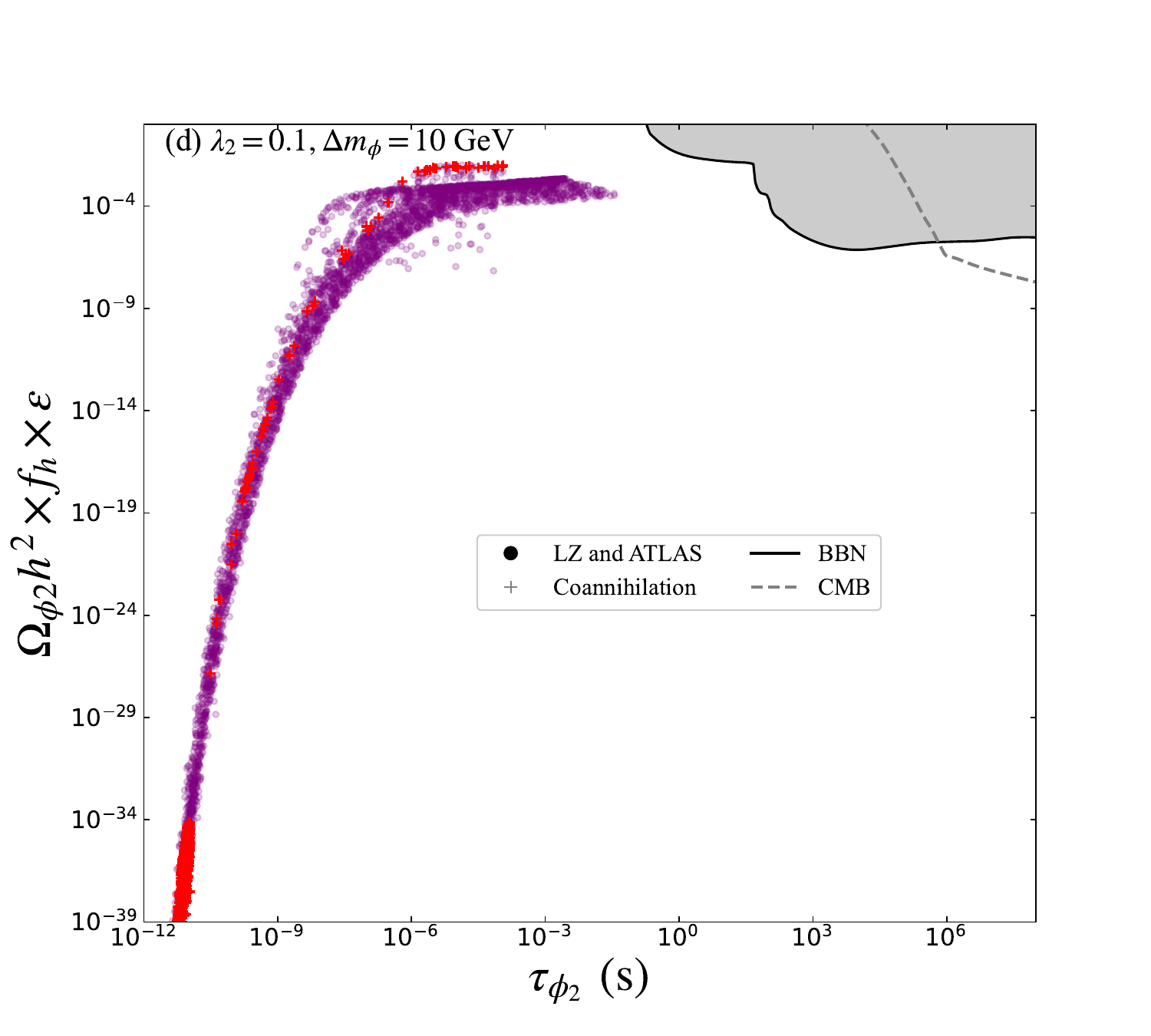}
	\end{center}
	\caption{CMB and BBN constraints on the dark partner $\phi_2$ in the Higgs portal scenario. The horizontal axis $\tau_{\phi_2}$ is the lifespan of $\phi_2$, while the vertical axis represents the product of relic density $\Omega_{\phi_2}h^2$, hadronic branching ratio $f_h$, and the energy transfer factor $\epsilon$. Panels (a)-(d) and  internal samples have the same definition as those in Figure~\ref{FIG:fig4}.  Here, we use the black curves to represent the current BBN constraint of hadronic final states~\cite{Kawasaki:2017bqm}. The gray dashed lines stand for the upcoming CMB results of purely electromagnetic decay~\cite{Lucca:2019rxf}.
	}
	\label{FIG:fig5}
\end{figure}

Firstly, we discuss the cosmological constraints arising from long-lived $\phi_2$ decay. In the Higgs portal scenario, the hadronic as well as the electromagnetic decays of $\phi_2$ occur simultaneously. The corresponding branching ratios $f_h$ and $f_e$ are inextricably linked to the mass splitting $\Delta{m_\phi}$.  Specifically, $f_h\simeq0.74$ and $f_e\simeq0.26$ when $\Delta{m_\phi}=1$ GeV. Nevertheless, $f_h\simeq0.66$ and $f_e\simeq0.34$ if $\Delta{m_\phi}=10$ GeV.  As the hadronic decay is always dominant in these scenarios, we illustrate the constraint of BBN on hadronic decay in Figure~\ref{FIG:fig5}. In order to align with the BBN bound, it is essential for all samples to be multiplied by $f_h$ and $\epsilon$, ,  where $\epsilon=(m_{\phi_2}^2-m_{\phi_1}^2)/2m_{\phi_2}^2$ is the fraction of the energy of $\phi_2$ that has been transferred to SM fermions.

One can find that BBN imposes obvious limits on $\tau_{\phi_2}\gtrsim50$ s from Figure~\ref{FIG:fig5}. Therefore, some samples at panels  (a) and (c) with $\tau_{\phi_2}\in[10^2,10^6]$ s are at risk. In reality, due to the prior constraints of ATLAS and LZ, only the remaining  samples in panel (a) will be effectively excluded, which mainly occurs in the coscattering region. Meanwhile, the ATLAS and LZ allowed samples in panel (c) are distributed below 10 s of $\tau_{\phi_2}$, completely avoiding the constraint of BBN. In panels (b) and (d) with $\Delta{m_\phi}=10$ GeV, the increased mass splitting significantly reduces $\tau_{\phi_2}$ according to Equation~\eqref{Eqn:HG2}. As a result, all samples are situated below 0.1~s and evidently smaller than the sensitive space of BBN.

Furthermore, the CMB  constraints primarily pertain to electromagnetic final states. The current influence space generally situates above $10^{12}$ s of $\tau_{\phi_2}$~\cite{Lucca:2019rxf,Acharya:2019uba}. The future CMB outcomes  may impose constraints on smaller $\tau_{\phi_2}$~\cite{Lucca:2019rxf}, which is displayed as the gray dashed lines in Figure~\ref{FIG:fig5}.  It is noteworthy that the originally CMB constraint  acts on the electromagnetic final state, so this bound needs to be multiplied by $f_h/f_e$ numerically to match the BBN limit. Compared to the BBN bound, the future CMB  can probe a smaller relic density of $\phi_2$ with $\tau_{\phi_2}\gtrsim\mathcal{O}(10^5)$ s.  From the results, it is challenging to verify the permissible samples through the future CMB under the exclusion of BBN.

\begin{figure} 
	\begin{center}
		\includegraphics[width=0.45\linewidth]{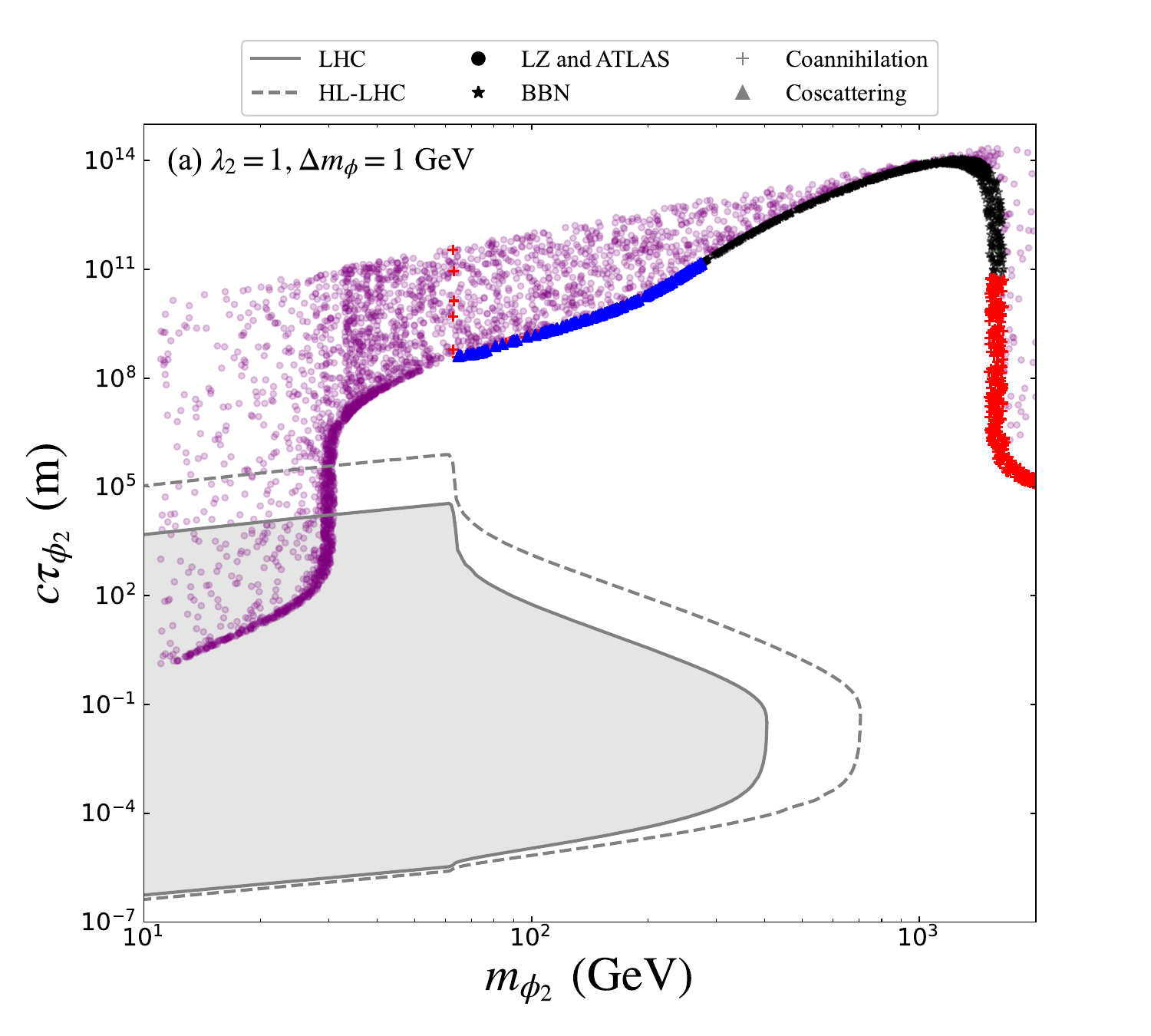}
		\includegraphics[width=0.45\linewidth]{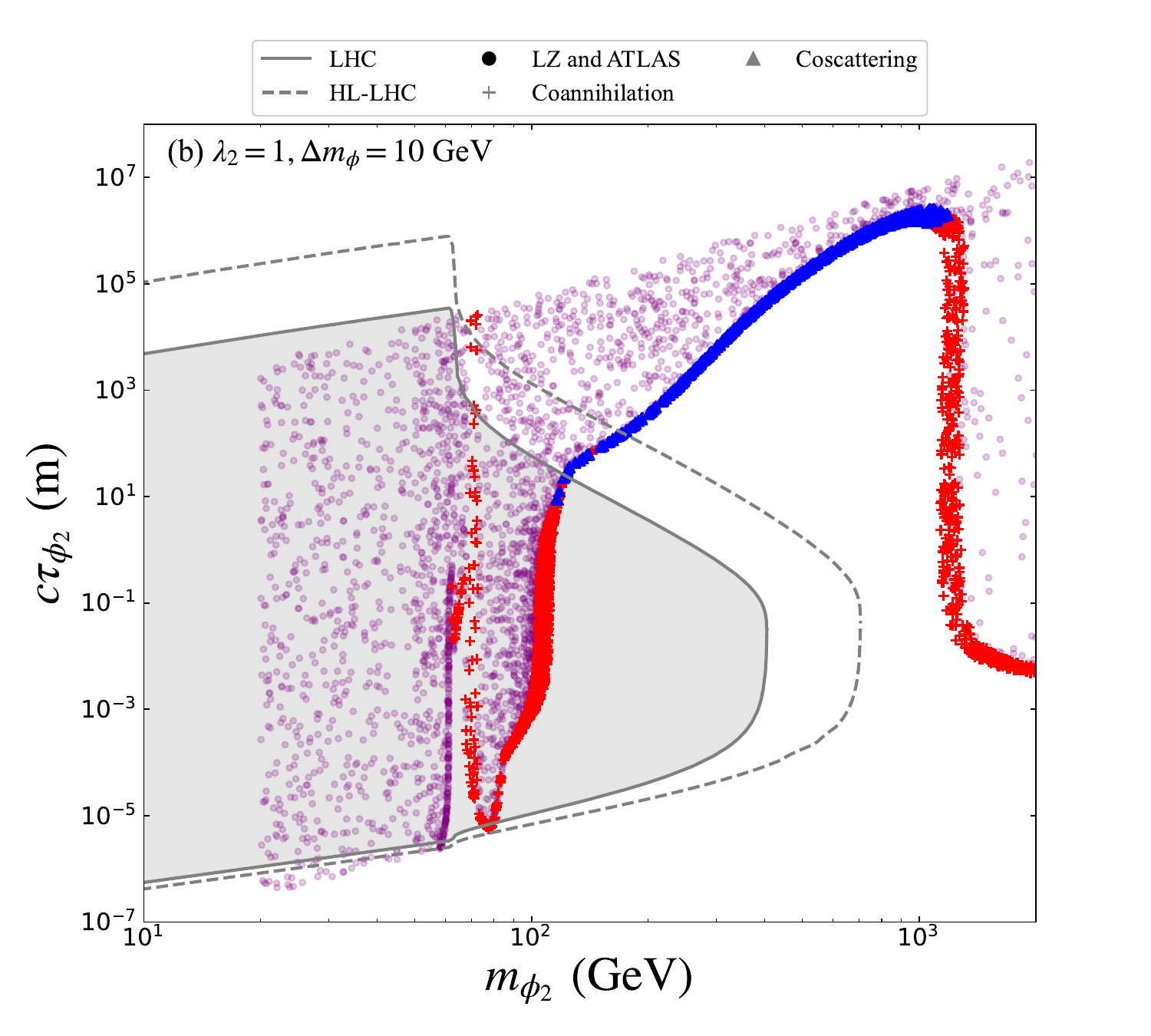}
		\includegraphics[width=0.45\linewidth]{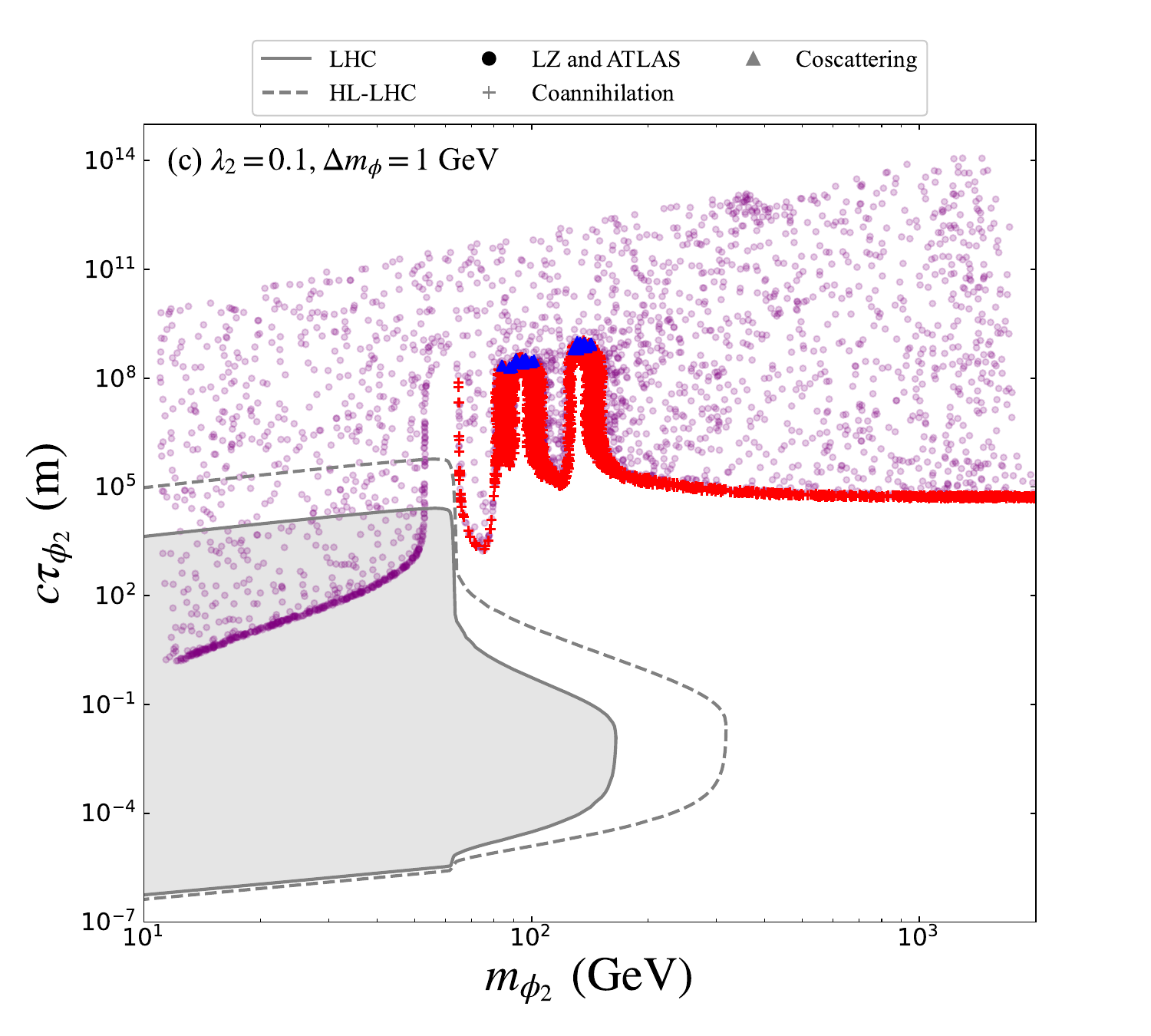}
		\includegraphics[width=0.45\linewidth]{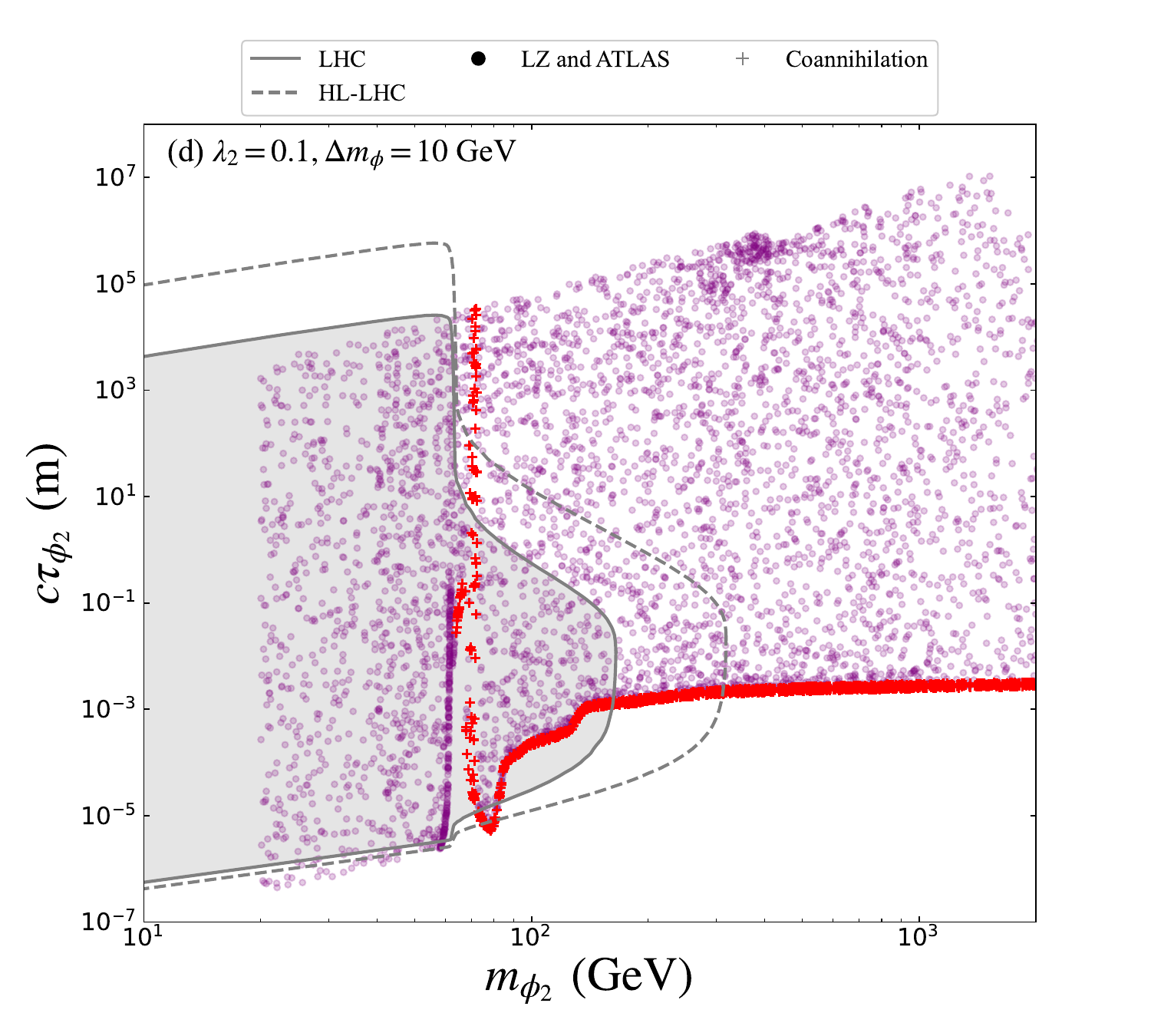}
	\end{center}
	\caption{The sensitive region of the one DV signature at LHC (gray solid lines) and future HL-LHC (gray dashed lines) in the Higgs portal scenario.  Panels (a)-(d), as well as the  internal samples, have the same representation as those in Figure~\ref{FIG:fig4}.
	}
	\label{FIG:fig6}
\end{figure}

The long-lived $\phi_2$ is determined by two crucial conditions: $\lambda_{12}$ must be small enough, and $\Delta {m_\phi}$ is not greater than $m_h$. In this way, the three-body decay process $\phi_2\to\phi_1h^\star\to\phi_1\bar{f}f$ ensures a long lifetime for $\phi_2$. 
When the lifetime of the dark partner $\tau_{\phi_2}$ is too large, $\phi_2$ becomes invisible at colliders. Under the constraints from ATLAS searches for Higgs decays \cite{ATLAS:2023tkt,ATLAS:2022vkf}, $m_{\phi_2}>m_h/2$ is required for the benchmark scenarios in this paper.  Then the invisible scalars can  be generated via the off-shell Higgs at colliders, which is only promising at LHC when  $\lambda_{1,2}\sim1$ and $m_{\phi_{1,2}}\lesssim100$ GeV \cite{Craig:2014lda,Ruhdorfer:2019utl}. However, this scale is  ruled out by direct detection experiments for dark matter $\phi_1$, whereas the long-lived $\phi_2$ is viable corresponding to the coscattering samples.  Meanwhile, the invisible scalars could be produced via the decay of doublet fermion $\Psi$, which results in the promising signature $pp\to \psi^+\psi^-\to \ell^+\phi_{1,2}+\ell^-\phi_{1,2}\to \ell^+\ell^- + \cancel{E}_T$. Currently, the direct search of this signature has excluded the region with $m_{\psi^\pm}\lesssim700$ GeV and $m_{\phi_{1,2}}\lesssim400$ GeV \cite{CMS:2020bfa,ATLAS:2019lff}.   For simplicity, we assume $m_{\psi^\pm}\gtrsim 1$ TeV to satisfy current constraints. 

For the proper lifetime of dark partner $\tau_{\phi_2}$, $\phi_2$ could induce the displaced vertex signature at colliders. We notice that the previous study only discusses the decay length of dark partner \cite{DiazSaez:2024nrq}. To obtain the promising region more precisely, the production cross section should  also be considered. In this section, we focus on the Higgs portal channels
\begin{eqnarray}\label{Eqn:Hpp22}
		pp\to h^{(\star)}&\to&\phi_1\phi_2\to \phi_1 \phi_1 \bar{f}f,\\
		pp\to h^{(\star)}&\to&\phi_2\phi_2\to\phi_1\phi_1\bar{f}f\bar{f}f,
\end{eqnarray}
which depend on the coupling $\lambda_{12}$ and $\lambda_2$.
The displaced vertex signature of dark partner $\phi_2$ through the Yukawa portal interaction will be considered in Section \ref{SEC:YP2}. 

The number of events for the one displaced vertex $N_{\rm DV}$ is calculated as \cite{Urquia-Calderon:2023dkf}
\begin{eqnarray}\label{Eqn:Hn1}
	N_{\rm DV}&=& ~L_{\rm int}\times\sigma(pp\to\phi_1\phi_2)\times P_{\rm dec}\times \BR_{\rm vis}\times\kappa_1
	 \\ \nonumber
	& +& 2L_{\rm int}\times\sigma(pp\to\phi_2\phi_2)\times P_{\rm dec}\times \BR_{\rm vis}\times\kappa_2,
\end{eqnarray}
where $L_{\rm int}$ is the integrated luminosity. We set $L_{\rm int}=139~\text{fb}^{-1}$  for LHC and $L_{\rm int}=3000~\text{fb}^{-1}$ for HL-LHC. At the hadron collider, the SM Higgs $h$ is dominantly generated through the gluon fusion process. The cross section $\sigma(pp\to\phi_1\phi_2,\phi_2\phi_2)$ is calculated with MadGraph5$\_$aMC@NLO \cite{Alwall:2011uj}. $P_{\rm dec}$ is the probability that $\phi_2$ will decay within the fiducial volume. In this Higgs portal scenario, the branching ratio of $\phi_2$ into visible final states is BR$_\text{vis}\simeq1$. For an optimistic estimation, the detection efficiency $\kappa_{1,2}$ is fixed to one. Assuming vanishing background, the sensitive region is derived with $N_{\rm DV}=3$, which corresponds to the 95\% exclusion limit.

The sensitive regions of the one DV signature at LHC and future HL-LHC are displayed in Figure \ref{FIG:fig6}, where we have assumed $\lambda_{12}=\lambda_2$ for illustration. In panel (a), $\lambda_{2}=1$ and $\Delta m_\phi=1$ GeV are fixed. The decay length  $c \tau_{\phi_2}$ of the coannihilation sample is less than $10^8$ m, and that of the coscattering sample is slightly larger, approximately $\mathcal{O}(10^9)$ m. Therefore, these allowed samples are far beyond the reach of current LHC and future HL-LHC. The DV signature of HL-LHC is sensitive to the parameter space with $m_{\phi_{1,2}}<m_h/2$ and $c\tau_{\phi_2}\lesssim10^5$ m. However, such a region is already disallowed by current ATLAS searches of Higgs decays due to a too large value of $\lambda_2$.

As shown in panel (b) of Figure \ref{FIG:fig6}, increasing the mass splitting $\Delta m_\phi$  to 10 GeV can significantly decrease the decay length, which results in the allowed samples located in the sensitive region of DV signature at LHC. The allowed samples with $m_{\phi_2}\lesssim110$ GeV can be tested at LHC, most of which correspond to the coannihilation scenario. In the future, the HL-LHC would expand this scope to $m_{\phi_2}\simeq200$ GeV, thus some coscattering samples will be probed. Although the TeV scale dark partner also predicts proper decay length $c \tau_{\phi_2}\sim\mathcal{O}(10^{-2})$ m, the cross section via off-shell Higgs is too small, so such heavy samples are beyond the scope of LHC.

In panel (c) of Figure \ref{FIG:fig6}, the decrease of $\lambda_{2}$ to 0.1 causes the reduction in the sensitive space of the DV signature. Similar to panel (a), it is also difficult to detect any surviving samples as the predicted decay length $c\tau_{\phi_2}\gtrsim 10^4$ m.  However, it has a turnaround when $\Delta m_\phi=10$ GeV in panel (d). LHC is sensitive to samples with $m_{\phi_2}\lesssim150$ GeV, and will increase to 300 GeV at HL-LHC.  Certainly, coannihilation  is the only mechanism responsible for their generation. In summary, the mass splitting $\Delta m_\phi=1$ GeV is not favored by the search for long-lived $\phi_2$ due to too large decay length $c\tau_{\phi_2}$. Nevertheless, this situation will improve conveniently if increasing $\Delta m_\phi=10$ GeV.

\section{Yukawa Portal Scenario}\label{SEC:YS}

In this alternative scenario, the contributions of the Yukawa portal interactions are predominant, which can be realized with $y_{i\alpha} \gg \lambda_{ij}$.  The needed free parameters are
\begin{eqnarray}
	\{m_{\phi_1},\Delta{m_\phi},\Delta{m_F},y_1,y_2\},
\end{eqnarray}
where $\Delta{m_F}=m_F-m_{\phi_2}$. When neglecting the final state lepton masses, the annihilation cross sections of dark scalars depend on the product of Yukawa couplings as $\sum_{\alpha \beta} |y_{i\alpha} y_{i\beta}^{*}|^2$ \cite{Kubo:2006yx}. For convenience, we define the effective Yukawa coupling $y_{i}=\sqrt{|y_{ie}|^2+|y_{i\mu}|^2+|y_{i\tau}|^2}$.

\subsection{Relic Density}\label{SEC:YRD}

In this scenario, the annihilation and conversion processes of dark scalars $\phi_{1,2}$ are mediated by the dark fermion $\Psi$ through  the Yukawa interactions in Equation \eqref{Eqn:Yuk}. Therefore, the involved SM particles consist solely of leptons $\ell$, i.e., three flavors of charged leptons and neutrinos. The related Boltzmann equations are as follows:
\begin{eqnarray}\label{Eqn:YBE1}
	\frac{dY_{\phi_1}}{dx} &= & -\frac{s}{\mathcal{H}x}\bigg[ \left<\sigma v\right>_{\phi_1\phi_1\to \bar\ell\ell}\left(Y_{\phi_1}^2-(Y_{\phi_1}^{\eq})^2\right)+\left<\sigma v\right>_{\phi_1\phi_2\to\bar\ell\ell}\left(Y_{\phi_1}Y_{\phi_2}-Y_{\phi_1}^{\eq}Y_{\phi_2}^{\eq}\right)\nonumber \\
	&-& \left<\sigma v\right>_{\phi_2\ell\to\phi_1\ell}\left(Y_{\phi_2}Y_{\ell}^{\eq}-\frac{Y_{\phi_2}^{\eq}}{Y_{\phi_1}^{\eq}}Y_{\phi_1}Y_{\ell}^{\eq}\right)
	-\frac{\Gamma_{\phi_2\to\phi_1\bar{\ell}\ell}}{s}\left(Y_{\phi_2}-\frac{Y_{\phi_2}^{\eq}}{Y_{\phi_1}^{\eq}}Y_{\phi_1}\right)\bigg],
\end{eqnarray}
\begin{eqnarray}\label{Eqn:YBE2}
	\frac{dY_{\phi_2}}{dx} &= & -\frac{s}{\mathcal{H}x}\bigg[ \left<\sigma v\right>_{\phi_2\phi_2\to\bar\ell\ell}\left(Y_{\phi_2}^2-(Y_{\phi_2}^{\eq})^2\right)+\left<\sigma v\right>_{\phi_1\phi_2\to\bar\ell\ell}\left(Y_{\phi_1}Y_{\phi_2}-Y_{\phi_1}^{\eq}Y_{\phi_2}^{\eq}\right)\nonumber \\
	&+& \left<\sigma v\right>_{\phi_2\ell\to\phi_1\ell}\left(Y_{\phi_2}Y_{\ell}^{\eq}-\frac{Y_{\phi_2}^{\eq}}{Y_{\phi_1}^{\eq}}Y_{\phi_1}Y_{\ell}^{\eq}\right)
	+\frac{\Gamma_{\phi_2\to\phi_1\bar{\ell}\ell}}{s}\left(Y_{\phi_2}-\frac{Y_{\phi_2}^{\eq}}{Y_{\phi_1}^{\eq}}Y_{\phi_1}\right)\bigg],
\end{eqnarray}
where the definitions of the various variables are consistent with those in Equation~\eqref{Eqn:HBE1} and Equation~\eqref{Eqn:HBE2}.
Slightly different from the Higgs portal scenario, here the conversion process $\phi_2\phi_i\to\phi_1\phi_j$ disappears. The  decay width of $\phi_2\to\phi_1\bar{\ell}\ell$ mediated by the dark fermion $\Psi$ is estimated as:
\begin{eqnarray}\label{Eqn:YG2}
	\tilde{\Gamma}_{\phi_2\to\phi_1\bar{\ell}\ell}\simeq\frac{y_1^2 y_2^2  (\Delta{m_\phi})^5}{240\pi^3(m_{\phi_1}+\Delta{m_\phi}+\Delta{m_F})^4}\times\theta^\prime(\Delta{m_\phi}-2m_\ell).
\end{eqnarray}

\begin{figure}
	\begin{center}
		\includegraphics[width=0.45\linewidth]{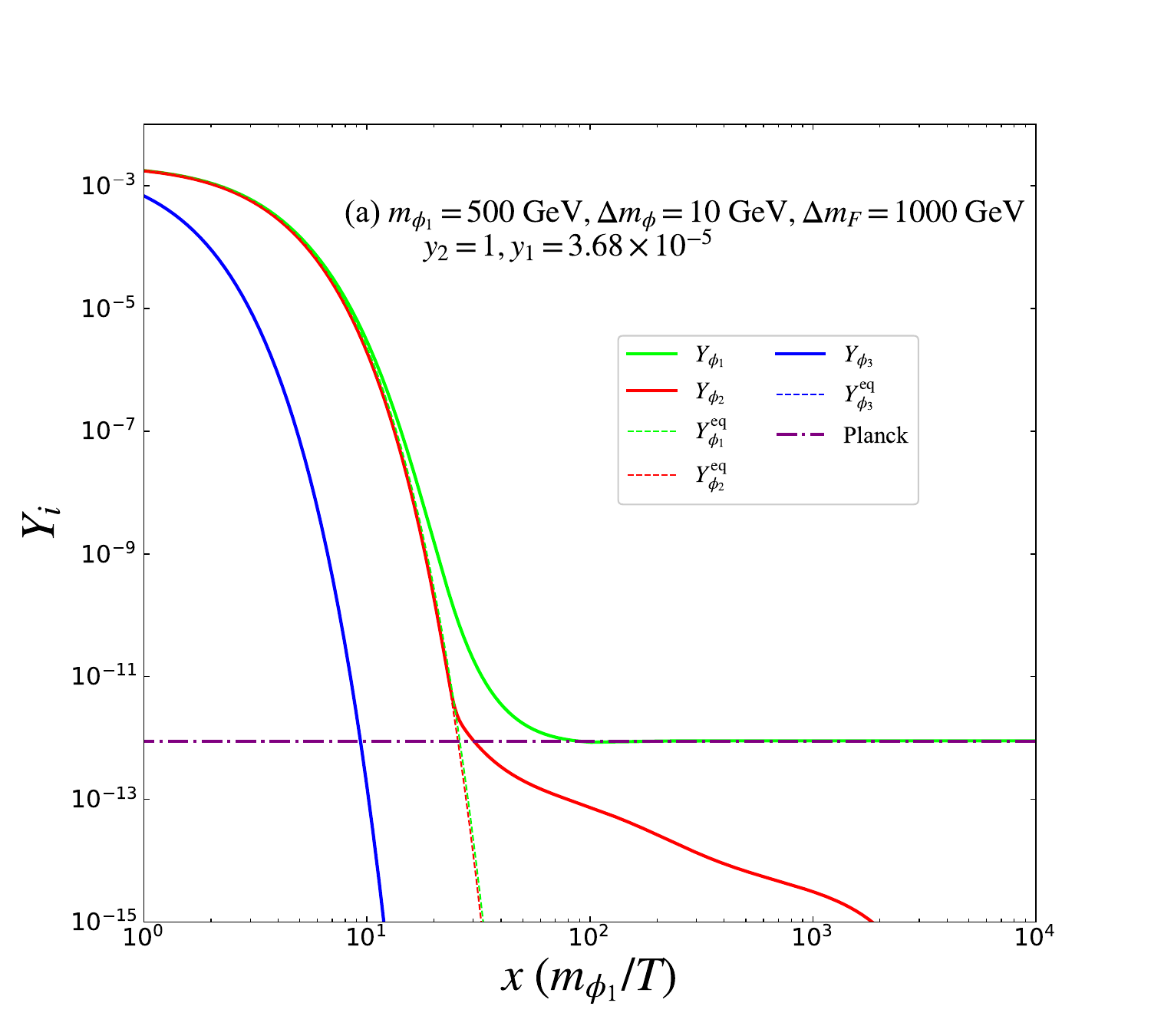}
		\includegraphics[width=0.45\linewidth]{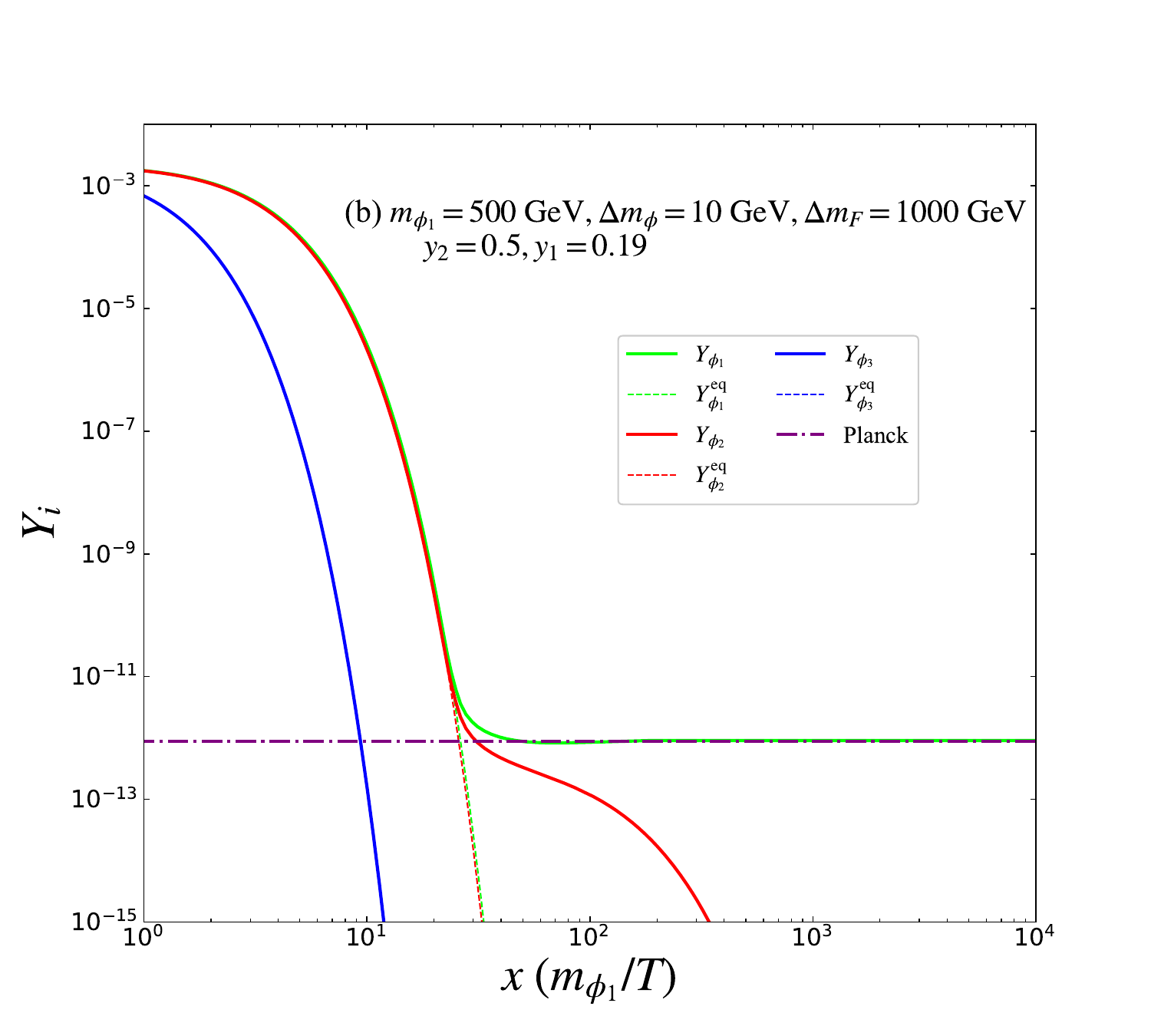}
		\includegraphics[width=0.45\linewidth]{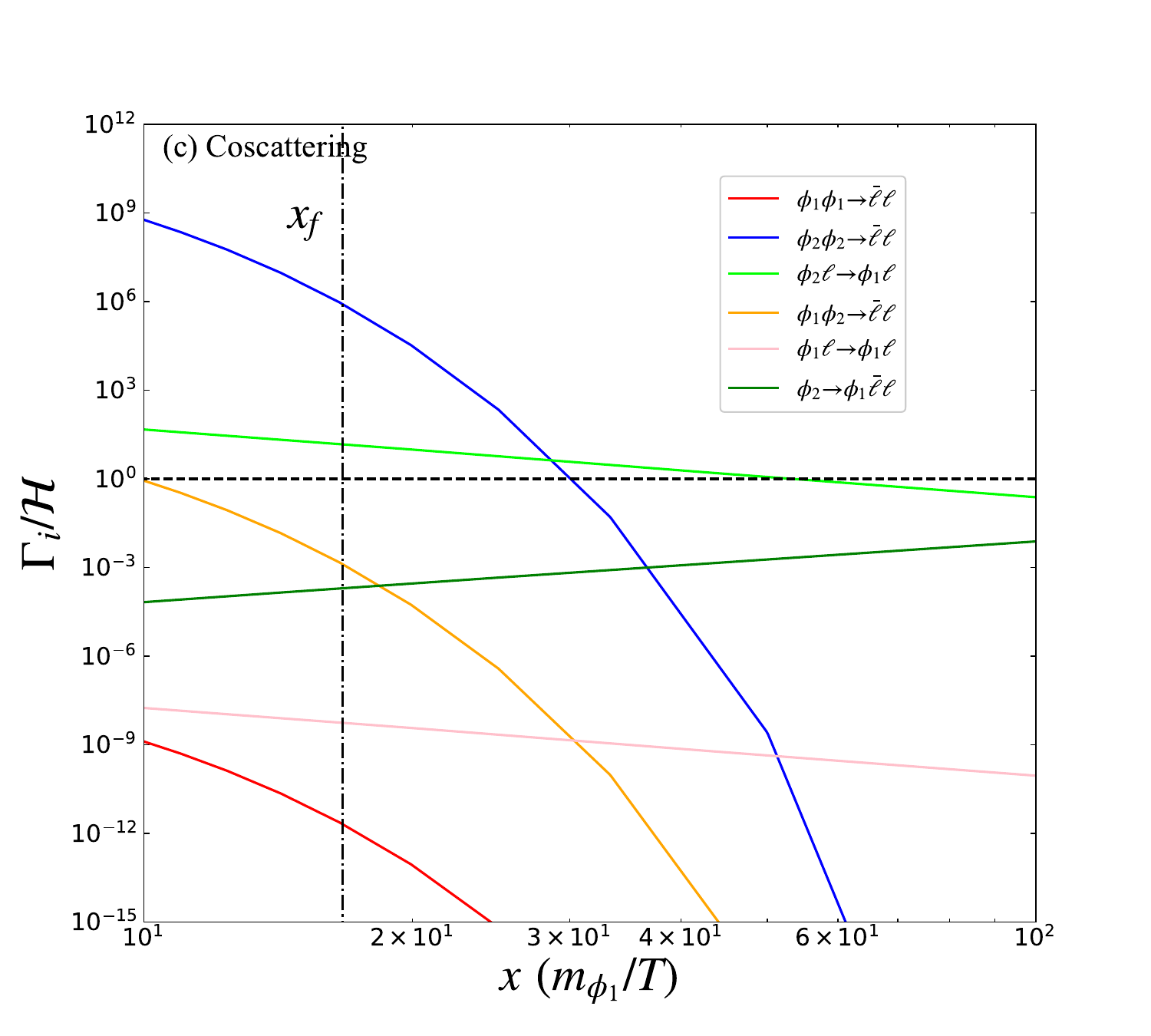}
		\includegraphics[width=0.45\linewidth]{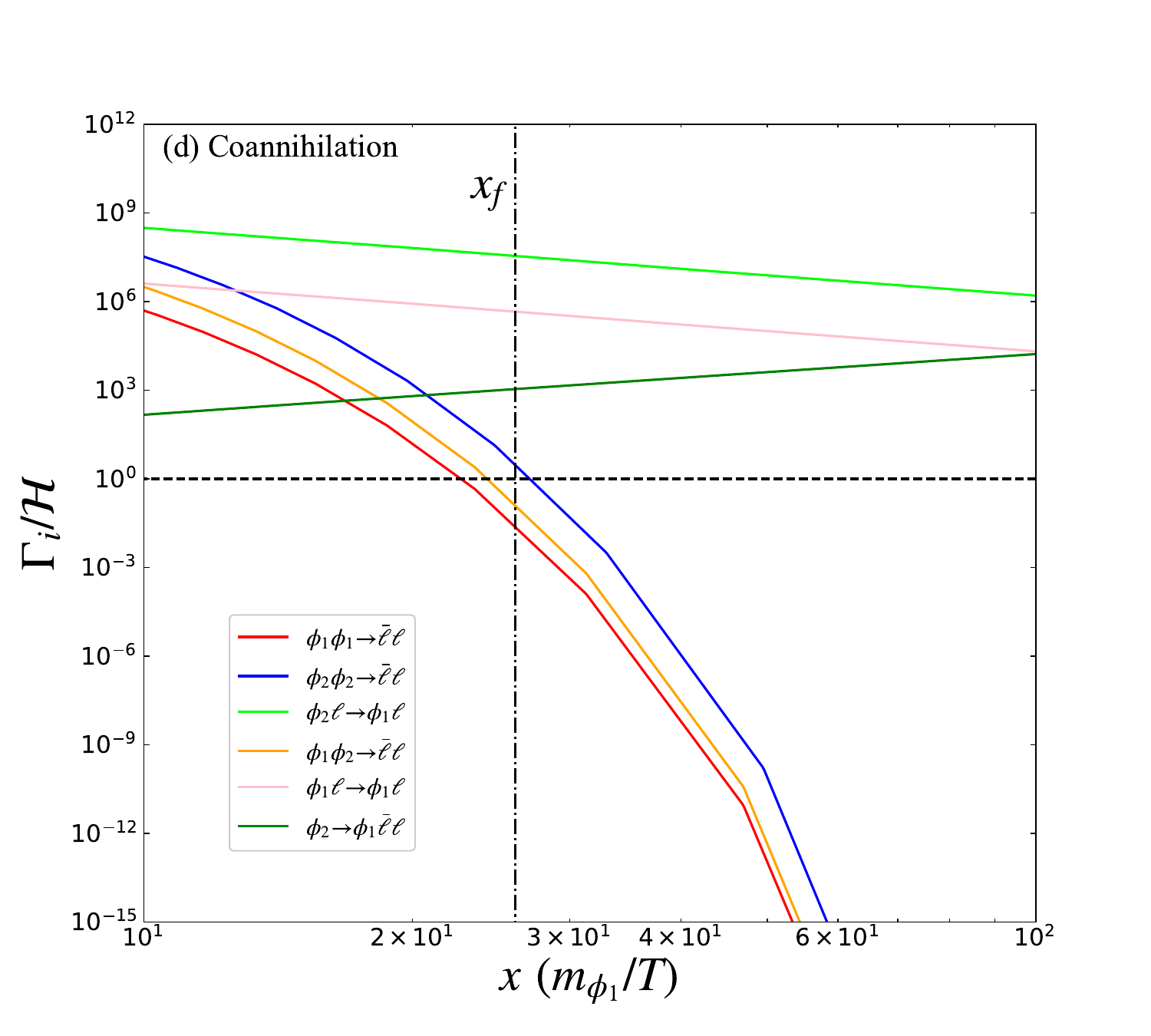}
	\end{center}
	\caption{Coscattering (a) and coannihilation (b) benchmarks for the Yukawa portal scenario. The thermal rates in panels (c) and (d) correspond to the results in panels (a) and (b), respectively. The representations of various symbols can be referenced in Figure~\ref{FIG:fig1}.
	}
	\label{FIG:fig7}
\end{figure}

In analogy to the Higgs portal scenario, we first utilize two benchmark points to describe the evolution of two different phases in Figure~\ref{FIG:fig7},   where the Boltzmann Equations \eqref{Eqn:YBE1} and \eqref{Eqn:YBE2} are evolved numerically up to $x=10^4$ to include the late-time $\phi_2\to \phi_1$ conversions.   For relatively small mass splitting $\Delta m_F\lesssim $ 10~GeV, the conversion process $\Psi\to \ell \phi_{1,2}$ could have a great impact on the evolution of dark scalars \cite{Heeck:2022rep}. A relatively large hierarchy $\Delta m_F=1000$ GeV is then implemented to avoid such interference, which also satisfies the LHC dilepton exclusion limit \cite{CMS:2020bfa}. In panel (a) of Figure~\ref{FIG:fig7}, it is obvious that coscattering $\phi_2 \ell \to \phi_1 \ell$  occurs when the interaction rate $\Gamma_{\phi_2\ell\to\phi_1\ell}$ is situated between $\Gamma_{\phi_2\phi_2\to\bar \ell\ell}$ and $\Gamma_{\phi_1\phi_1\to\bar\ell\ell}$.  Different from the Higgs portal scenario, the tiny $y_1$ not only influences the inelastic scattering $\phi_2\ell\to\phi_1\ell$, but also determines the elastic scattering $\phi_1\ell\to\phi_1\ell$.  Thus we have $\Gamma_{\phi_1\ell\to\phi_1\ell}\ll\mathcal{H}$ during the thermal decoupling.  Hence  we assume that the kinetic equilibrium of $\phi_1$ is determined by $\Gamma_{\phi_2\ell\to\phi_1\ell}\gg\mathcal{H}$.  Next, we reduce the annihilation rate of $\phi_2$ by fixing $y_2=0.5$ in panel (b) of Figure~\ref{FIG:fig7}. The insufficient $\phi_2\phi_2\to\bar \ell\ell$ prevents $\phi_1$ from meeting the observation through the conversion process. As a result, $y_1=0.19$ quickly increases to the coannihilation regime $\phi_1 \phi_2 \to \bar{\ell}\ell$. The pair annihilation $\phi_1\phi_1\to \bar{\ell}\ell$ will become the dominant contribution when decreasing the Yukawa coupling $y_2$ to 0.1, which corresponds to the canonical WIMP phase.   Additionally, the heaviest dark scalar $\phi_3$ has a Yukawa coupling at $\mathcal{O}(1)$ with hierarchical $m_{\phi_3}=1600~\GeV$, so the extremely strong interaction almost keeps it in the thermal equilibrium throughout, thus having negligible impact on the production of conversion dark matter. 

 As will shown in Section \ref{SEC:YP1}, the constraints from lepton flavor violation favor a hierarchical Yukawa coupling structure as $|y_{ie}|\ll |y_{i\mu}|\lesssim |y_{i\tau}|\sim \mathcal{O}(1)$. In terms of neutrino masses, such a highly hierarchical $y_{i\alpha}$ can also successfully account for the neutrino-oscillation constraints.  By requiring $y_\chi=0.87$ and  $\omega_{12}=0.72+1.87i$, $\omega_{13}=2.6+1.1i$, $\omega_{23}=10+10.6i$, the benchmark Yukawa coupling for the coscattering case in Figure \ref{FIG:fig7} is
\begin{align}
	|y_{i\alpha}| \simeq \begin{pmatrix}
		7.6\times10^{-6} && 1.6\times10^{-5} && 3.3\times10^{-5} \\
		9.7\times10^{-4} && 2.6\times10^{-1} && 9.6\times10^{-1} \\
		1.4\times10^{-3}  && 3.9\times10^{-1} && 1.4
	\end{pmatrix},
\end{align}	
meanwhile, the benchmark Yukawa coupling for the coannihilation case in Figure \ref{FIG:fig7} is 
\begin{align}
	|y_{i\alpha}| \simeq \begin{pmatrix}
		3.1\times10^{-2} && 5.7\times10^{-3} && 1.9\times10^{-1} \\
		8.4\times10^{-2} && 1.5\times10^{-2} && 4.9\times10^{-1} \\
		1.3\times10^{-1} && 2.4\times10^{-2} && 7.9\times10^{-1}
	\end{pmatrix},
\end{align}
by fixing $\omega_{12}=0.01+1.6i$, $\omega_{13}=0.01+9.8i$, $\omega_{23}=3+1.7i$.
In principle, other combinations of $y_{i\alpha}$ can be achieved by adjusting the three mixing angles. With relatively large Yukawa couplings $y_{i\alpha}$ and $y_\chi$, tiny neutrino mass is obtained due to certain structural cancellation of the Yukawa coupling $y_{i\alpha}$ for the benchmarks \cite{Kersten:2007vk}.  Therefore, the naive estimation of neutrino mass scale in Equation \eqref{Eqn:mves} is not accurate in the Yukawa scenario.  We then substitute these two complete Yukawa matrices into Equation \eqref{Eqn:Nm}. Due to the structural cancellations,  the resulting neutrino mass matrix is at the order of $\mathcal{O}(0.01)~\eV$. The explicit values of the resulting neutrino mass matrix are
\begin{align}
	|m^\nu| = \begin{pmatrix}
		1.3\times10^{-2} && 1.6\times10^{-2} && 1.4\times10^{-2} \\
		1.6\times10^{-2} && 2\times10^{-2} && 1.5\times10^{-2} \\
		1.4\times10^{-2} && 1.5\times10^{-2} && 2.6\times10^{-2}
	\end{pmatrix}~\eV,
\end{align}
which are equivalent to the results from direct neutrino oscillation predictions as $m^\nu = U\hat{m}_\nu U^T$.
Moreover, since both Yukawa matrices are obtained by the Casas-Ibarra parametrization in Equation \eqref{Eq:CIy}, the resulting $m^\nu$ is identical, when the corresponding inputs of neutrino oscillation parameters are the same as in Equation~\eqref{Eqn:Bfnm}.

\begin{figure}
	\begin{center}
		\includegraphics[width=0.45\linewidth]{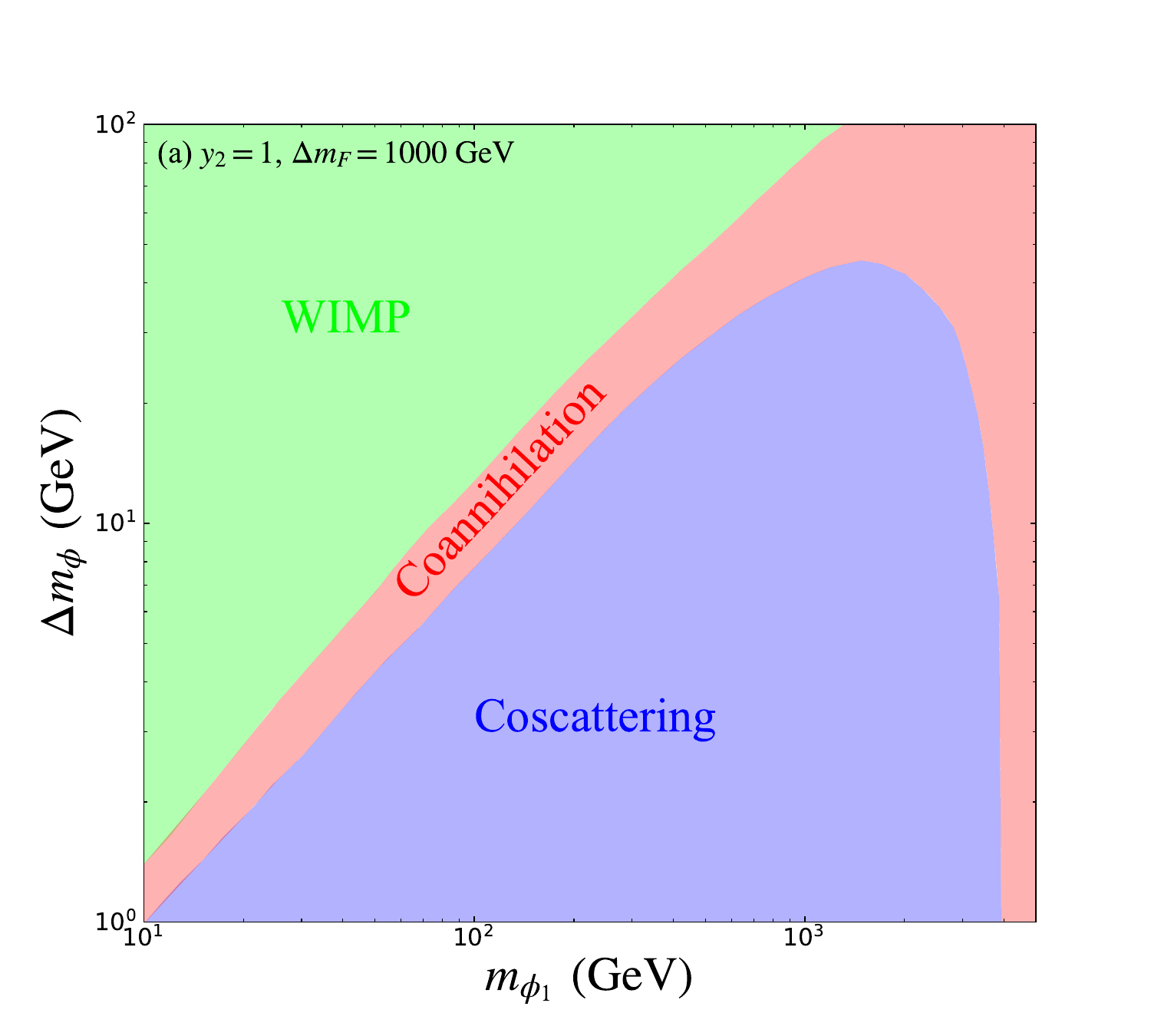}
		\includegraphics[width=0.45\linewidth]{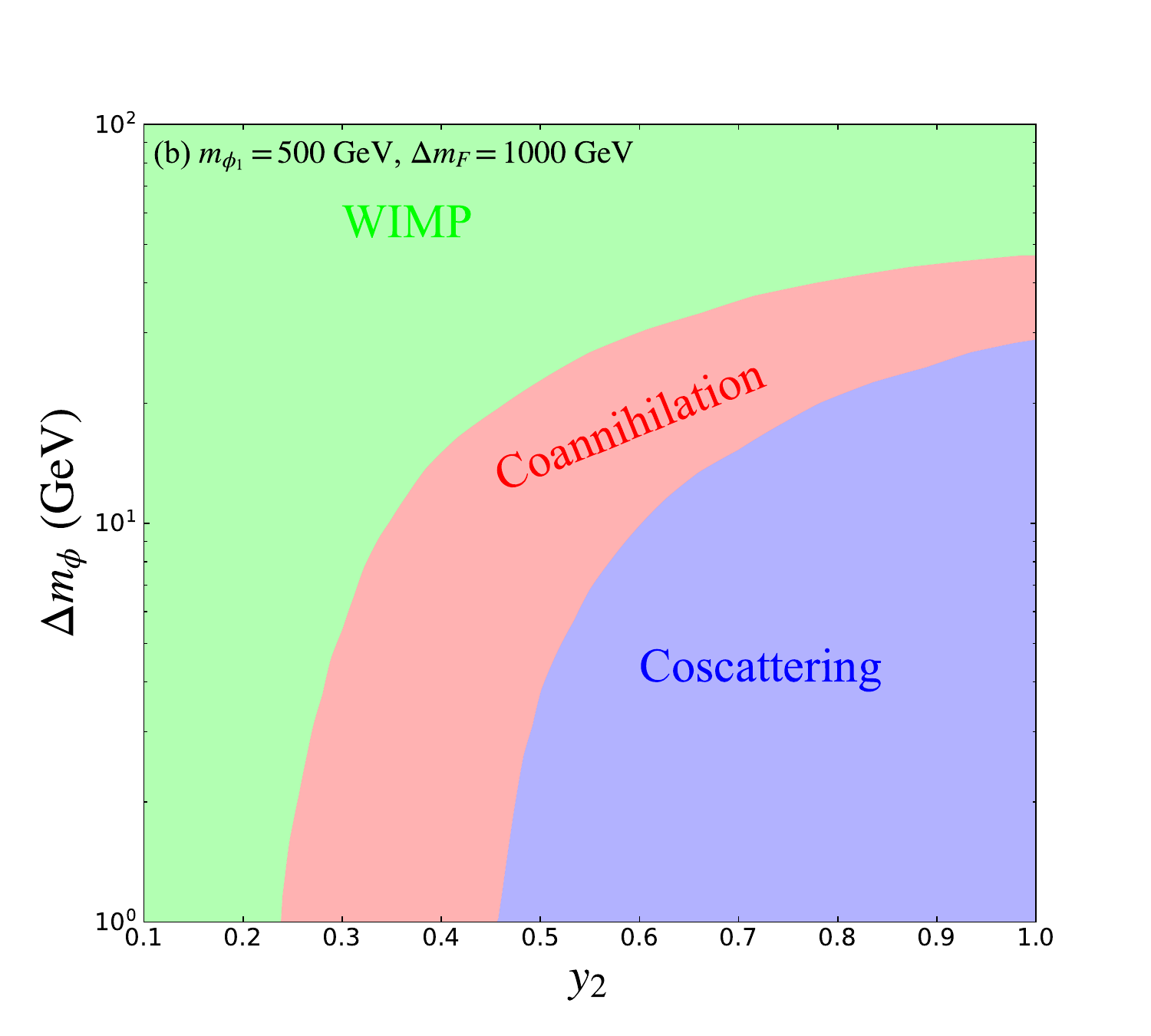}
	\end{center}
	\caption{Freeze-out phase diagrams in the parameter spaces of  $\Delta{m_\phi}-m_{\phi_1}$ in panel (a) and  $\Delta{m_\phi}-y_{2}$ in panel (b) for the Yukawa portal scenario. Definitions of different colors are consistent with those presented in Figure~\ref{FIG:fig2}.
	}
	\label{FIG:fig8}
\end{figure}

In Figure~\ref{FIG:fig8}, we investigate the specific distribution of different phases.   With $y_2=1$ and $\Delta m_F=1000$~GeV in panel (a), the coscattering region dominated by $\phi_2 \ell \to \phi_1 \ell$  in the Yukawa portal scenario is explicitly larger than that in the Higgs portal scenario. We find that the mass splitting $\Delta m_\phi$ for coscattering exhibits rapid growth for sub-TeV $m_{\phi_1}$, even reaching  $\Delta m_\phi\simeq40$ GeV, which is nearly twice as large as the upper limit of the Higgs portal scenario. For sub-TeV dark matter, the coannihilation scenario  $\phi_1 \phi_2 \to \bar{\ell}\ell$ is confined to a narrow region, which  roughly satisfies the condition $\Delta m_\phi/m_{\phi_1}\simeq0.1$. Above the TeV scale, the coscattering region  declines as $m_{\phi_1}$ increases. The upper limit $m_{\phi_1}\lesssim4$ TeV of the coscattering region in the Yukawa portal scenario is much higher than that in the Higgs portal scenario, which is mainly due to the more intense  reaction of $\phi_2$ for the continuous conversion to $\phi_1$ through the Yukawa interaction. For the larger ratio of $\Delta m_\phi/m_{\phi_1}\gtrsim0.15$, WIMP  annihilation $\phi_1 \phi_1 \to \bar{\ell}\ell$ becomes the dominant contribution.

We then explore the dependence of $y_2$ on various phases in panel (b) of Figure~\ref{FIG:fig8}. Analogous to the Higgs portal scenario, coscattering  $\phi_2 \ell \to \phi_1 \ell$  distributes over the region $y_2\gtrsim0.4$. The WIMP  $\phi_1 \phi_1 \to \bar{\ell}\ell$ exhibits a significant dependence on $y_2$, which is different from the Higgs portal scenario with fixed $\lambda_1$. Furthermore, the space with $y_2\lesssim 0.2$ will be entirely dominated by WIMP. The coannihilation  $\phi_1 \phi_2 \to \bar{\ell}\ell$ can only be distributed in the band between WIMP and coscattering. The Yukawa coupling $y_1$ determines  the interaction rates of both pair annihilation $\phi_1 \phi_1\to \bar{\ell}\ell$ and coannihilation $\phi_1\phi_2\to \bar{\ell}\ell$. A smaller $y_2$ typically requires a larger $y_1$ to satisfy the observed relic density in the Yukawa portal scenario, which leads to an extreme expansion of the WIMP domain and ultimately compresses the space of coannihilation. In comparison, the pair annihilation $\phi_1 \phi_1\to \SM \SM$ and coannihilation $\phi_1\phi_2\to \SM\SM$ are separately determined by the coupling $\lambda_1$ and $\lambda_{12}$ in the Higgs portal scenario, thus coannihilation can occupy a relatively larger parameter space when $\lambda_{12}\gg\lambda_1$.

In summary, coscattering $\phi_2 \ell \to \phi_1 \ell$ favors relatively large $y_2\sim1$ but tiny $y_1\lesssim10^{-4}$ with degenerate $m_{\phi_1}\simeq m_{\phi_2}$. Coannihilation $\phi_1 \phi_2 \to \bar{\ell}\ell$ is dominant in the narrow region with $\Delta m_\phi/m_{\phi_1}\sim0.1$ for proper $y_{1,2}$. And larger mass splitting results in WIMP $\phi_1 \phi_1 \to \bar{\ell}\ell$ when $y_1\sim\mathcal{O}(0.1)$. The origins of coupling hierarchy are similar to those in the Higgs portal scenario. However, it is noteworthy that,  here two BSM couplings $y_1$ and $y_2$ are involved. In certain parameter regions, the minimal attainable value of $y_1$ can be smaller than that of $\lambda_{12}$, this behavior is associated with both the relative magnitude of $y_2$ compared to the SM coupling in the Higgs portal scenario, and the mass of the mediator.

Based on the results in Figure \ref{FIG:fig8}, we also choose four specific scenarios: $\Delta{m_\phi}=1$ GeV or 10 GeV combined with $y_{2}=0.3$ or 1. The equilibrium condition of dark matter through the two-body decay $\Psi\to \ell \phi_1$ is $\Gamma_{\Psi\to\ell \phi_1}\simeq y_1^2 m_F/(16\pi)\gtrsim \mathcal{H}(T=m_F)$, which implies $y_1\gtrsim3\times10^{-7}$ when $m_F$ is at the TeV scale. It should be mentioned that varying the mass splitting $\Delta m_F$ could also affect the required Yukawa coupling. So the scanning parameters and ranges in the Yukawa portal scenarios are:
\begin{eqnarray}
	\begin{aligned}
		m_{\phi_1}\in[10,5000]~\GeV, y_{1}\in[10^{-6},1], \Delta {m_F}\in[100,2000]~\GeV.
	\end{aligned}
\end{eqnarray}
The large hierarchy $\Delta {m_F}$ is established to mitigate the influence of conversion from $\Psi$ to $\phi_{1,2}$.

\subsection{Phenomenology of Dark Matter $\phi_1$}\label{SEC:YP1}

In the previous Higgs portal scenario, the constraints from lepton flavor violating (LFV) are overlooked due to the relatively small Yukawa coupling. However, as shown in Figure \ref{FIG:fig8}, the coscattering region favors a relatively large value of $y_2$ in the Yukawa portal scenario. So it is essential to discuss the influence of LFV on the Yukawa couplings. The branching ratio of $\mu\to e\gamma$ can be expressed as \cite{Esch:2016jyx}
\begin{eqnarray}\label{Eqn:brm}
	\BR_{\mu\to e\gamma}=\frac{3\alpha_{\rm {em}}}{64\pi G_f^2m_{F}^4}\left|\sum\limits_{i}y_{ie}^{*}y_{i\mu }G(m_{\phi_i}^2/m_{F}^2)\right|^2,
\end{eqnarray}
where Fermi constant $G_f=1.17\times10^{-5}~\GeV^{-2}$ and fine structure constant $\alpha_{\rm {em}}=1/137$. And 
\begin{eqnarray}
	G(a)= \frac{2-3a-6a^2+a^3+6a\log a}{6(1-a)^4}.
\end{eqnarray}
Currently, the most stringent constraint from MEG experiment is $\BR_{\mu\to e\gamma}<1.5\times10^{-13}$~\cite{MEGII:2025gzr}, which roughly requires the condition $|\sum y_{ie} y_{i\mu}|\lesssim10^{-3}$ for TeV scale $m_F$. To simultaneously satisfy the LFV constraint and DM relic density, we need hierarchical Yukawa couplings as $|y_{ie}|\ll |y_{i\mu}|\lesssim |y_{i\tau}|\sim \mathcal{O}(1)$ \cite{Vicente:2014wga}. This implies that the dark scalars predominantly couple to the second and third generations of leptons in the Yukawa portal scenario.

Another constraint is the direct detection of DM $\phi_1$. The DM $\phi_1$ and nucleon scattering is mediated by the SM Higgs, whose cross section is determined by the coupling $\lambda_1$ as shown in Equation~\eqref{Eq:Hdd}. In the Yukawa portal scenario, the coupling $\lambda_1$ is assumed to be small enough ($\lambda_1\lesssim10^{-3}$) to satisfy the direct detection limit. Meanwhile, the mass splitting $\Delta m_\phi\geq1$ GeV is large enough to avoid the inelastic scattering constraints.

\begin{figure}
	\begin{center}
		\includegraphics[width=0.45\linewidth]{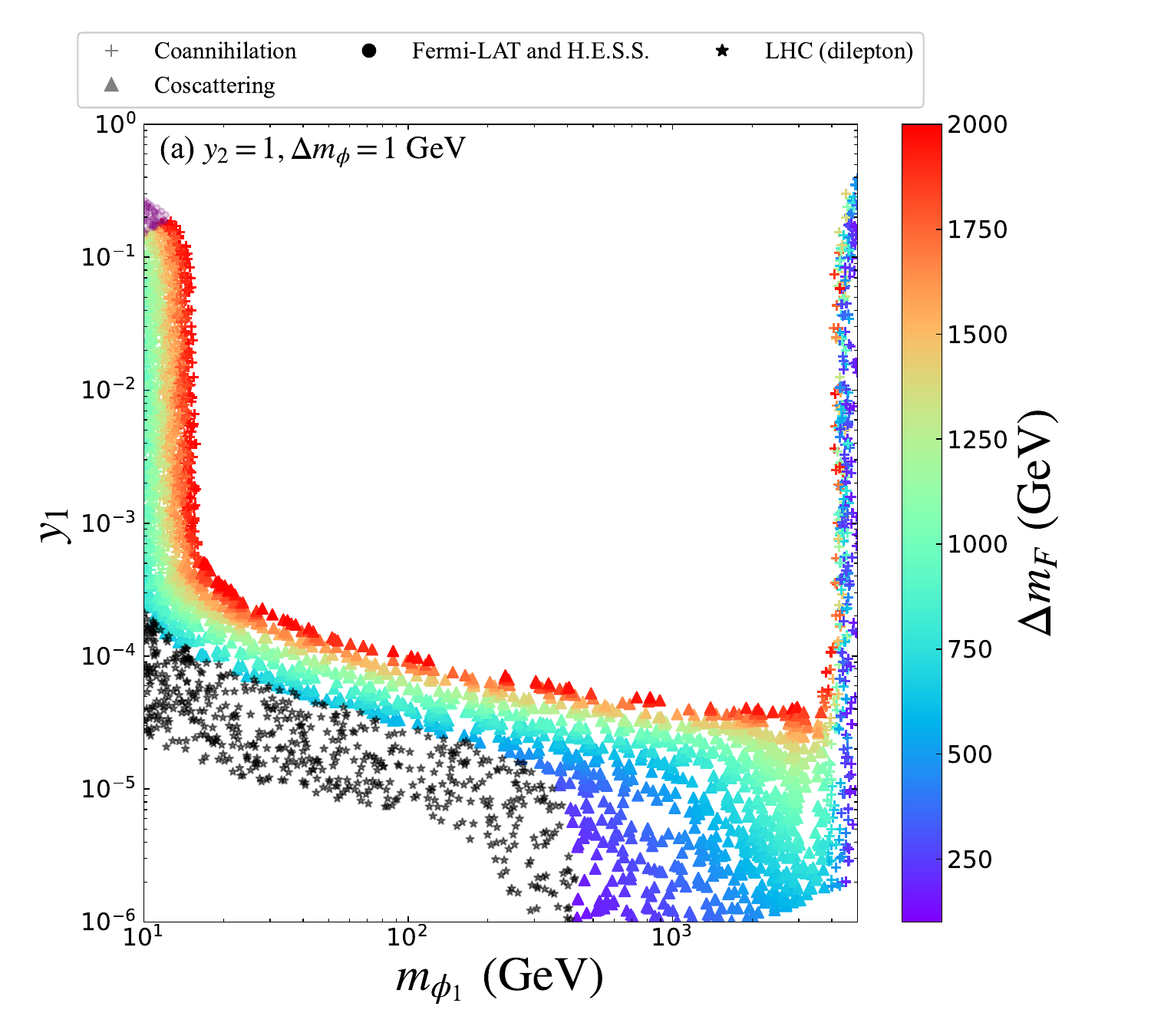}
		\includegraphics[width=0.45\linewidth]{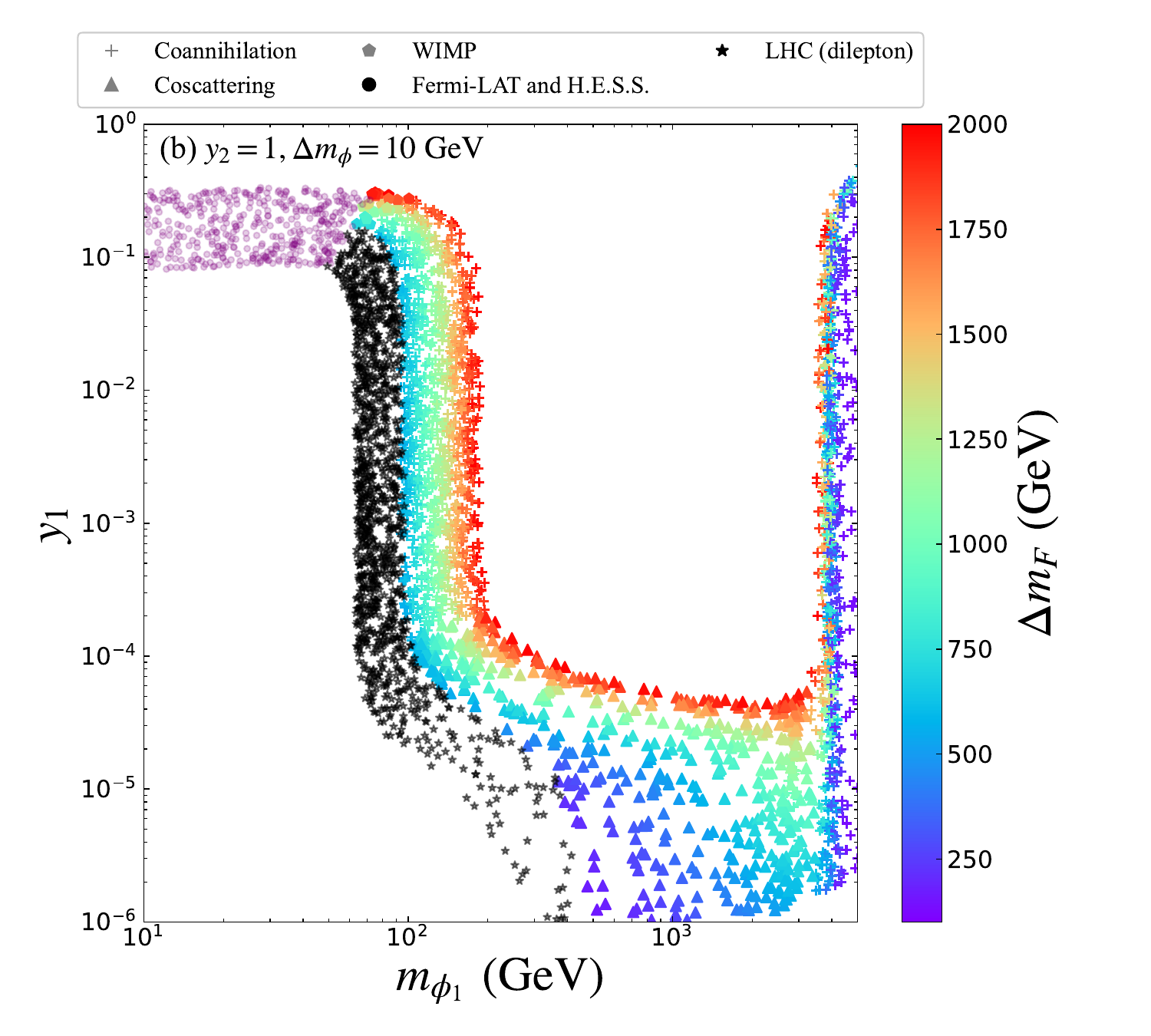}
		\includegraphics[width=0.45\linewidth]{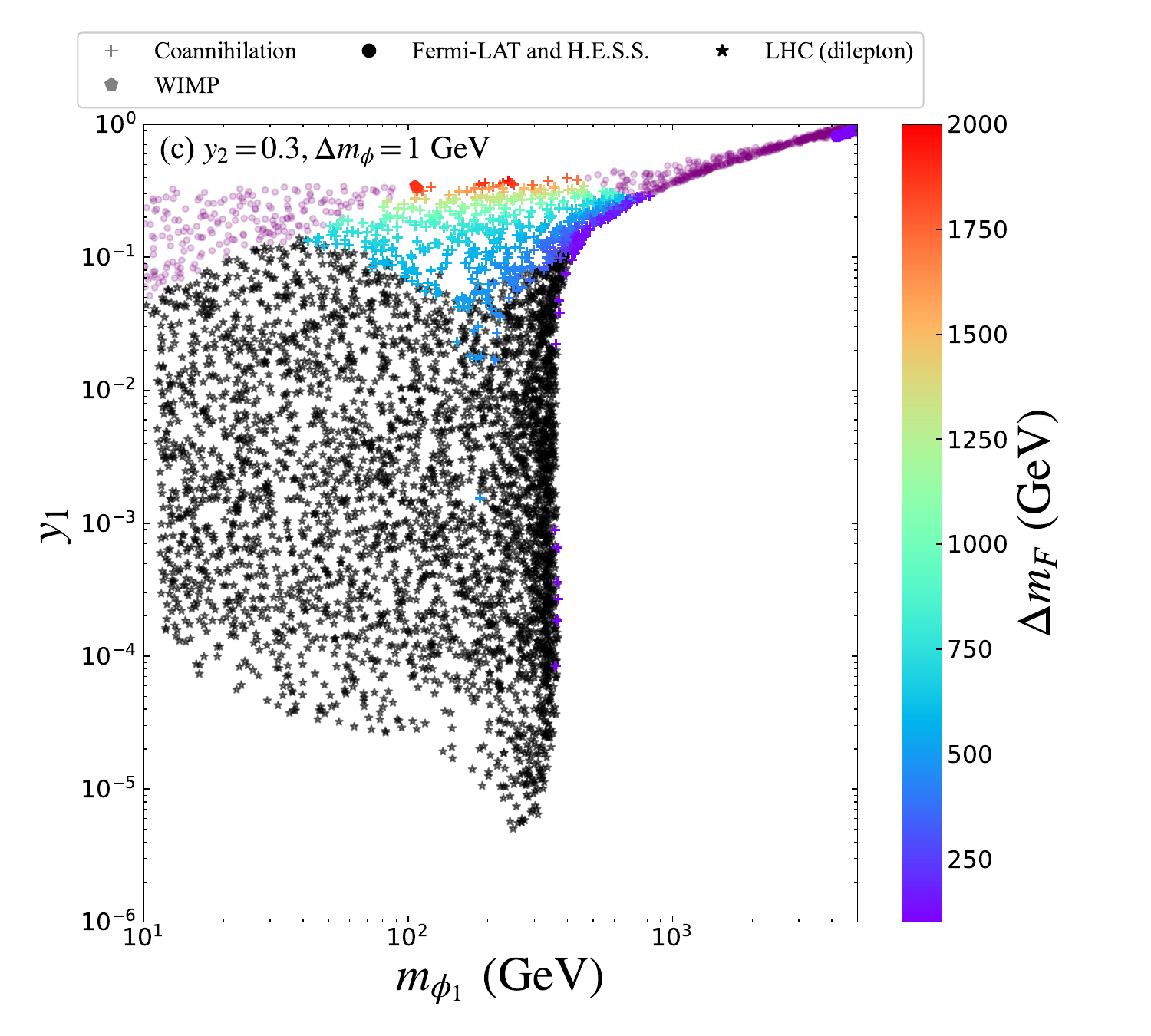}
		\includegraphics[width=0.45\linewidth]{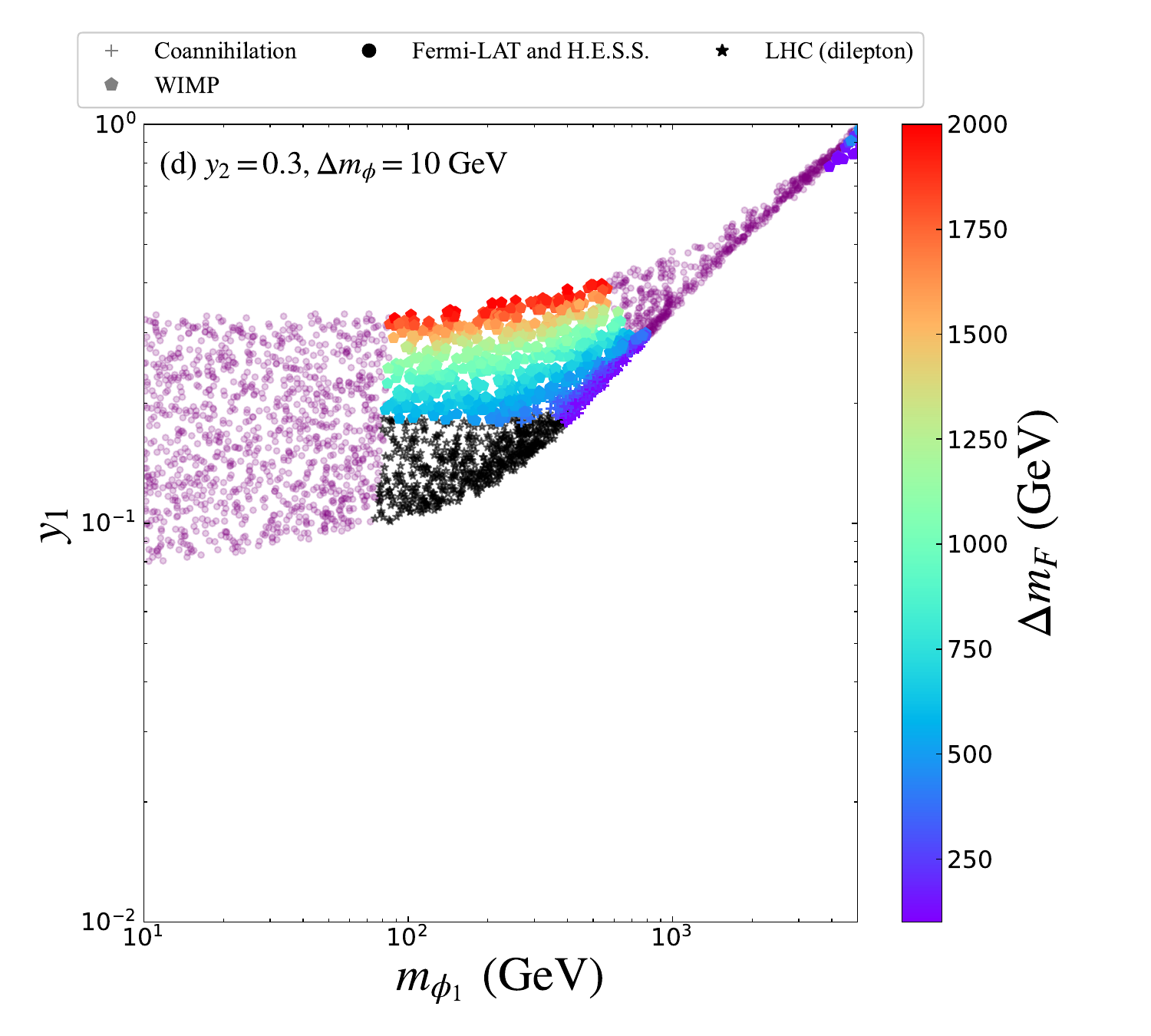}
	\end{center}
	\caption{Sample distribution in the $y_{1}-m_{\phi_1}$ space, which is colored by $\Delta m_F$. The shapes of coscattering and coannihilation are identical to those in Figure~\ref{FIG:fig3}, while additional WIMP samples  are shaped as pentagons. The purple samples are excluded by Fermi-LAT \cite{Fermi-LAT:2015att} and H.E.S.S. ~\cite{HESS:2016mib} experiments. The black samples are disfavored by the current LHC dilepton search \cite{Das:2020hpd}. 
	}
	\label{FIG:fig9}
\end{figure}

The scanning results are illustrated in Figure~\ref{FIG:fig9}. In panel (a) with $y_2=1$ and $\Delta m_\phi=1$ GeV, coscattering is within the region of $y_1\lesssim6\times10^{-4}$ and $m_{\phi_1}\lesssim4$ TeV, meanwhile, coannihilation is dominant the region of $y_{1}\gtrsim10^{-3}$ with $m_{\phi_1}\lesssim25$ GeV or $m_{\phi_1}\gtrsim4$ TeV. It is obvious that the upper and lower bounds of the samples correspond to the maximum and minimum value of $\Delta m_F$, which is mainly because a larger $\Delta m_F$ requires a larger coupling $y_1$ for the same value of interaction rate. As $\Delta m_\phi$ increases to 10 GeV in panel (b), the magnitude of $y_1$ possessed by coscattering remains relatively unchanged, but the mass distribution shifts towards larger $m_{\phi_1}$, namely above 70 GeV. Below 100 GeV, WIMP is viable with $y_{1}\sim\mathcal{O}(0.1)$, but most WIMP samples are excluded by Fermi-LAT \cite{Fermi-LAT:2015att}. With a fixed value of $\Delta m_F$, the coupling $y_1$ of coannihilation sharply decreases from the WIMP-favor value to the coscattering-favor value around 100 GeV or 4 TeV. In both panels (a) and (b), the black samples with $m_F\lesssim660$ GeV are excluded by the current LHC search of dilepton signature $\ell^+\ell^-+\cancel{E}_T$ \cite{Das:2020hpd}.

When $y_2$ drops to 0.3 with $\Delta m_\phi=1$ GeV in panel (c) of Figure~\ref{FIG:fig9}, we report that coscattering is compressed into the region with $4\times10^{-6}\lesssim y_1\lesssim2\times10^{-4}$, $\Delta m_F\lesssim500$ GeV and $m_{\phi_1}\lesssim 360$~GeV. However, such a region is completely excluded by the current LHC dilepton search \cite{Das:2020hpd}. Most allowed samples are dominated by coannihilation when $y_1\simeq10^{-1}$. With relatively small coupling $y_2$, the WIMP is dominant when $\Delta m_F\gtrsim1000$ GeV or $m_{\phi_1}\gtrsim1000$ GeV, which is almost excluded by indirect detection. In panel (d) with $y_2=0.3$ and $\Delta m_\phi=10$ GeV, there is no coscattering dominant sample. Nearly all allowed points belong to WIMP, with the corresponding coupling $y_{1}$ being greater than 0.2.  Coannihilation is confined to a small area with $m_{\phi_1}\sim600$ GeV near the lower boundary.

\begin{figure}
	\begin{center}
		\includegraphics[width=0.45\linewidth]{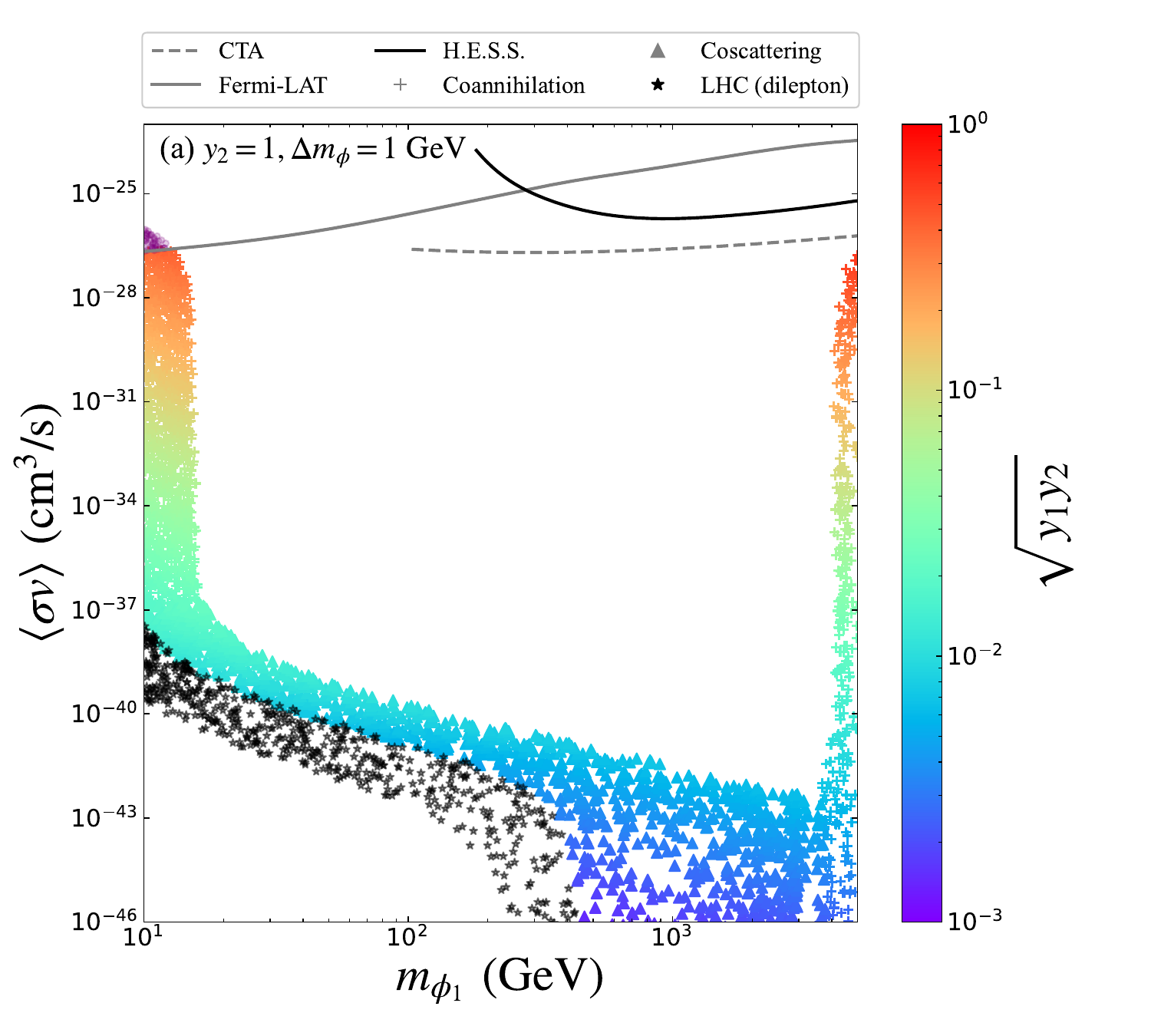}
		\includegraphics[width=0.45\linewidth]{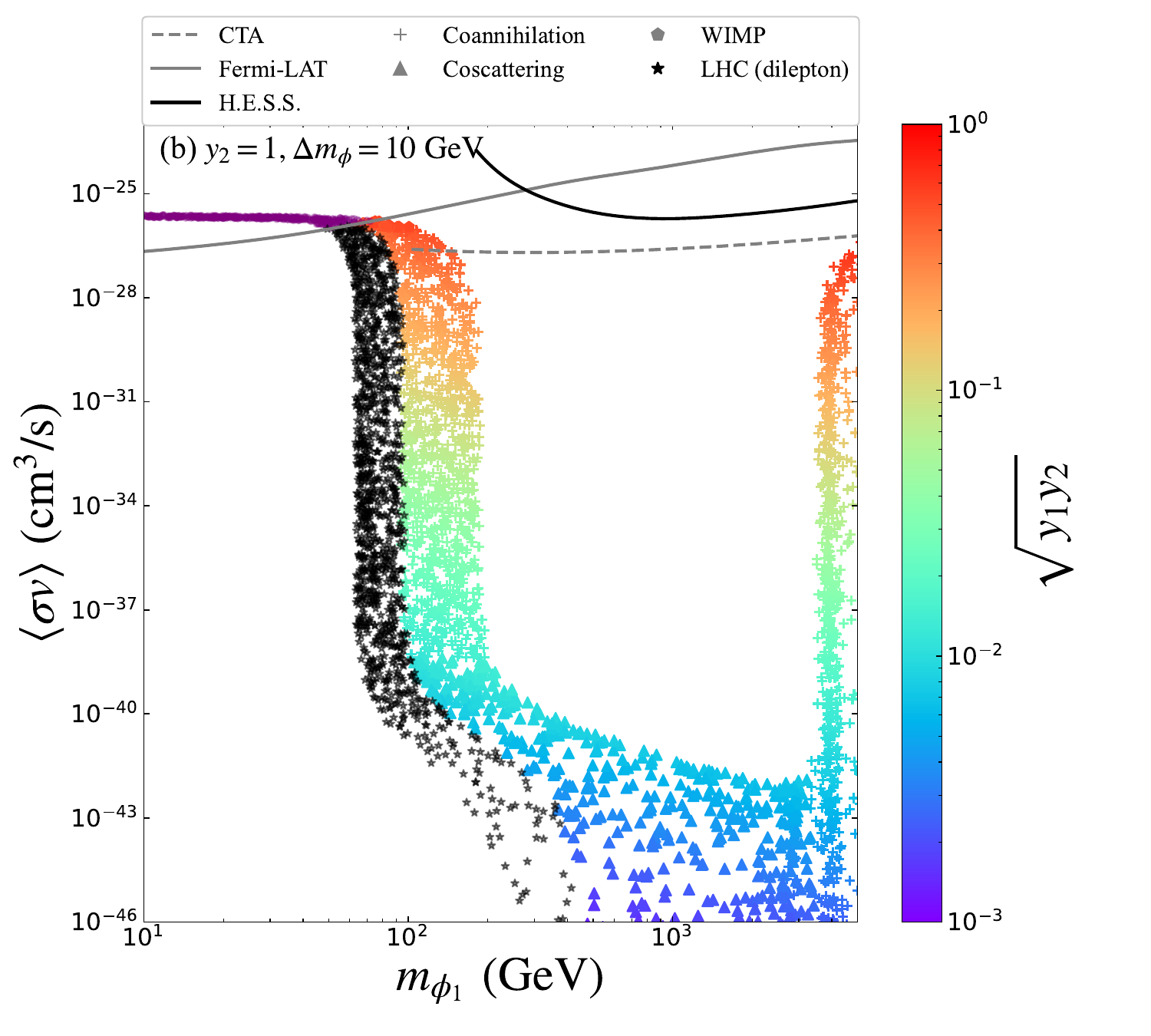}
		\includegraphics[width=0.45\linewidth]{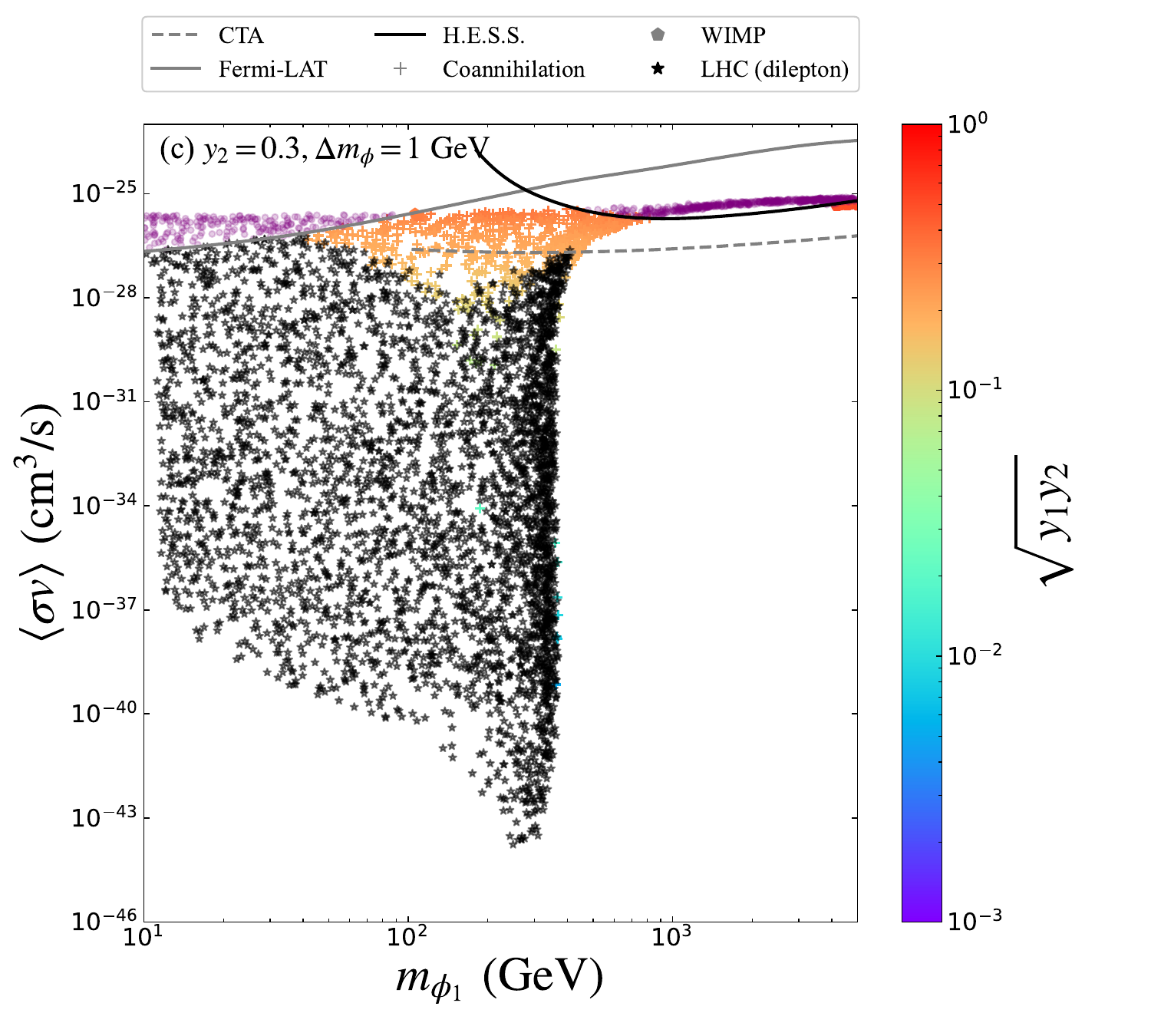}
		\includegraphics[width=0.45\linewidth]{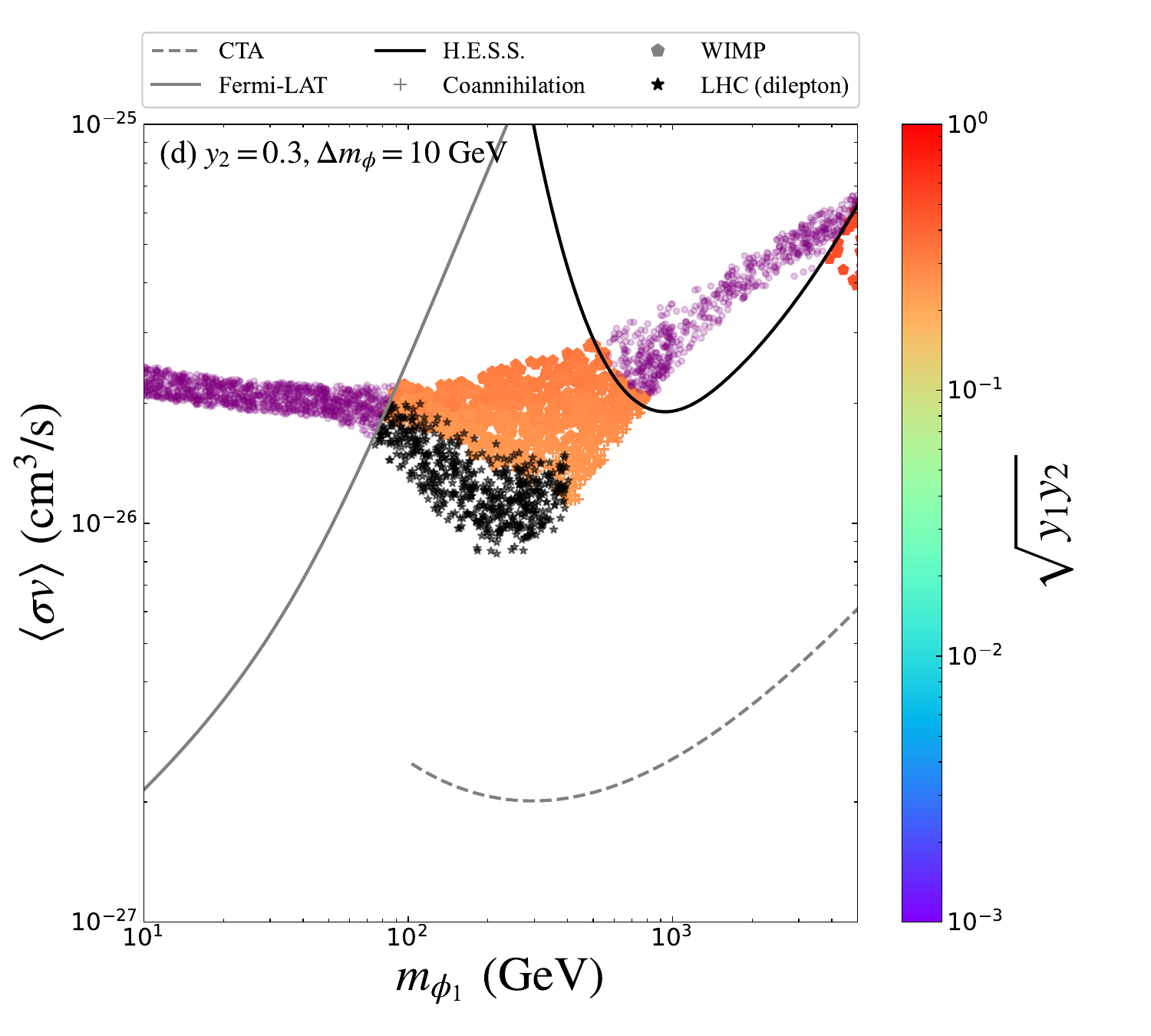}
	\end{center}
	\caption{Constraints from the indirect detection experiments in the Yukawa portal scenario. The gray solid, black solid, and gray dashed lines stand for the Fermi-LAT \cite{Fermi-LAT:2015att}, H.E.S.S.~\cite{HESS:2016mib}, and future CTA \cite{CTA:2020qlo}  limit on the $\tau^+\tau^-$ final sate. other labels are the same as Figure \ref{FIG:fig9}.
	}
	\label{FIG:fig10}
\end{figure}

Another important aspect is the indirect detection of dark matter. The primary annihilation products of DM are charged leptons and neutrinos in the Yukawa portal scenario. The conventional WIMP dark matter mainly annihilates into third generation leptons under the constraints from LFV \cite{Vicente:2014wga}. Therefore, we focus on experimental limits on the $\tau^+\tau^-$ final state in this section. The theoretically predicted annihilation cross section $\left<\sigma v\right>$ and experimental constraints are shown in Figure~\ref{FIG:fig10}. Panel (a) is the results with $y_2=1$ and $\Delta m_\phi=1$ GeV. As already shown in panel (a) of Figure \ref{FIG:fig9}, coscattering samples have relatively small coupling $y_1\lesssim6\times10^{-4}$, thus the pair annihilation of dark matter at present is heavily suppressed. The typical annihilation cross section of coscattering is less than $\mathcal{O}(10^{-37})~\rm{cm^3/s}$, which is far beyond the scope of even future CTA reach. A few coannihilation samples with $y_1\gtrsim0.2$ are excluded by the current Fermi-LAT limit. In panel (b) with $\Delta m_\phi$ increased to 10 GeV, the annihilation cross section of light WIMP below 100 GeV is about $2\times10^{-26}~\rm{cm^3/s}$, hence most of the WIMP samples are disallowed by Fermi-LAT. The projected  CTA could detect coannihilation samples with relatively large annihilation around $m_{\phi_1}\lesssim200$~GeV. In panel (c) with $y_2=0.3$ and $\Delta m_\phi=1$ GeV, we find that almost all the WIMP samples are disfavored by current indirect detection limits. In panel (d) with $y_2=0.3$ and $\Delta m_\phi=10$ GeV, certain WIMP and coannihilation samples in the sub-TeV region are still allowed by the current limit. However, these samples are all within the sensitivity of future CTA.

\subsection{Phenomenology of Dark Partner $\phi_2$}\label{SEC:YP2}

There exists a discrepancy with respect to the Higgs portal scenario in the dominant decay of $\phi_2$, which is the totally leptonic decay $\phi_2\to \Psi^\star\bar{\ell}\to\phi_1\bar{\ell}\ell$ in the Yukawa portal scenario. According to Equation \eqref{Eqn:YG2}, the decay width of $\phi_2\to\phi_1\bar{\ell}\ell$ is usually suppressed by the small mass splitting $\Delta m_\phi$ and tiny Yukawa coupling $y_1$ favored by coscattering. Therefore, $\phi_2$ is also long-lived in this scenario, which would lead to observable cosmological and collider signatures.

For the cosmological constraints of BBN, the limitations on $\tau_{\phi_2}$ induced by the leptonic final states are one order of magnitude weaker than those arising from the hadronic final states \cite{Kawasaki:2017bqm}. In our validation, the samples predict $\Omega_{\phi_2} h^2\times f_e\times \epsilon\lesssim\mathcal{O}(10^{-2})$ and $\tau_{\phi_2}\lesssim{O}(10^{3})$ s, which is below the exclusion limits of BBN and CMB \cite{Kawasaki:2017bqm}. The neutrinos from delayed decay will contribute to the effective number of relativistic neutrino species $N_{\rm eff}$.  The current Planck results require $(f_\nu\epsilon~\Omega_{\phi_2}/\Omega_{\phi_1})^2\tau_{\phi_2}\lesssim5\times10^{9}$~s ~\cite{Hambye:2021moy}, where $f_\nu$ denotes the branching ratio for decay into neutrinos. The scanned samples satisfy $(f_\nu\epsilon~\Omega_{\phi_2}/\Omega_{\phi_1})^2\tau_{\phi_2}\lesssim\mathcal{O}(10)$ s, which is far below the current Planck limit. Taking all factors into consideration, this Yukawa portal scenario holds little promise for being captured in terms of cosmological experiments. Therefore, we omit the predictive figures.

From Figure \ref{FIG:fig9}, it is obvious that coscattering and coannihilation samples require $y_2\gg y_1$, which indicates that the $\Psi\to \phi_2 \ell$ is the dominant decay mode of the dark fermion $\Psi$. Regarding the displaced vertex signature of $\phi_2$ at LHC, we consider the dominant processes as 
\begin{eqnarray}\label{Eqn:Ypp22}
	pp &\to &\psi^+\psi^-\to \ell^+ \phi_2  + \ell^- \phi_2  \to \ell^+ \phi_1 \ell^+\ell^- + \ell^- \phi_1 \ell^+ \ell^-, \\  
	pp&\to & \bar{\psi}^0\psi^0\to \bar{\nu} \phi_2  + \nu \phi_2  \to \bar{\nu} \phi_1 \ell^+\ell^- + \nu \phi_1 \ell^+ \ell^-, \\ 
	pp&\to & \psi^\pm \psi^0 \to \ell^\pm \phi_2 + \nu \phi_2 \to \ell^\pm \phi_1 \ell^+ \ell^- + \nu \phi_1 \ell^+ \ell^-.
\end{eqnarray}

The number of events for the one displaced vertex $N_{\rm DV}$ is calculated as
\begin{eqnarray}
	N_\text{DV} &=& 2 L_\text{int} \times \sigma(pp\to \psi^+\psi^-,\bar{\psi}^0 \psi^0)\times \text{BR}^2_{\Psi\to \phi_2 \ell} \times P_{\rm dec}\times \BR_{\rm vis}\times\kappa_1^{\prime}\\ \nonumber
	& + & 2 L_\text{int} \times \sigma(pp\to \psi^\pm\psi^0)\times \text{BR}^2_{\Psi\to \phi_2 \ell} \times P_{\rm dec}\times \BR_{\rm vis}\times\kappa_2^{\prime}.
\end{eqnarray}
For the coscattering and coannihilation regime, we typically have BR$_{\Psi\to \phi_2 \ell}\simeq1$ as the hierarchy Yukawa coupling $y_2\gg y_1$. The  branching ratio of visible dark partner decay $\phi_2\to \phi_1 \ell^+\ell^-$ depends on the mass splitting when considering the masses of final state leptons, e.g., BR$_\text{vis}\simeq0.175$ for $\Delta m_\phi=1$~GeV and BR$_\text{vis}\simeq0.454$ for $\Delta m_\phi=10$~GeV.  We also assume the detection efficiency $\kappa^{\prime}_{1,2}=1$ for an optimistic estimation. The sensitive region is also derived with $N_{\rm DV}=3$ for vanishing background.

\begin{figure}
	\begin{center}
		\includegraphics[width=0.45\linewidth]{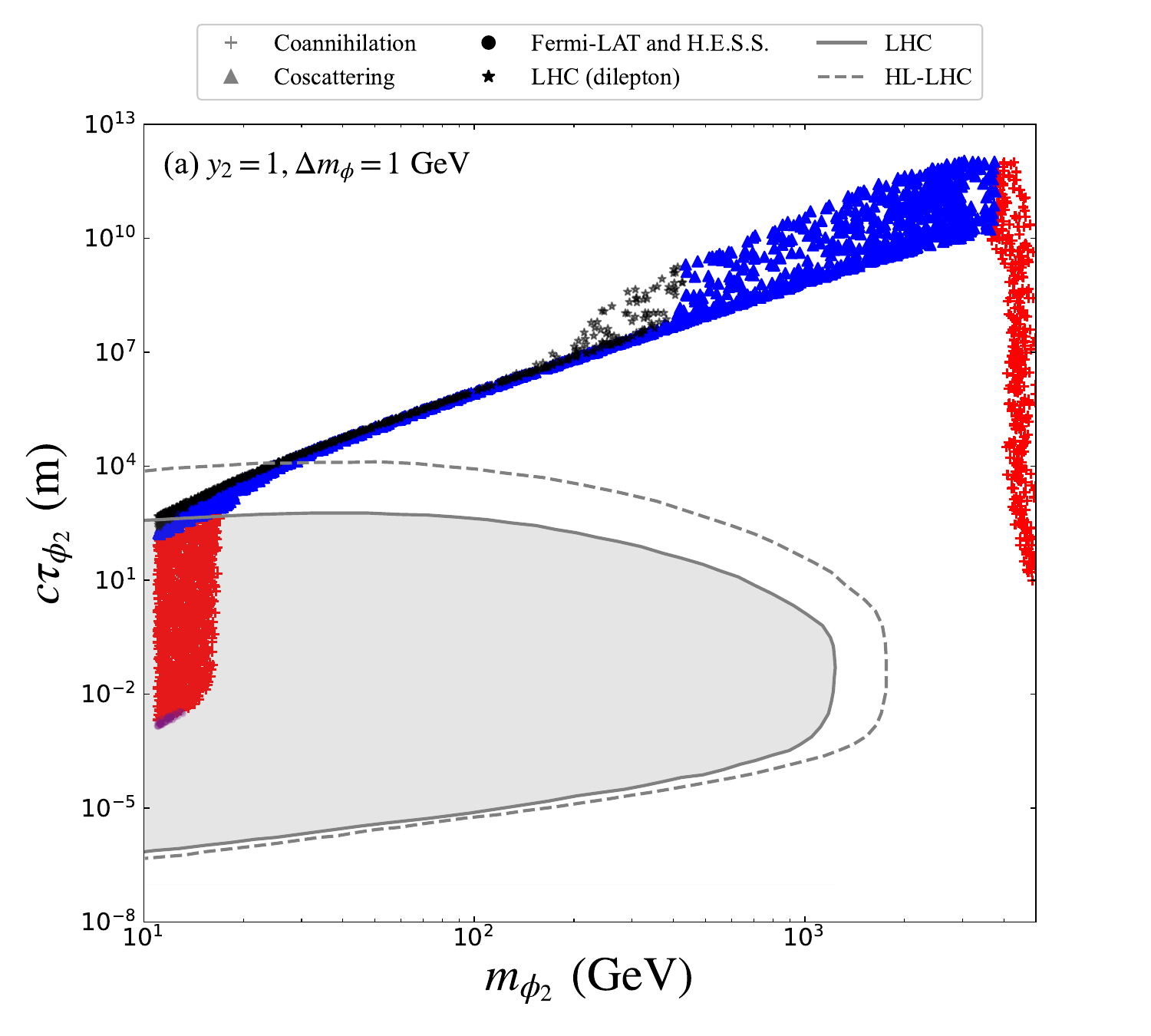}
		\includegraphics[width=0.45\linewidth]{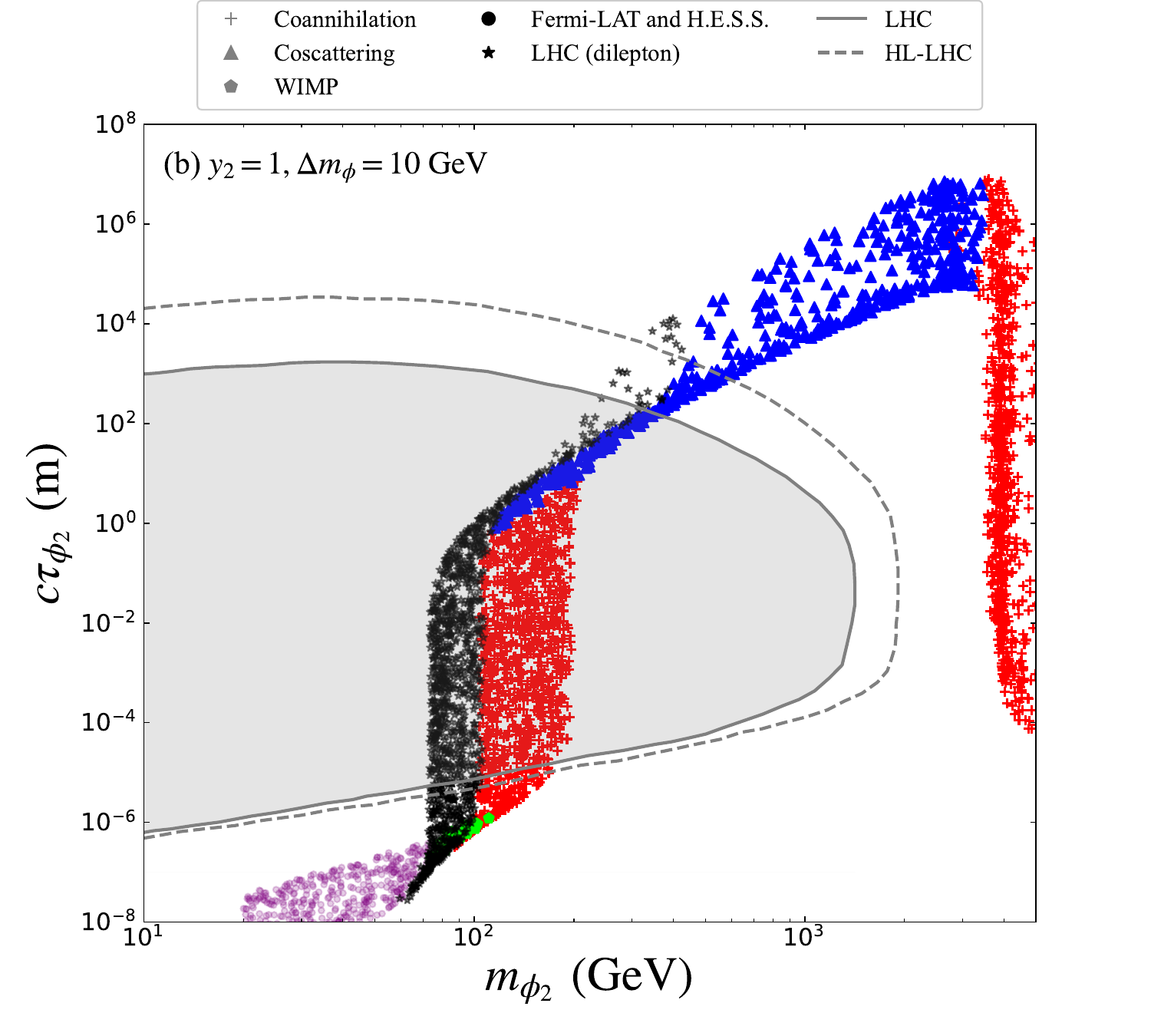}
		\includegraphics[width=0.45\linewidth]{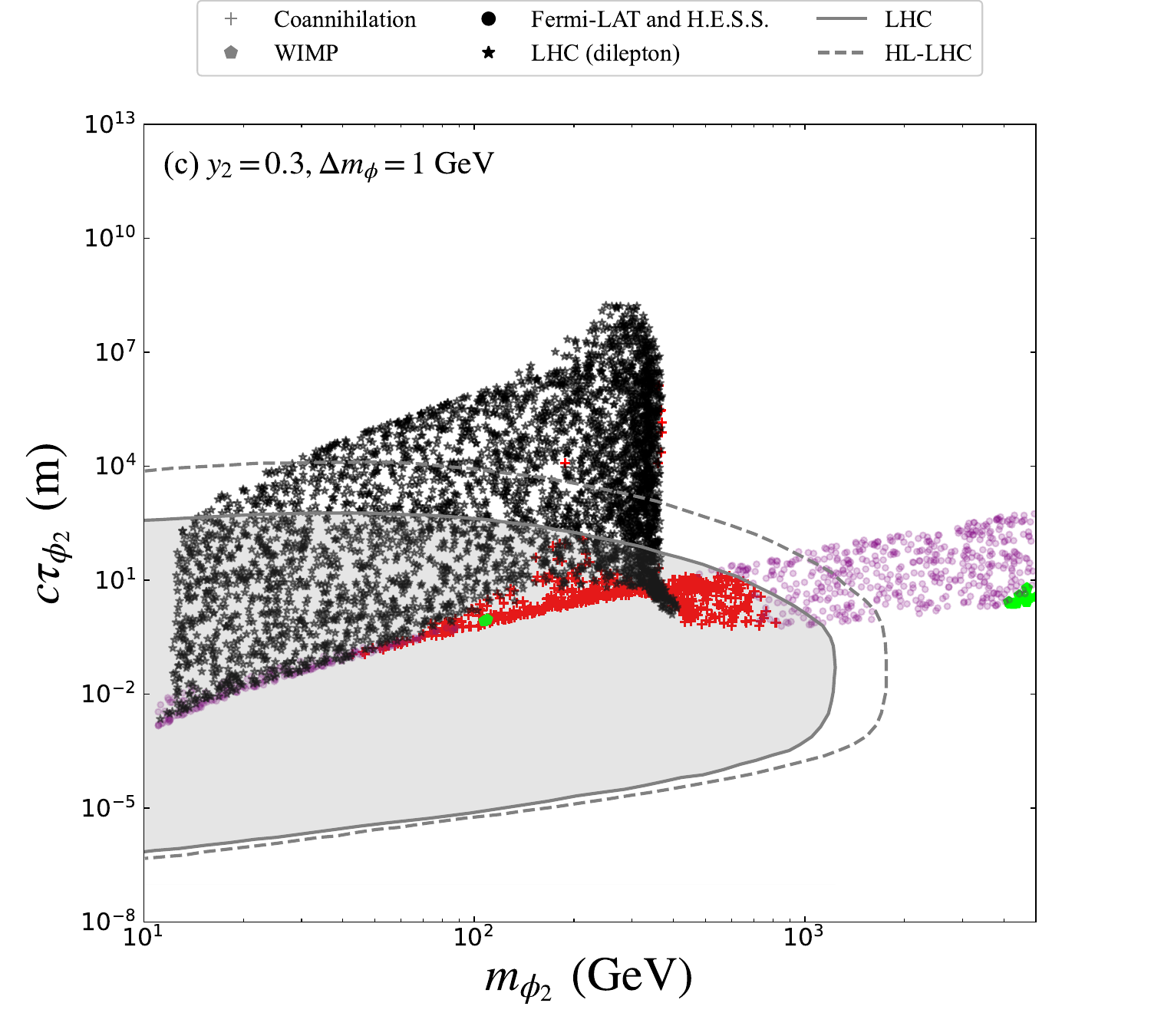}
		\includegraphics[width=0.45\linewidth]{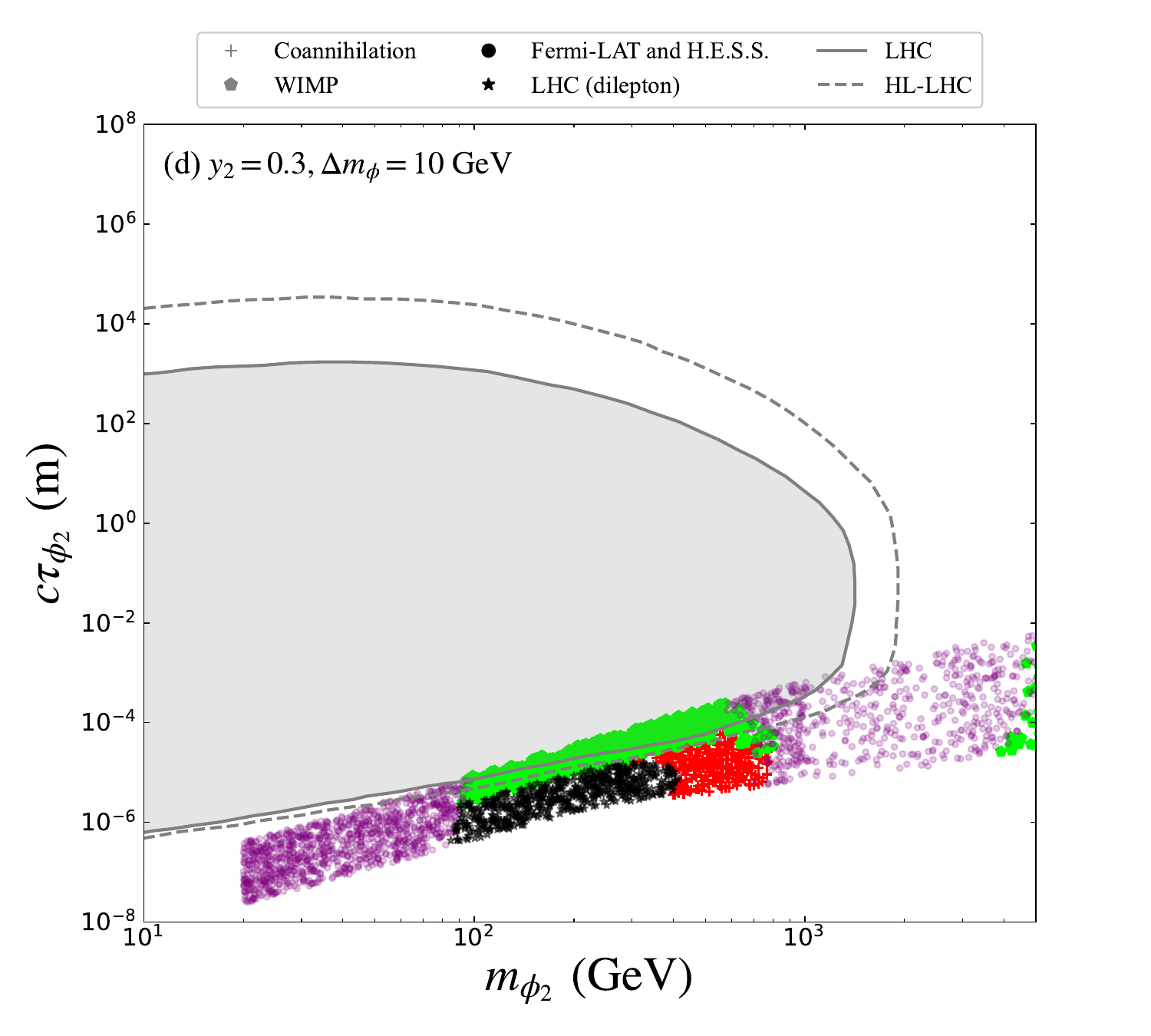}
	\end{center}
	\caption{The sensitive region of the one DV signature at LHC (gray solid lines) and future HL-LHC (gray dashed lines) in the Yukawa portal scenario. To facilitate readability, coannihilation, coscattering and WIMP are highlighted in red, blue and green, respectively. Other labels of samples are consistent with those shown in Figure~\ref{FIG:fig9}.
	}
	\label{FIG:fig11}
\end{figure}

The predicted results are shown in Figure~\ref{FIG:fig11}, where the one DV still exhibits the highest sensitivity with $\Delta m_F =100$ GeV. In principle, LHC could probe $m_{\phi_2}\lesssim$ 1 TeV via the DV signature. In panel (a) with $y_2=1$ and $\Delta m_\phi=1$ GeV,  all samples are distributed in $c\tau_{\phi_2}\in[10^{-3},10^{12}]$ m. Those above  $c\tau_{\phi_2}\sim \mathcal{O}(10^2)$~m are dominated by coscattering, while below is coannihilation. The current LHC is capable of capturing nearly all light coannihilation samples. The future HL-LHC can detect a small part of the light coscattering sample with the corresponding $c\tau_{\phi_2}$ not exceeding $10^4$ m. The decay length is reduced by increasing the mass splitting $\Delta m_\phi$ to 10 GeV in panel (b). With too large Yukawa coupling of $y_1$, the decay length of WIMP is typically less than $c\tau_{\phi_2}\lesssim \mathcal{O} (10^{-6})$ m, thus $\phi_2$ decays promptly in the detector. DV signatures from coannihilation and coscattering samples are promising at LHC with $m_{\phi_{2}}\lesssim400$ GeV. The future HL-LHC could extend the upper limit to $m_{\phi_{2}}\sim600$ GeV. In panel (c) with $y_2=0.3$, the allowed  coannihilation samples around the electroweak scale are within the reach of LHC. In panel (d), the sensitive region of HL-LHC covers part of the WIMP regime below 1 TeV.

\subsection{Phenomenology of Dark Fermion $\Psi$}\label{SEC:YPF}

\begin{figure}
	\begin{center}
		\includegraphics[width=0.45\linewidth]{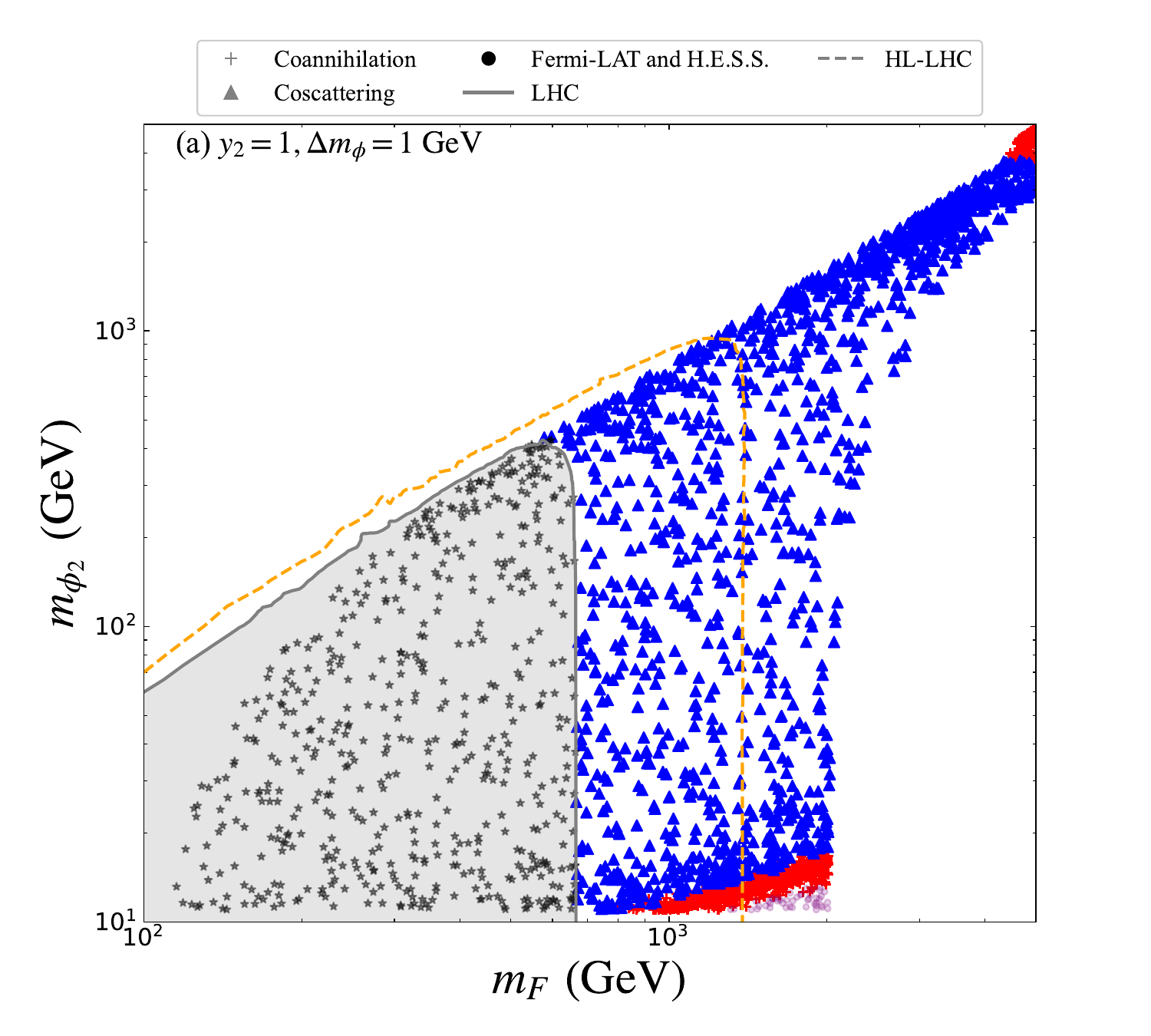}
		\includegraphics[width=0.45\linewidth]{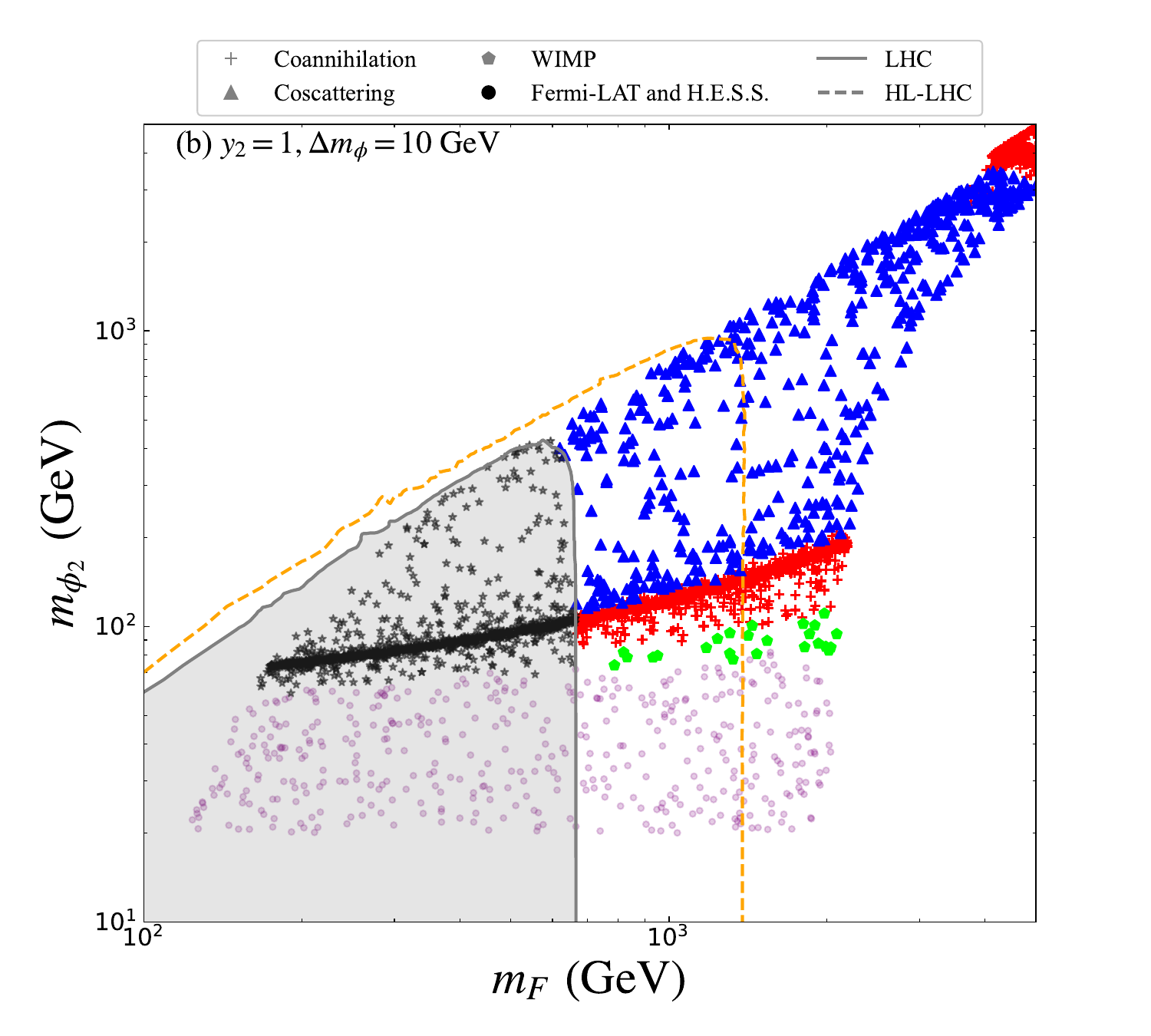}
		\includegraphics[width=0.45\linewidth]{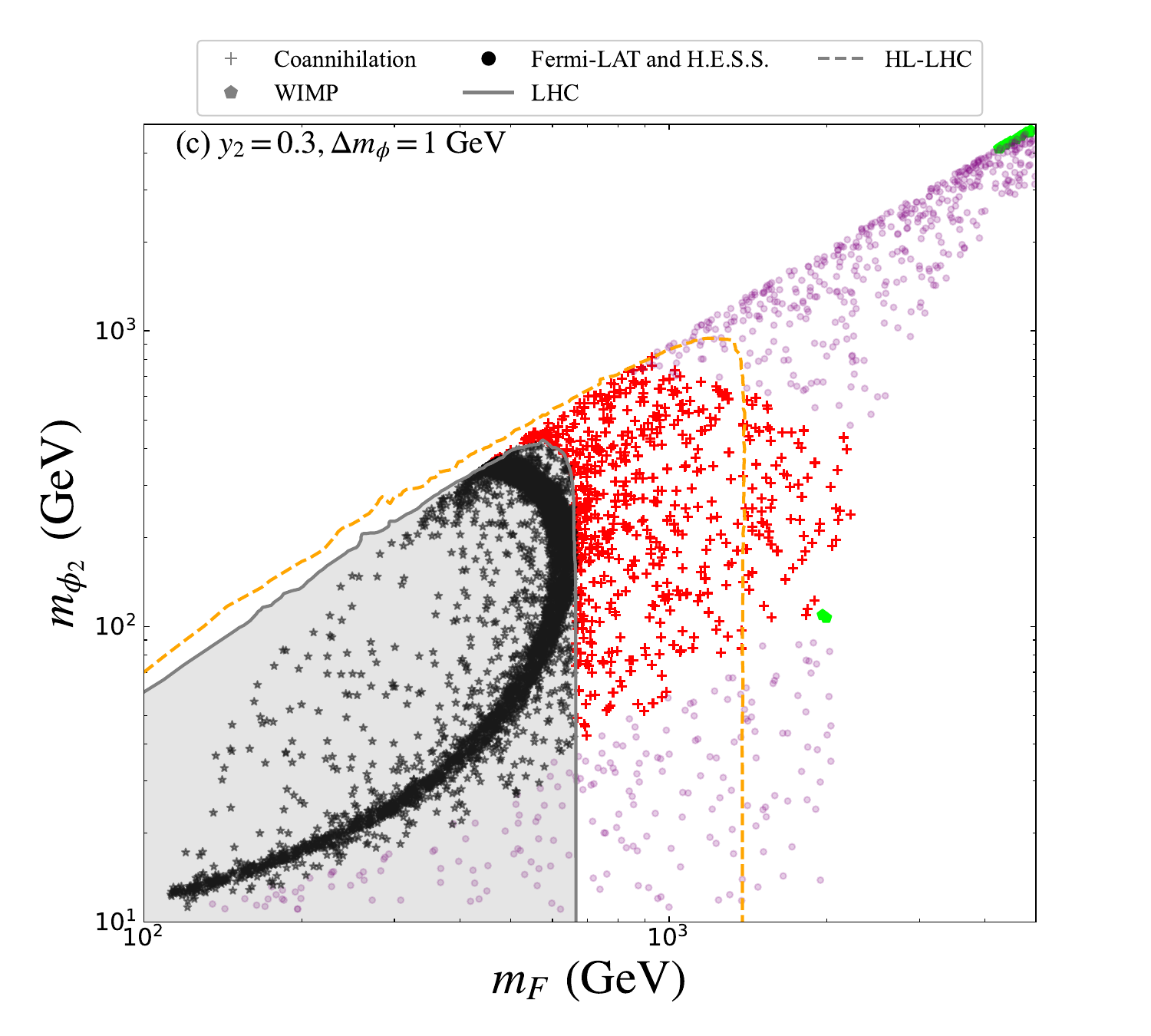}
		\includegraphics[width=0.45\linewidth]{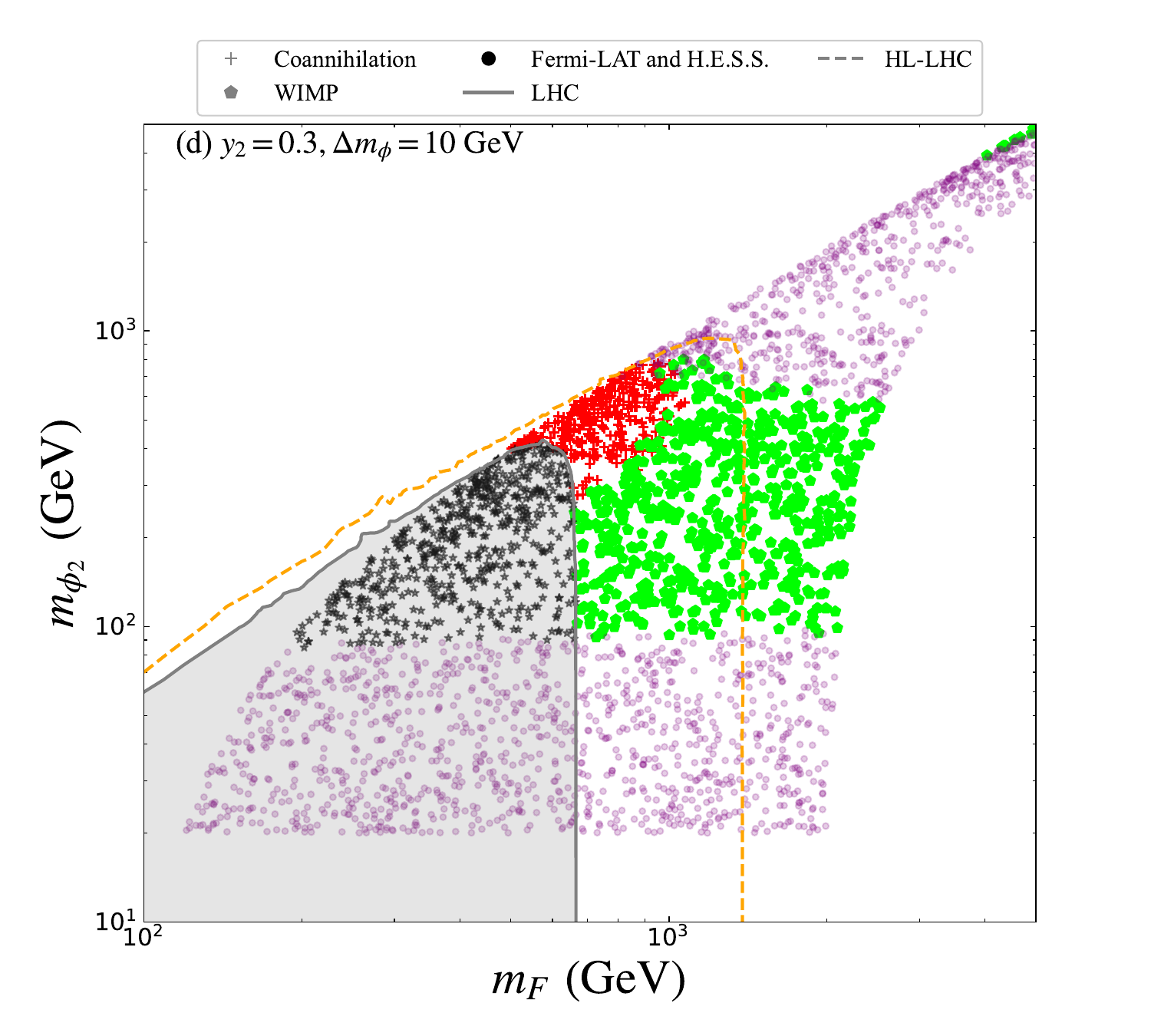}
	\end{center}
	\caption{The sensitivities of dilepton signature $\ell^+\ell^-+\cancel{E}_T$ at LHC in the Yukawa portal scenario. The solid and dashed gray lines are the current LHC and future HL-LHC exclusion limits \cite{Das:2020hpd}, respectively. The labelings of samples are consistent with those shown in Figure~\ref{FIG:fig11}.
	}
	\label{FIG:fig12}
\end{figure}

The neutral fermion $\psi^0$ mediates the decay mode $\phi_2\to \phi_1 \bar{\nu}\nu$, which is invisible at colliders. As shown in Figure \ref{FIG:fig11}, the decay length of most coscattering samples is quite large, so that the decay mode $\phi_2\to \phi_1 \ell^+\ell^-$ is also invisible at LHC. In this case, the promising signature becomes $pp\to \psi^+\psi^-\to \ell^+\phi_2+\ell^-\phi_2\to \ell^+\ell^-+\cancel{E}_T$. In Figure \ref{FIG:fig12}, we show the sensitive region of the dilepton signature $\ell^+\ell^-+\cancel{E}_T$ at LHC by assuming $\phi_2$ is totally invisible. Currently, LHC has excluded the region with $m_F\lesssim660$ GeV and $m_{\phi_2}\lesssim 400$ GeV. In the future, the HL-LHC could probe the region with $m_F\lesssim1400$ GeV and $m_{\phi_2}\lesssim 900$~GeV.

In panel (a) of Figure~\ref{FIG:fig12} with $y_2=1$ and $\Delta \phi=1$ GeV, we report that the coscattering samples with $M_F\lesssim660$ GeV are already excluded by the current LHC. In the HL-LHC sensitive region, most samples are coscattering and a few samples are coannihilation with $m_{F}\gtrsim 1$ TeV  and $m_{\phi_2}\sim \mathcal{O}(10)$ GeV. In the subsequent panel (b) with $\Delta m_\phi=10$ GeV, the HL-LHC could probe coscattering and coannihilation samples with $m_{\phi_2}\gtrsim80$~GeV, since the light WIMP samples are already excluded by indirect detection. In panel (c) with $y_2=0.3$ and $\Delta \phi=1$ GeV, we find that the coscattering samples require $m_F\lesssim600$ GeV, which is already excluded by the current LHC. So the future HL-LHC is only promising to test coannihilation samples. Finally, in panel (d) with $y_2=0.3$ and $\Delta m_\phi=10$ GeV, HL-LHC can capture all coannihilation as well as most of the WIMP samples.

 When the $\tau_{\phi_2}$ is very small which corresponds to certain coannihilation and WIMP  samples in Figure~\ref{FIG:fig11}, the prompt decay $\phi_2\to \ell^+\ell^- \phi_1$ leads to multilepton signatures from Equation \eqref{Eqn:Ypp22} in colliders. With small mass splitting $\Delta m_\phi<10$ GeV for the benchmark scenarios in this paper, the leptons from $\phi_2\to \ell^+\ell^- \phi_1$ are soft. The search results at LHC restrict $m_{\psi^\pm}\sim m_{\phi_2}$ smaller than 250 GeV in the multi soft lepton channel \cite{ATLAS:2019lng,CMS:2024gyw}. On the other hand, when the mass splitting is large enough, i.e., $\Delta m_{\phi} > $ 10 GeV, the multilepton signature could exclude $m_{\psi^\pm}\sim m_{\phi_2}<1200$ GeV \cite{ATLAS:2021yyr}, which is clearly out of the parameter space considered in this study. For the dark scalar $\phi_3$, more leptons in the signatures  are possible form the cascade decay chain $\psi^\pm \to \ell^\pm \phi_3 \to \ell^\pm \phi_2 \ell^+\ell^- \to \ell^\pm \phi_1 \ell^+\ell^- \ell^+\ell^- $ when $m_{\phi_3}<m_F$. For simplicity, we assume $m_{\phi_3} > m_F$, thus $\psi^\pm\to \ell^\pm \phi_3$ is not allowed.

\section{Conclusion} \label{SEC:CL} 

In this paper, we investigate the coscattering mechanism of scalar dark matter in the Scotogenic model,  meanwhile, coannihilation and WIMP regimes are also included for comparison. Within the framework of inverse Scotogenic  seesaw, this model  contains a fermion doublet   $\Psi\equiv(\psi^0,\psi^-)^T$ and singlet  $\chi$, as well as three real singlet scalars $\phi_{i}(i=1,2,3)$. Considering that all new particles are odd under $Z_2$ symmetry, the lightest $\phi_1$ can serve as a DM candidate. We consider nearly degenerate dark scalars $m_{\phi_1}\lesssim m_{\phi_2}$, so that the coscattering process is dominated by the dark partner $\phi_2$ through the Higgs portal or Yukawa portal interactions.

For the Higgs portal scenario, in order to provide a clearer illustration, we categorize the analysis into four distinct cases: (a) $\lambda_2=1$ and $\Delta{m_\phi}=1$ GeV, (b) $\lambda_2=1$ and $\Delta{m_\phi}=10$ GeV, (c) $\lambda_2=0.1$ and $\Delta{m_\phi}=1$ GeV, (d) $\lambda_2=0.1$ and $\Delta{m_\phi}=10$ GeV. Qualitatively speaking, coscattering favors small mass splitting $\Delta m_\phi$ and large coupling $\lambda_2$. Coscattering mainly exists in cases (a) and (b), which favors the parameter spaces with $\lambda_{12}\sim\mathcal{O}(10^{-4})$ and $m_{\phi_1}\lesssim$ 1.1 TeV. The upcoming direct detection experiment DARWIN could test the coscattering samples at the electroweak scale  as well as the TeV coannihilation points in these two cases. Cases (c) and (d) are dominated by coannihilation, which have a broad mass range from dozens of GeV to TeV of $m_{\phi_1}$. Such coannihilation samples with $\lambda_1\gtrsim10^{-3}$ are also within the reach of DARWIN. Under the stringent constraint from LZ, the allowed samples are not promising for indirect detection experiments.  For the search of long-lived $\phi_2$, it is difficult to discover permissible samples for the future CMB S4 experiment.  Consequently, redirecting hope towards  the DV search at LHC and HL-LHC is essential. We find that LHC is sensitive to coannihilation in cases (b) and (d), and future HL-LHC  is expected to capture the light coscattering regime.

In the Yukawa portal scenario, we also consider four options: (a) $y_2=1$ and $\Delta{m_\phi}=1$ GeV, (b) $y_2=1$ and $\Delta{m_\phi}=10$ GeV, (c) $y_2=0.3$ and $\Delta{m_\phi}=1$ GeV, (d) $y_2=0.3$ and $\Delta{m_\phi}=10$ GeV. Coscattering favors a large Yukawa coupling $y_2$ with a small Yukawa coupling $y_1$ in this scenario. To avoid tight constraints from LFV, a hierarchical Yukawa structure as $|y_{ie}|\ll |y_{i\mu}|\lesssim |y_{i\tau}|\sim \mathcal{O}(1)$ is required. Cases (a), (b) and (c) are favored by coscattering. The corresponding $y_1$ is mostly below $\mathcal{O}(10^{-4})$. However, there is a significant variation in the $m_{\phi_1}$ distribution. Notably, case (a) exhibits a broad range of $m_{\phi_1}$, while case (b) and (c) only meet $m_{\phi_1}\gtrsim 70$ GeV and $m_{\phi_1}\lesssim 360$ GeV, respectively. Under various current constraints, we find that indirect detection  is sensitive only to coannihilation and WIMP. The one DV signature is promising for the coannihilation samples. Meanwhile, the dilepton signature $\ell^+\ell^-+\cancel{E}_T$ has excluded $m_F\lesssim660$ GeV, which totally excludes the coscattering region in case (c). 

In summary, both the Higgs portal and Yukawa portal can achieve the coscattering dark matter. With relatively small coupling $\lambda_1$ or $y_1$, the coscattering samples can naturally satisfy the constraints from dark matter detection. Meanwhile, we can probe the coscattering regime with DV signature from long-lived $\phi_2$ decay for not too small mass splitting. These two scenarios can be easily distinguished because hadronic decay $\phi_2\to q\bar{q}\phi_1$ is the dominant mode in the Higgs portal scenario, and leptonic decay $\phi_2\to \ell^+\ell^-\phi_1$ is the dominant one in the Yukawa portal scenario.

\section*{Acknowledgments}
This work is supported by the National Natural Science Foundation of China under Grant No. 12125503, No. 12305103,  No. 12375074 and No. 12505112, Natural Science Foundation of Shandong Province under Grant No. ZR2024QA138, and State Key Laboratory of Dark Matter Physics.

%%%%%%%%%%%%%%%%%%%%%%%%%%%%%

\end{document}